\documentclass[aps, pra, amsfonts, amssymb, amsmath, twocolumn]{revtex4-1}

\pdfoutput=1 

\usepackage{amsfonts}
\usepackage{amssymb}
\usepackage{amsmath}
\usepackage{hyperref}           
\hypersetup{colorlinks = True, 
            allcolors = {blue}
}
\usepackage{mathtools}
\usepackage{natbib}
\usepackage{relsize}            
\usepackage{qcircuit}           
\usepackage{graphicx}                  
\usepackage[caption=false]{subfig}     
\usepackage{soul}
\usepackage{multirow}           

\newcommand{\bra}[1]{ \langle{#1}| }
\newcommand{\ket}[1]{ |{#1}\rangle }

\def\rVac{ |\text{vac} \rangle }
\def\rsVac{ | \mathbf{0} \rangle }

\def\rAGP{|\text{AGP}\rangle}
\def\lAGP{\langle \text{AGP} |}

\def\AGPnorm{\langle \text{AGP}|\text{AGP} \rangle}
\def\rBCS{|\text{BCS}\rangle}
\def\lBCS{\langle \text{BCS} |}

\newcommand{\adag}[1]{ {a}^{\dagger}_{#1} }
\newcommand{\an}[1]{{a}_{#1}}
\newcommand{\cdag}[1]{ {c}^{\dagger}_{#1} }
\newcommand{\cn}[1]{{c}_{#1}}

\def\GamD{ \mathlarger{{{\Gamma}}}^{\dagger} }

\newcommand{\Pdag}[1]{\mathbf{P}^{\dagger}_{#1} }
\newcommand{\N}[1]{ \mathbf{N}_{#1} }
\newcommand{\Ntot}[1]{\mathbf{N}}
\newcommand{\Pp}[1]{ \mathbf{P}_{#1} }

\newcommand{\Sx}[1]{ {X}_{#1} }
\newcommand{\Sy}[1]{ {Y}_{#1} }
\newcommand{\Sz}[1]{ {Z}_{#1} }

\newcommand{\E}[1]{\langle {#1} \rangle}
\newcommand{\Eagp}[1]{\lAGP {#1} \rAGP}

\newcommand{\Ebcs}[1]{\lBCS {#1} \rBCS}

\newcommand{\Proj}{\mathcal{P}_N}

\newcommand{\icol}[1]
    {\left(\begin{smallmatrix}#1\end{smallmatrix}\right)}

\newcommand{\bigO}[1]{$\mathcal{O}(#1)$}

\newcommand{\Ham}{H} 

\newcommand{\prob}[1]{P(#1)}
\DeclarePairedDelimiter\ceil{\lceil}{\rceil}

\newcommand{\Eq}[1]{Eq.~({#1})}
\newcommand{\Sec}[1]{Sec.~{#1}}
\newcommand{\Fig}[1]{Fig.~{(#1)}}
\newcommand{\Reference}[1]{Ref.~{#1}}
\newcommand{\Appx}[1]{Appendix~{#1}}

\newcommand{\ansatz}{\textit{Ansatz}~}
\newcommand{\ansatze}{\textit{Ans\"atze}~}
\newcommand{\saa}{\ensuremath{^{\alpha\alpha}}}
\newcommand{\sab}{\ensuremath{^{\alpha\beta}}}
\newcommand{\sbb}{\ensuremath{^{\beta\beta}}}

\begin{document}

\title{AGP-based unitary coupled cluster theory for quantum computers}
\author{Armin Khamoshi}
    \email{armin.khamoshi@rice.edu}
    \affiliation{Department of Physics and Astronomy, Rice University, Houston, Texas 77005, USA}

\author{Guo P. Chen}
    \email{guo.chen@rice.edu}
    \affiliation{Department of Chemistry, Rice University, Houston, Texas 77005, USA}
    
\author{Francesco A. Evangelista}
    \affiliation{Department of Chemistry and Cherry Emerson Center for
      Scientific Computation, Emory University, Atlanta, Georgia, 30322, USA}

\author{Gustavo E. Scuseria}
    \email{guscus@rice.edu}
    \affiliation{Department of Chemistry, Rice University, Houston, Texas 77005, USA}
    \affiliation{Department of Physics and Astronomy, Rice University, Houston, Texas 77005, USA}

\date{\today}

\begin{abstract} 
Electronic structure methods typically benefit from symmetry breaking and restoration, specially in the strong correlation regime. The same goes for  \textit{Ans\"atze} on a quantum computer. We develop a unitary coupled cluster method based on the antisymmetrized geminal power (AGP)---a state formally equivalent to the number-projected Bardeen--Cooper--Schrieffer wavefunction. We demonstrate our method for the single-band Fermi--Hubbard Hamiltonian in one and two dimensions. We also explore post-selection as a state preparation step to obtain correlated AGP and prove that it scales no worse than $\mathcal{O}(\sqrt{M})$ in the number of measurements, thereby making it a less expensive alternative to gauge integration to restore particle number symmetry. 
\end{abstract}


\maketitle

\section{Introduction} 

One of the most sought after applications of quantum computers is to solve the
strong correlation problem in electronic structure theory 
\cite{cao_quantum_2019, mcardle_quantum_2020}. 
Electronic
structure methods often start from a mean-field state, typically a 
Hartree--Fock (HF) Slater determinant, and are systematically improved using a
variety of techniques \cite{helgaker_molecular_2000}. When electronic 
correlation is weak, single reference coupled cluster theory (CC) 
is known to be sufficiently 
accurate in quantum chemistry and is often regarded as the 
\textit{gold-standard}
\cite{scuseria_closedshell_1987, bartlett_coupled-cluster_2007}. 
However, in the strongly correlated regime, many Slater determinants become 
equally dominant and single-reference methods often struggle 
\cite{bulik_can_2015}.

Strong correlation in finite systems typically manifest itself by breaking one 
or more symmetries at the mean-field level.
In this case, one could either make the mean-field state retain desired
symmetries and correlate it in a \textit{symmetry-adapted} manner, or, one could
allow the symmetries to break and correlate the \textit{symmetry-broken}
mean-field state \cite{ring_nuclear_1980, blaizot_quantum_1986}.
The trade-off between these two strategies is that the broken symmetry 
wavefunction gives a lower energy at
the expense of a less accurate wavefunction due to the 
lack of correct symmetries. This is known as the \textit{symmetry dilemma} 
in electronic structure theory \cite{lowdin_quantum_1955,
lykos_discussion_1963, scuseria_projected_2011,
jimenez-hoyos_projected_2012}
and extends to traditional and unitary coupled
cluster theory alike \cite{evangelista_alternative_2011,
duguet_symmetry_2015, qiu_projected_2017}.

In recent years, there has been extensive interest in using physically
inspired \ansatze in variational quantum algorithms 
\cite{cerezo_variational_2021}. In electronic structure
calculations, the disentangled form of unitary CC (uCC) is perhaps the most 
natural extension of physically inspired \ansatze that can be implemented on a 
quantum computer 
\cite{shen_quantum_2017, tilly_variational_2021, anand_quantum_2022}. 
For example, one can write down the uCC \ansatz as
\begin{align}\label{eq:d-uCC}
    \ket{\psi(\vec{t})}
    = \prod_k e^{t_k \left(\hat{\mathlarger{\tau}}_k - \hat{\mathlarger{\tau}}^
    {\dagger}_k \right)} \ket{\phi},
\end{align}
where $\ket{\phi}$ is a mean-filed reference, $\hat{\mathlarger{\tau}}_k$ is 
the $k{\text{-th}}$ excitation operator which can be 
symmetry-adapted or broken, and $t_k$ is the corresponding amplitude
\cite{evangelista_exact_2019, grimsley_is_2020, izmaylov_order_2020}.
We can variationally minimize the energy,
\begin{align}
    E(\vec{t}) = \bra{\psi(\vec{t})} \Ham \ket{\psi(\vec{t})},
\end{align}
using the variationally quantum
eigensolver (VQE) in a hybrid quantum--classical optimization manner
\cite{peruzzo_variational_2014, mcclean_theory_2016}. 
To address the ambiguity in the ordering of
the operators and the depth of the uCC \textit{Ansatz}, 
heuristics such as the ADAPT-VQE \cite{grimsley_adaptive_2019}
have been developed and extended over the years 
\cite{tang_qubit-adapt-vqe_2021, grimsley_adapt-vqe_2022, romero_solving_2022}. 
Other notable works along this
direction are the qubit coupled cluster approach \cite{ryabinkin_qubit_2018, ryabinkin_iterative_2020},
cluster-Jastrow \ansatz 
\cite{matsuzawa_jastrow-type_2020}, k-UpCCGSD 
\cite{lee_generalized_2019}, and the projective quantum
eigensolver (PQE) \cite{stair_simulating_2021} among others
\cite{kandala_hardware-efficient_2017,
dallaire-demers_low-depth_2019, 
izmaylov_order_2020, xia_qubit_2020, 
anselmetti_local_2021}.
Combining VQE \ansatze with quantum Monte Carlo has been recently proposed as 
well \cite{huggins_unbiasing_2022}.

An important consideration among all methods 
for noisy intermediate--scale 
quantum (NISQ) devices is how deep an \ansatz needs to be in order 
to achieve sufficient accuracy 
\cite{preskill_quantum_2018, leymann_bitter_2020} 
in the strongly correlated regime. 
For example, when using a symmetry-adapted \textit{Ansatz}, it is conceivable 
that more collective excitations are needed to recover the same level of
accuracy in energy as those of symmetry-broken \ansatze in the regimes where 
symmetries break. This could require a larger number of variational parameters 
and deeper circuits to implement, which, in some cases, could scale 
exponentially. 

A remedy for this problem could be to combine symmetry restoration with CC 
theory 
\cite{duguet_symmetry_2015, 
qiu_projected_2017, 
duguet_symmetry_2017, 
wahlen-strothman_merging_2017,
qiu_particle-number_2019, 
song_power_2022}.
Symmetry breaking and restoration has long been studied in nuclear physics 
and electronic structure theory \cite{ring_nuclear_1980, 
scuseria_projected_2011, jimenez-hoyos_projected_2012, 
sheikh_symmetry_2021}.
It has also gained attention for applications on quantum computers in recent 
years. Restoration of parity, $S^2$, $S_z$, 
translational, and other symmetries on a quantum computer have been proposed 
\cite{izmaylov_construction_2019, yen_exact_2019, 
tsuchimochi_spin-projection_2020, seki_symmetry-adapted_2020, 
lacroix_symmetry-assisted_2020, siwach_filtering_2021, seki_spatial_2022,
ruiz_guzman_accessing_2022}.
Some of the 
authors of this paper have shown how to efficiently restore particle number 
symmetry over the BCS wavefunction on a quantum computer
\cite{khamoshi_correlating_2020}.
The projected BCS wavefunction is known as the antisymmetrized geminal power 
(AGP) whose claim to fame is its ability to capture off-diagonal long-range 
order (ODLRO) without breaking number symmetry 
\cite{yang_concept_1962, coleman_structure_1965, sager_cooper-pair_2022}.
Common electronic structure methods, including HF and density 
functional theory, fail to capture ODLRO. Combining uCC with AGP is the central 
goal of this paper.

Number symmetry often breaks spontaneously for systems where the effective 
two-body interaction is attractive.
The prime example is superconductivity or the superfluid phase in condensed 
matter and nuclear physics 
\cite{bardeen_theory_1957, brink_nuclear_2005, sedrakian_superfluidity_2019}. 
Number symmetry does not typically break in molecules 
\cite{bach_generalized_1994}. However the AGP 
wavefunction has a long history in quantum chemistry as well, and it is 
connected to the concept of bonding \cite{surjan_introduction_1999}. 
AGP, by itself, is usually insufficient to accurately describe molecular 
systems, but it could be a good starting point for correlated methods such 
as configuration interactions or coupled cluster theory 
\cite{henderson_geminal-based_2019, khamoshi_efficient_2019, 
dutta_geminal_2020, henderson_correlating_2020,  
dutta_construction_2021, khamoshi_exploring_2021}.
AGP-based quantum Monte Carlo methods have been used extensively over the years 
for molecules and solids
\cite{casula_geminal_2003, casula_correlated_2004, wei_reduced_2018, 
genovese_assessing_2019, genovese_nature_2020}. Neuscamman has 
shown that variational Jastrow correlators on AGP is fully size-consistent and 
highly accurate \cite{neuscamman_jastrow_2013, neuscamman_communication_2013}.
Ref. \cite{matsuzawa_jastrow-type_2020} developed a unitary 
analogue of this theory, although the authors were not 
aware of an efficient implementation of AGP on a quantum computer. Other 
geminal-based methods include using Richardson-Gaudin wavefunction
\cite{fecteau_richardson-gaudin_2020, johnson_richardsongaudin_2020, 
johnson_transition_2021, fecteau_reduced_2020, 
fecteau_near-exact_2022, moisset_density_2022},
which although are not done on a quantum computer, they are relevant to AGP.

In Ref. \cite{khamoshi_correlating_2020} we demonstrated an efficient 
correlated AGP method on a quantum
computer in the seniority-zero space. Our goal in this paper is to show how our
previous formalism can be extended beyond seniority-zero systems and onto
general \textit{ab initio} Hamiltonians. To this end, we develop a unitary
coupled cluster \ansatz atop of AGP and benchmark our numerical method against
the ground state of the single-band Fermi--Hubbard model. The Hubbard
Hamiltonian breaks number symmetry when the on-site interaction is
attractive; and, when the on-site interactions is repulsive, it features strong
correlation prototypical of those that occur in molecules. Our intent is to
demonstrate our method in both regimes of this Hamiltonian. 

Another contribution of this paper is to exploit post-measurement selection, or
\textit{post-selection} in short, as an alternative to the $U(1)$ gauge
integral in restoring number symmetry. Post-selection has been widely used as a
crucial error mitigation technique for calculations performed on NISQ
hardware \cite{endo_practical_2018, bonet-monroig_low-cost_2018, 
kandala_error_2019, endo_hybrid_2021, huggins_efficient_2021}.
We show, however, how post-selection can be used in conjunction with
uCC to sample over correlated AGP. Our primary contribution is to show that
restoring number symmetry with post-selection could scale more favorably than 
gauge integration as judged by the circuit depth and number of measurements. 
We analytically prove an asymptotic bound of 
\bigO{\sqrt{M}} measurements and confirm it with 
computations based on a noise-free quantum emulator. 

Restoring symmetries via post-selection is a distinct possibility that is not
viable on classical computers but can be conveniently performed on a quantum
computer. Indeed, post-selection can be combined with the integral or phase
estimation \cite{lacroix_symmetry-assisted_2020} to restore multiple 
symmetries. Our general philosophy
for methods in strong correlation is to allow multiple symmetries to break and
later restore them in the presence of the correlator. This could pave the way
for methods that work equally well in repulsive, attractive, weak, and strong 
correlation regimes.

This paper is organized as follows: We begin \Sec{\ref{sec:theory}}
by first reviewing some previous work and basic concepts, then we build our way 
to seniority nonzero methods in \Sec{\ref{subsec:beyond-sen-zero}}. 
In \Sec{\ref{subsec:agp-optimization}}, we discuss the optimization of AGP 
state on classical and quantum computers and review a formal connection with
number projected HFB. \Sec{\ref{subsec:uCC-agp}} is devoted to disentangled uCC
on AGP. \Sec{\ref{sec:post-selection}} discusses post-selection; the procedure 
is discussed in \Sec{\ref{subsec:ps-procedure}} and the numerical benchmarks 
are presented in \Sec{\ref{subsec:ps-experiment}}. In 
\Sec{\ref{sec:application}}
and the subsections therein, we demonstrate different uCC correlators 
based-on AGP and discuss our results for 
the Hubbard Hamiltonian. We conclude the paper with 
some final remarks in \Sec{\ref{sec:discussions}}. 

\section{Theory} \label{sec:theory}

Define $\GamD$ to be the geminal creation operator,
\begin{align}
    \GamD = \frac{1}{2} \sum_{pq} \eta_{pq} \cdag{p}\cdag{q},
    \label{eq:gemenial_in_ao}
\end{align}
where $\cdag{p}$ is a fermionic creation operator in a spin-orbital $p$ and
$\eta$ is an anti-symmetric matrix whose elements are known as the
\textit{geminal coefficients}.
An AGP with $N$ pairs, or $2N$ electrons, can then be defined as 
\cite{coleman_structure_1965, surjan_introduction_1999}
\begin{align}
    \label{eq:agp}
    \rAGP = \frac{1}{N!} \left( \GamD \right)^{N} \rVac.
\end{align}
Unless otherwise specified, we choose to work in the natural orbital basis of
the geminal wherein the matrix $\eta$ is brought to a
quasi-diagonal form  \cite{Hua_theory_1944},
\begin{align}\label{eq:eta_canonical}
    \bar{\eta}
    = \bigoplus_{p=1}^{M}
    \begin{pmatrix}
        0 & \eta_p \\
        -\eta_p & 0 \\ 
    \end{pmatrix}
    = D^\dag \eta D^*,
\end{align}
via the orbital rotation
\begin{equation}\label{eq:natorb}
  \adag{p} = \sum_{q=1}^{2M} D_{qp} \cdag{q},
\end{equation}
where $D$ is the unitary matrix of natural orbital coefficients and
$M$ is the number of spatial orbitals. As such, the geminal operator can
be written as
\begin{align}\label{eq:gemenial_in_natmo}
    \GamD = \sum_{p=1}^{M} \eta_{p} \adag{p}\adag{\bar{p}},
\end{align}
where the natural orbitals $p$ and $\bar{p}$ are ``paired" 
in the sense of \Eq{\ref{eq:eta_canonical}}.
As shown in \Appx{\ref{sec:bm}}, a $(p,\, \bar{p})$ pair
reduces to a spin $(\uparrow,\, \downarrow)$ pair if the spins of the
natural orbitals are collinear. With this pairing scheme,
the relationship between AGP and the BCS wavefunction can be easily seen from
\begin{subequations}
    \begin{align}
        \rBCS
        &= \prod_{p=1}^{M}(u_{p} + v_{p}\adag{p}\adag{\bar{p}})\rVac\\
        &\propto \prod_{p=1}^{M} (1+\eta_{p}\adag{p}\adag{\bar{p}})\rVac \\
        &= \sum_{N=0}^{M} \frac{1}{N!} \left( \GamD \right)^{N} \rVac,
    \end{align}
\end{subequations}
where the BCS coefficients $\{u_p, v_p\}$ relate to $\{\eta_p\}$ through
$\eta_{p}=v_p/u_p$.
Clearly the BCS wavefunction is a linear combination of AGPs with all possible
number of electron pairs (up to an inconsequential normalization factor).
The AGP of $N$ pairs, \Eq{\ref{eq:agp}}, corresponds to
the projected BCS (PBCS) state with the correct particle number.
The AGP ground state can be optimized
on a classical computer at mean-field cost
(see \Sec{\ref{subsec:agp-optimization}}).

\subsection{Beyond zero seniority}\label{subsec:beyond-sen-zero}
Seniority is defined as the number of unpaired electrons
and is denoted by $\Omega$ \cite{ring_nuclear_1980}.
Any two-body \textit{ab initio} Hamiltonian can be written as 
\cite{henderson_pair_2015}
\begin{align}
     H = H^{\delta{\Omega}=0} + H^{\delta{\Omega}=\pm2} 
     + H^{\delta{\Omega}=\pm4},
\end{align}   
where the superscript differentiates between sectors 
of the Hamiltonian that change seniority by $0,\pm2$ and $\pm4$. 
The seniority-zero subspace features a remarkable simplicity
\cite{bytautas_seniority_2011, 
stein_seniority_2014, 
henderson_seniority-based_2014, 
bytautas_seniority_2015, shepherd_using_2016}, which makes it 
particularly convenient for quantum algorithms 
\cite{khamoshi_correlating_2020,elfving_simulating_2021}.
One can define a $su(2)$ \textit{pairing} algebra with the generators
\begin{subequations} \label{eqs:generators}
    \begin{align}
        \Pdag{p}&= \adag{p}\adag{\bar{p}}, \\
        \N{p}   &= \adag{p}\an{p}+\adag{\bar{p}}\an{\bar{p}},
    \end{align}
\end{subequations}
and posit a \textit{paired} encoding, 
$\Pdag{p} \mapsto \frac{1}{2} (\Sx{p} - i\Sy{p})$ and $\N{p} \mapsto 1-\Sz{p}$,
such that the qubit states $\ket{1}=\icol{0\\1}$ and $\ket{0}=\icol{1\\0}$ 
correspond to an electron-pair being present and absent respectively; here, 
$X, Y$, and $Z$ are Pauli matrices. 
Any seniority-zero Hamiltonian can be encoded as
\begin{align} \label{eq:senzero-ham}
    H^{\delta\Omega=0}
    &\mapsto \sum_{p}h_p\Sz{p}
    +\sum_{pq}w_{pq}\Sz{p}\Sz{q}\nonumber\\
    &\quad+ \sum_{pq} v_{pq}\left(\Sx{p}\Sx{q} + \Sy{p}\Sy{q} \right),
\end{align}
where $w$ and $v$ are Hermitian matrices. For most Hamiltonians, seniority is 
not a good quantum number, and the restriction to the 
seniority-zero subspace leads to approximate energies. 
Indeed, for \textit{ab initio} Hamiltonians, 
doubly-occupied configuration interaction 
\cite{veillard_complete_1967, couty_generalized_1997, kollmar_new_2003} 
(an exact diagonalization method in the seniority-zero subspace) 
is not sufficient to reach chemical accuracy. Therefore it becomes necessary 
to incorporate broken pair excitations.  

\begin{figure*}[t]
    \centering
    \subfloat[][]{
        \Qcircuit @C=2em @R=1.5em {
            \lstick{\ket{0}_{a}}       &\gate{\text{H}}      &\qw      & \ctrl{1}                   &\qw \\
            \lstick{\ket{0}_1}         &\gate{R_y(\theta_1)} &\ctrl{1} & \multigate{3}{{\mathcal{R}_i}} &\qw \\
            \lstick{\ket{0}_{\bar{1}}} &\qw                  &\targ    & \ghost{{\mathcal{R}_i}}        &\qw \\
            \lstick{\ket{0}_2}         &\gate{R_y(\theta_2)} &\ctrl{1} & \ghost{{\mathcal{R}_i}}        &\qw \\
            \lstick{\ket{0}_{\bar{2}}} &\qw                  &\targ    & \ghost{{\mathcal{R}_i}}        &\qw 
            \gategroup{2}{2}{3}{3}{.9em}{--} 
            \gategroup{4}{2}{5}{3}{.9em}{--}            
        }
    }
    \hspace{0.1\textwidth}
    \subfloat[][]{
        \Qcircuit @C=1em @R=1.5em {
            \lstick{{anc.}}    & \ctrl{1}                       &\qw &&          && & \ctrl{1}          &\qswap      &\qw        &\qw                &\qw        &\qw        &\qw        &\qw\\
            \lstick{1}         & \multigate{3}{{\mathcal{R}_i}} &\qw &&          && &\gate{R_z(\phi_i)} &\qswap\qwx  &\qswap     &\qw                &\qw        &\qw        &\qw      &\qw\\
            \lstick{{\bar{1}}} & \ghost{{\mathcal{R}_i}}        &\qw && \implies && &\qw                &\qw         &\qswap\qwx &\ctrl{1}           &\qswap     &\qw        &\qw      &\qw\\
            \lstick{{2}}       & \ghost{{\mathcal{R}_i}}        &\qw &&          && &\qw                &\qw         &\qw        &\gate{R_z(\phi_i)} &\qswap\qwx &\qswap     &\qw      &\qw\\
            \lstick{{\bar{2}}} & \ghost{{\mathcal{R}_i}}        &\qw &&          && &\qw                &\qw         &\qw        &\qw                &\qw        &\qswap\qwx &\qw      &\qw}
    }    
    \caption{Circuits to implement the number projected BCS wavefunction for 
    seniority nonzero applications using the integral. Here it is shown for a 
    system with 4 spin-orbitals ($M=2$) for simplicity. (a) Implementation of 
    the BCS wavefunction and circuit to the do Hadamard test for 
    every grid angle 
    $\phi_i$. The dashed box implements 
    $\exp{\left(i\theta_p \left(\Sx{\bar{p}}\Sx
    {p}+\Sy{\bar{p}}\Sy{p}\right)/2 \right)}$ acting on the vacuum state after simplification. (b) Implementation of the 
    controlled $R_z(\phi_i)$ using nearest neighbors connectivity. 
    Since the BCS orbitals $p$ and
    $\bar{p}$ are either simultaneously present or absent, the 
    controlled $R_z(\phi_i)$ can be chosen to apply only to the ``no-bar" 
    qubits. The swap gates move the logical qubit of the ancilla 
    qubit to the last qubit to measure $\E{X_a} + i\E{Y_a}$ 
    needed for the Hadamard test.
    }
    \label{fig:prep_AGP}
\end{figure*}
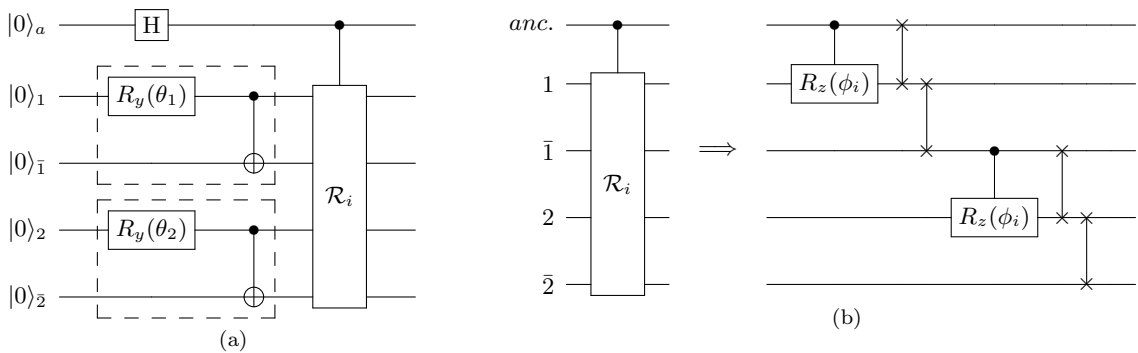

To allow pairs to break, we first need to map individual spin-orbitals to
qubits. Our goal is to follow our formalism 
for seniority-zero \cite{khamoshi_correlating_2020}
closely to take advantage of its low asymptotic scaling and
other desirable properties. We want to implement AGP, again, in 
its natural orbital basis where all electrons are paired. Broken-pair
excitations then come from the correlator that couples the reference to all 
seniority sectors of the Hamiltonian. 

We interlace the qubits associated with natural orbitals
$p$ and $\bar{p}$ (i.e. logical qubits of $p$ ($\bar{p}$) correspond to odd (even) physical qubits), and adopt the Jordan--Wigner (JW) encoding of fermions 
\cite{jordan_uber_1928},
so that $a_{p} \mapsto
\frac{1}{2}\left(\Sx{p}+i\Sy{p}\right)\Sz{\bar{p}-1}\Sz{p-1}
\cdots\Sz{\bar{1}}\Sz{1}$ and
$a_{\bar{p}} \mapsto 
\frac{1}{2}\left(\Sx{\bar{p}}+i\Sy{\bar{p}}\right)\Sz{p}
\Sz{\bar{p}-1}\Sz{p-1}\cdots\Sz{\bar{1}}\Sz{1}$. Note that the 
paired encoding of \Eq{\ref{eqs:generators}} 
is embedded in this mapping. The BCS state can 
be implemented at \bigO{1} depth with $M$ two-qubit gates acting on every
$p,\bar{p}$ qubits \cite{jiang_quantum_2018,lacroix_symmetry-assisted_2020, khamoshi_correlating_2020}, 
\begin{align}\label{eq:bcs-qc}
    \rBCS = \prod_{p=1}^{M} e^{i\theta_p(\Sx{\bar{p}}\Sx{p} +
    \Sy{\bar{p}}\Sy{p})/2} \rsVac,
\end{align}
where we have defined {$\theta_p = 2\arctan(\eta_p)$}. If one chooses 
to obtain AGP from gauge integration, 
\begin{align}
  \rAGP = \frac{1}{2\pi} \int_0^{2\pi} d\phi \:e^{i\phi {(\hat{N}- 2N)} } \rBCS,
\end{align}
where $\hat{N} = \sum_p (\adag{p}\an{p} + \adag{\bar{p}}\an{\bar{p}}),$
we can follow the procedure outlined in 
Ref. \cite{khamoshi_correlating_2020},
whereby we introduce an ancilla qubit to compute grid-point overlaps
$\Ebcs{\hat{O} \mathcal{R}_i}$ via a Hadamard test, 
where $\hat{O}$ is an arbitrary observable. 
For every gauge angle, $\phi_i$, the operator $\mathcal{R}_i$ takes the form
\begin{align}
    \mathcal{R}_i
    = e^{i\phi_i {(\hat{N}- 2N)} }
    \mapsto e^{i \chi_i} \prod_{p=1}^{M} 
    e^{-i \phi_i \Sz{p}{/2}} 
    e^{-i \phi_i \Sz{\bar{p}}{/2}},
\end{align}
where $N$ is the number of pairs, and
$\chi_i$ is a constant phase that can be incorporated on a classical computer. Further simplification is
possible by noting that the state of qubit $p$ must be
identical to its conjugate pair $\bar{p}$ 
{in the BCS wavefunction}. Thus, it
suffices to construct $\mathcal{R}_i$ such that 
it runs either on $p$ or $\bar{p}$ orbitals only. This simplification is only 
possible if the $\mathcal{R}_i$ circuit is
applied immediately after the BCS circuit. A circuit diagram combining the BCS
implementation with number projection is depicted in \Fig{\ref{fig:prep_AGP}}
The depth of the circuit is \bigO{M} and we need \bigO{M} grid point to do the
integration exactly. In \Sec{\ref{sec:post-selection}}, we analyze an
alternative to gauge integration by using post-selection.

\subsection{AGP optimization}\label{subsec:agp-optimization}

Our discussion so far has focused on how to implement AGP when $\{\eta_p\}$ and 
the pairing scheme are given; very little has been said on how they can be 
obtained. Recall that for 
a generic \textit{ab initio} Hamiltonian, 
\begin{align}\label{eq:abinito}
    H
    = \sum_{pq} h_{pq} \cdag{p}\cn{q}
    + \frac{1}{2} \sum_{pqrs} v_{pqrs} \cdag{p}\cdag{q}\cn{s}\cn{r},
\end{align}
the AGP energy is not invariant under orbital rotation, so we need
to optimize the geminal coefficients and the natural orbitals 
together. The latter determines the pairing scheme of the geminal.
In what follows, we discuss two methods to optimize AGP for seniority nonzero 
Hamiltonians.

\subsubsection{Direct optimization}\label{subsec:direct-optimization}

We can always write down the Hamiltonian in the natural orbital basis as
\begin{align}\label{eq:abinito-ham-natorb}
    e^{-\sigma} H e^{\sigma}
    = \sum_{pq} \tilde{h}_{pq}(D) \adag{p}\an{q}
    + \frac{1}{2} \sum_{pqrs} \tilde{v}_{pqrs}(D) \adag{p}\adag{q}\an{s}\an{r},
\end{align}
where $\sigma$ is the number-conserving Thouless operator defined as
\cite{thouless_stability_1960}
\begin{align}\label{eq:thouless-op}
    \sigma
    = \sum_{p>q} [\log(D)]_{pq} \left(\cdag{p}\cn{q} - \cdag{q}\cn{p} \right);
\end{align}
$\tilde{h}(D)$ and $\tilde{v}(D)$ are the one- and two-electron integrals
transformed into the natural orbital basis.
To obtain an optimized AGP, we need to variationally minimize the energy with 
respect to $\{\eta_p\}$ together with $\{D_{pq}\}$, where the overlaps are 
taken with respect the AGP state of the form \Eq{\ref{eq:agp}} and 
\Eq{\ref{eq:gemenial_in_natmo}}. 
On a classical computer, all relevant reduced density 
matrices (RDMs) can be computed efficiently using the \textit{sumESP} algorithm 
and the reconstruction formulas outlined in 
Ref. \cite{khamoshi_correlating_2020}
or using generalized Wick's theorem and one-body BCS transition RDMs 
\cite{balian_nonunitary_1969, dutta_construction_2021}. 

We may also opt for optimizing AGP on a quantum computer. This is useful, for 
example, when we perform reference optimization in the presence of the uCC 
correlator (see \Sec{\ref{subsec:oo-uCC}}). The unitary Thouless rotation, 
$\exp(\sigma)$, can be decomposed into a product of 2-qubit gates using the QR 
decomposition and Givens rotations \cite{kivlichan_quantum_2018}, 
and so, it can be implemented exactly with linear circuit depth. This 
allows us to optimize the reference on a quantum computer without incurring any 
Trotterization error.  All gradients with respect to $\{D_{pq}\}$
and $\{\eta_p\}$ can be computed efficiently by the fermionic shift-rule 
\cite{schuld_evaluating_2019, kottmann_feasible_2021, izmaylov_analytic_2021}. 
Note that the Hamiltonian is measured in the atomic-orbital basis,
\Eq{\ref{eq:abinito}}, since the orbital rotation are absorbed into the 
circuit. Alternatively, it is also possible to work with 
\Eq{\ref{eq:abinito-ham-natorb}}, which is essentially the approach 
taken in Refs. \cite{mizukami_orbital_2020, sokolov_quantum_2020}. However, we 
adopt the former method in this work as it requires considerably fewer terms to 
measure. 

In analogy with the HF wavefunction being classified by different spin-symmetry
restrictions, i.e. restricted (RHF), unrestricted (UHF) , and generalized
(GHF) \cite{helgaker_molecular_2000}, 
we may choose to retain $S^2$ and/or $S_z$ symmetries in AGP. These
symmetries are reflected in the matrix elements of $D$, which we treat as
independent variables during the optimization if we decide to allow the
symmetries to break. Just as in HF theory, RAGP denotes spin-restricted
AGP; UAGP denotes spin-unrestricted AGP (broken $S^2$); and GAGP denotes
generalized AGP (broken $S^2$ and $S_z$).

\subsubsection{Number projected Hartree--Fock--Bogoliubov}

An alternative way to optimize AGP is through number projecting the
Hartree--Fock--Bogoliubov (HFB) wavefunction. The equivalence between
number-projected HFB (NHFB) and AGP with optimized natural orbitals is
guaranteed by the Bloch--Messiah theorem \cite{bloch_canonical_1962},
which states that the Bogoliubov
quasiparticle operators defining an HFB state can be constructed from the
physical fermionic operators through three consecutive Bogoliubov
transformations of special forms: An orbital rotation, a BCS transformation,
and a rotation amongst quasiparticles. 
They are represented by the $D$, $(\bar{U}, \bar{V})$, and $C$
transformations in \Appx{\ref{sub:bmthm}}, respectively.
The last $C$ transformation is not
physically significant since it only alters the global phase. Therefore, the
broken-symmetry HFB state in an NHFB is essentially a BCS state
with optimized orbitals, and
variationally minimizing the PBCS or AGP energy with respect to both
$\{D_{pq}\}$ and $\{\eta_p\}$ amounts to minimizing the NHFB energy in the
variation-after-projection scheme \cite{egido_symmetry_1982,
sheikh_symmetry-projected_2000, scuseria_projected_2011}.

An algorithm for computing the Bloch--Messiah decomposition is outlined
in \Appx{\ref{sub:bmcomp}}. 
This algorithm is used in \Sec{\ref{sec:application}}
to prepare the initial UAGP references, where we extract
the $\{D_{pq}\}$ and $\{\eta_p\}$
from the deformed HFB of the optimized number-projected spin-unrestricted
HFB (NUHFB) wavefunction.

\subsection{Disentangled uCC on AGP}\label{subsec:uCC-agp}
Given an optimized AGP for an \textit{ab initio} Hamiltonian,
we define the singles and doubles anti-Hermitian coupled cluster operators 
\cite{nooijen_can_2000, taube_new_2006}
\begin{subequations}
    \label{eq:T1T2}
    \begin{align}
        T_1 &= \sum_{p>q} t^{p}_{q} \left(\adag{p}\an{q} - h.c.\right), \\
        T_2 &= \sum_{pq>rs} t^{pq}_{rs} \left(\adag{p}\adag{q}\an{s}\an{r} - h.c. \right),
    \end{align}    
\end{subequations}
where the indices run over spin-orbitals in the natural orbital basis.
Higher order excitations can be defined similarly. Just as in HF
theory, the uCC operators can be spin-restricted (uRCC), spin-unrestricted
(uUCC), or spin-general (uGCC) \cite{bartlett_coupled-cluster_2007}. 
For HF-based uCC, the cluster operators may be
chosen over particles and holes only, or they can be general index, i.e. include
hole-hole and particle-particle excitations \cite{nooijen_can_2000}. 
In AGP, the separation between 
particles and holes is not well defined since the wavefunction inherently 
contains all possible ways $N$ pairs occupy $M$ orbitals 
\cite{khamoshi_efficient_2019}. Therefore the general 
correlator is most natural on AGP. (We shall return to defining particle--hole
excitations on AGP later in \Sec{\ref{subsec:oo-uCC}}.)

On a quantum computer, uCC needs to be expressed in terms of elementary gates. 
We resort to using the disentangled uCC \ansatz 
\cite{cao_quantum_2019, evangelista_exact_2019, anand_quantum_2022}, 
as in \Eq{\ref{eq:d-uCC}}. 
Note that the disentangled uCC is number conserving throughout the 
implementation; for example, it is easy to verify that every term in
\begin{align}
    \adag{p}\adag{q}\an{r}\an{s} - h.c. &\mapsto 
    \frac{-i}{8}\Big(Y_pX_qX_rX_s + X_pY_qX_rX_s \nonumber \\
    &-  X_pX_qY_rX_s - X_pX_qX_rY_s \nonumber \\
    &-  X_pY_qY_rY_s - Y_pX_qY_rY_s \nonumber \\
    &+  Y_pY_qX_rY_s + Y_pY_qY_rX_s\Big) \prod_n Z_n,
\end{align}
mutually commutes and together conserve particle number
even after writing it as a product of exponentials. Computing the gradient 
on the hardware could follow Refs. 
\cite{schuld_evaluating_2019, kottmann_feasible_2021, izmaylov_analytic_2021}. 
Overall, the implementation cost and 
asymptotic scaling of uCC is the same as that of HF. Indeed, recent 
advancements in finding a lower-rank representation of an \ansatz or
address the ordering ambiguity 
\cite{romero_strategies_2018, 
ogorman_generalized_2019,
matsuzawa_jastrow-type_2020,
takeshita_increasing_2020,
motta_low_2021,
anand_quantum_2022, 
rubin_compressing_2022,
kottmann_optimized_2022}
are applicable to AGP-based calculations as well. 

The Hamiltonian transformation can be done either using \Eq{\ref
{eq:abinito-ham-natorb}}, or we can
implement the Thouless rotations on the circuit and measure the Hamiltonian in
the on-site or atomic orbital basis. The trade-off between the two approaches
is well known \cite{babbush_quantum_2019, 
yen_measuring_2020, 
izmaylov_unitary_2020}. 
Typically the Hamiltonian is sparser in the on-site basis, hence
requiring fewer terms to measure; this comes at the expense of implementing the
orbital rotation on a quantum computer, which is often less expensive than
implementing uCC. Following the discussion in \Sec{\ref
{subsec:direct-optimization}} and the method in 
Ref. \cite{kivlichan_quantum_2018}, we can
implement the orbital rotations with \bigO{M} circuit depth with only nearest
neighbors Givens rotations appended to the end of the \ansatz circuit. 

\section{Post-selection}\label{sec:post-selection}

Computing the expectation value of an observable $\hat{O}$ with
respect to a wavefunction $\ket{\psi}$ on a quantum computer entails rotating
$\hat{O}$ to the computational basis by some unitary operator
$\hat{V}$ (i.e. $\hat{O}=\hat{V}^{\dagger}{\hat{\Lambda}}\hat{V}$), and
empirically estimating $\sum_a \lambda_a |\langle \lambda_a
|\hat{V}\ket{\psi}|^2$, where $\lambda_a$ is an eigenvalue of ${\hat{\Lambda}}$ 
\cite{nielsen_quantum_2010}. Often,
diagonalizing $\hat{O}$ could be expensive or even prohibitive, but
if $\hat{O}$ is expressible as a linear combination of Pauli matrices
(e.g. \textit{ab initio} Hamiltonians under the JW transformation), we can
sample the Pauli terms separately where the diagonalization is often
straightforward.

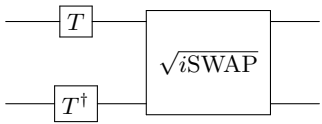
\begin{figure}[t]
  \centering
  \subfloat{
    \Qcircuit @C=2em @R=2em {
      & \gate{T} & \multigate{1}{\sqrt{i\mathrm{SWAP}}} & \qw\\
      & \gate{T^{\dagger}} & \ghost{\sqrt{i\mathrm{SWAP}}} & \qw
      }
  }    
  \caption{Number conserving circuit that diagonalizes $X_pX_q+Y_pY_q$
  \cite{google_ai_quantum_and_collaborators_hartree-fock_2020}.
  The T-gate is $R_z(\pi/4)$.}
  \label{fig:diag-xx-yy}
\end{figure}

Now suppose $\hat{O}$ commutes with some symmetry operator
$\hat{\mathcal{S}}$. We want
to measure $\hat{O}$ along with its symmetry eigenvalues in order to
project out samples that belong to the undesired symmetry sectors. To do this,
we need to diagonalize $\hat{O}$ in its shared eigenbasis with
$\hat{\mathcal{S}}$. This involves finding disjoint sets of Pauli matrices of
$\hat{O}$ that commute amongst each other as well as with
$\hat{\mathcal{S}}$, then diagonalize and measure each group separately. As an
example, consider measuring the energy of a general seniority-zero Hamiltonian,
\Eq{\ref{eq:senzero-ham}}, and suppose we want to post-select based on the particle number symmetry
i.e. $\hat{\mathcal{S}} = \sum_p \N{p}$. The first two terms,
$\sum_p h_p\Sz{p} + \sum_{pq} v_{pq}\Sz{p}\Sz{q}$, are readily in the
computational basis and conserve particle number; for the last term, one needs 
to diagonalize $\Sx{p}\Sx{q} +\Sy{p}\Sy{q}$ with a number conserving unitary,
$\hat{V}$, to get \cite{google_ai_quantum_and_collaborators_hartree-fock_2020}
\begin{align}
    \frac{1}{2}\E{\Sx{p}\Sx{q} + \Sy{p}\Sy{q}} 
    = \frac{1}{2} \E{\hat{V}^{\dagger} \left(\Sz{q}-\Sz{p}\right)\hat{V}}.
\end{align} 
See \Fig{\ref{fig:diag-xx-yy}}. Computing RDMs in seniority nonzero
systems follows the same idea and involves grouping Pauli terms into
{fragments} of mutually commuting and number conserving terms. 
{
Indeed, some of the most efficient strategies to find optimal fragments that minimize the scaling of the total number of measurements in VQE, such as the ``basis rotation grouping" of \Reference{\cite{huggins_efficient_2021}} and the ``full rank optimization" and its variants of \Reference{\cite{yen_cartan_2021}}, create fragments that conserve particle number by construction. Thus, we can readily adopt these strategies for AGP post-selection, thereby alleviating concerns about the significant overhead needed to obtain number conserving subsets.}

Post-selection has gained attention as a useful method to mitigate noise when
sampling from quantum hardware 
\cite{endo_practical_2018,
bonet-monroig_low-cost_2018,
mcardle_error-mitigated_2019,
kandala_error_2019,
google_ai_quantum_and_collaborators_hartree-fock_2020,arute_observation_2020,
obrien_error_2021,
huggins_efficient_2021}.
However, our goal in this section is to use
post-selection as an alternative to gauge integration to restore particle 
number and sample over the correlated AGP wavefunction. 
We present an analysis of the cost
and scaling of post-selection and show how to adjust $\{\eta_p\}$ to maximize
getting samples in the desired particle sector. It is noteworthy that
post-selection and the gauge integration are not necessarily mutually 
exclusive; to
restore multiple symmetries, one might choose to use a combination of the
integral and post-selection, should that provide an advantage. In this paper,
we concentrate on restoring number symmetry only and leave the analysis of
other symmetries for future work. 

We must mention that there are alternative ways to observing symmetries prior
to the final measurement, such as phase estimation on $\hat{\mathcal{S}}$ or
other circuits 
\cite{bonet-monroig_low-cost_2018,
mcardle_error-mitigated_2019, 
lacroix_symmetry-assisted_2020, 
siwach_filtering_2021,
tilly_variational_2021, 
ruiz_guzman_accessing_2022}.
In these methods, typically a specialized circuit is applied in
the  \textit{bulk} of the circuit and symmetries are observed with the help of
ancilla qubits. Our strategy to maximize sample outcomes and scaling analysis
of the number of measurements applies to those methods as well.

\subsection{Procedure}\label{subsec:ps-procedure}

\begin{figure*}[t]
  \centering
  \subfloat[]{{\includegraphics[width=\columnwidth]{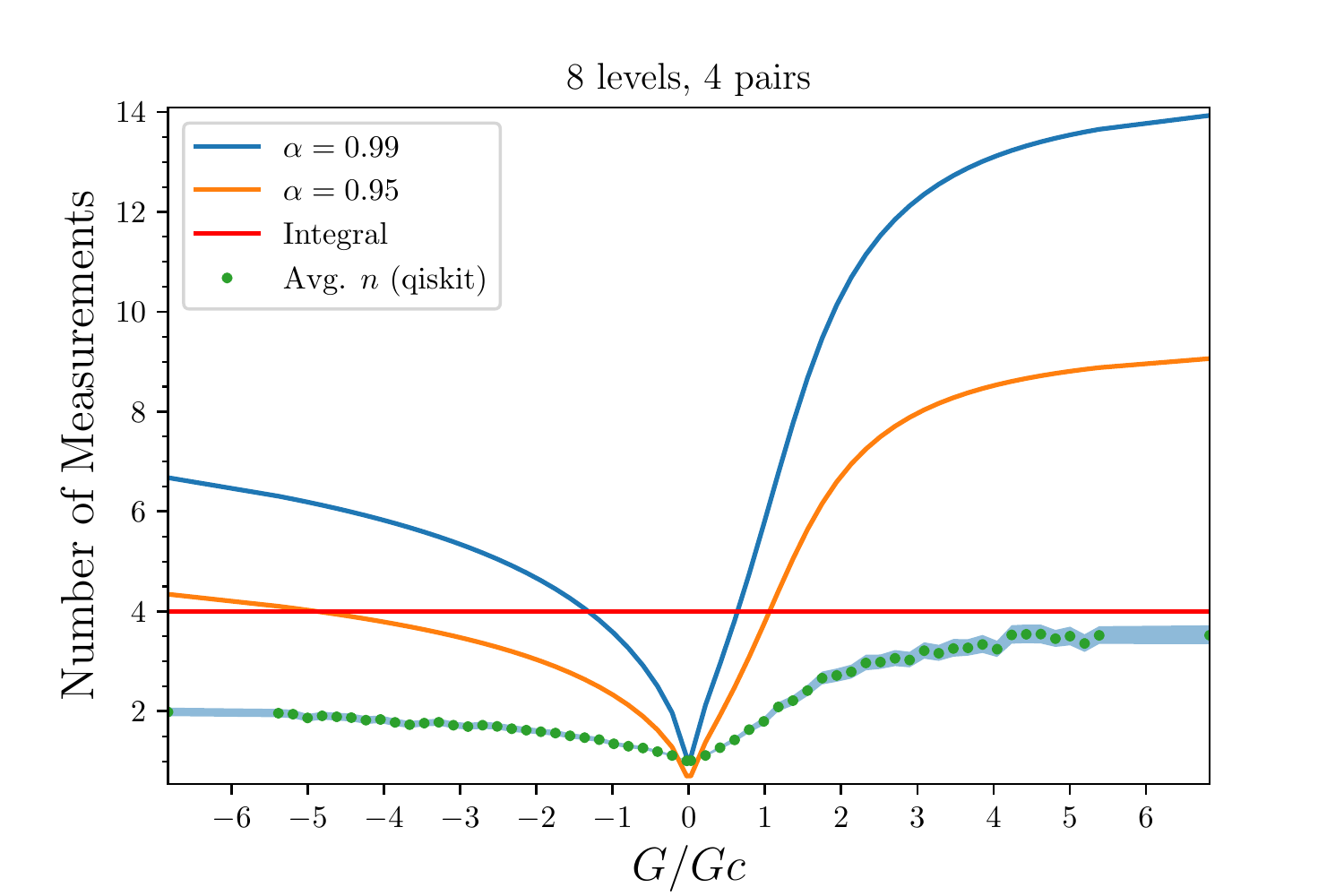}}}
  \subfloat[]{{\includegraphics[width=\columnwidth]{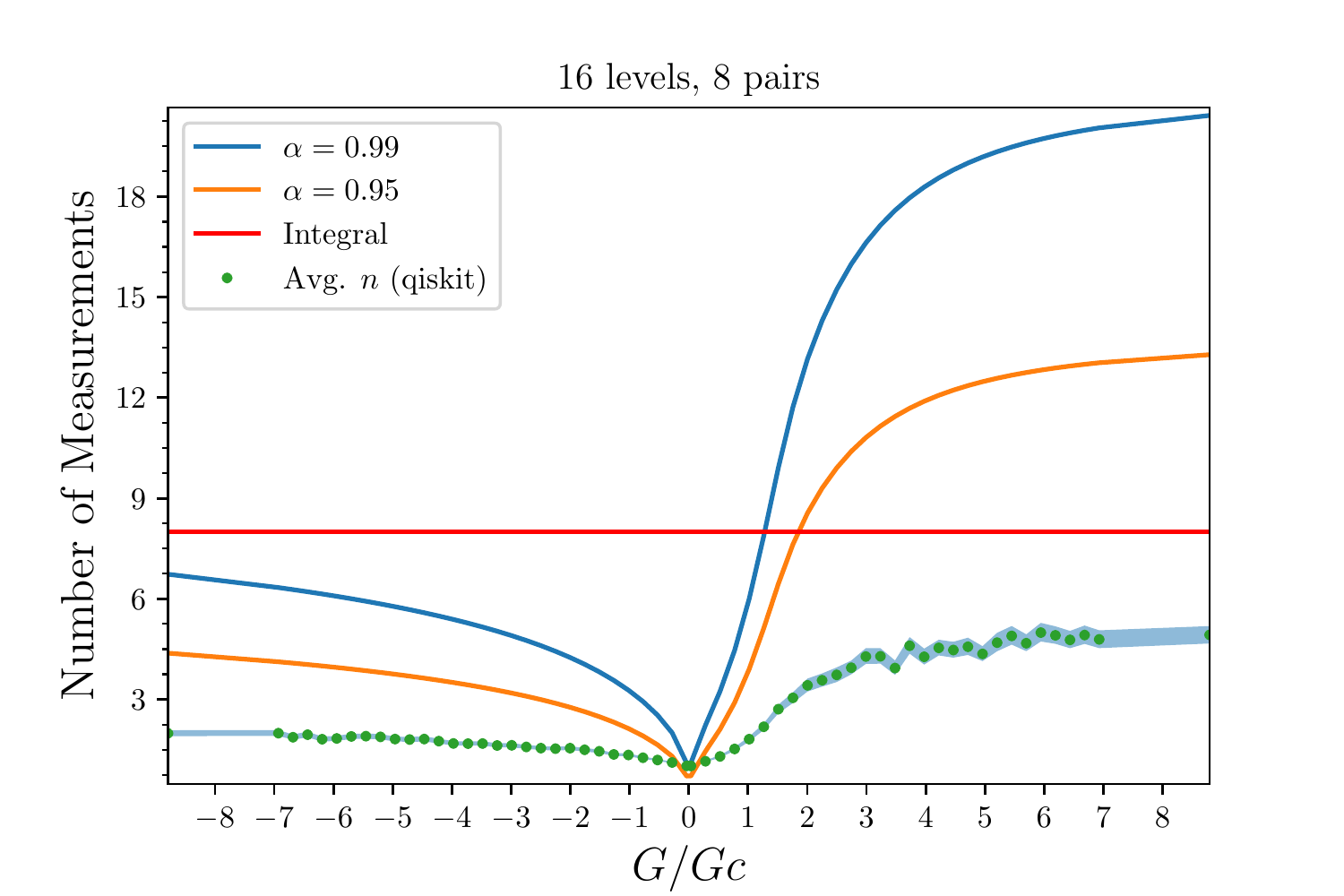}}}
  \caption{Estimated number of measurements, $n$, needed to get at least one
  observation with the correct particle number as a function of $G/G_c$ for
  the pairing Hamiltonian with $8$ (a) and $16$ (b) orbitals at half-filling.
  The red curve corresponds to the number of grid points to carry out number
  projection with gauge integration. The green dots are the average number for $n$
  obtained experimentally from $10^3$ parallel runs on a noise-less quantum
  simulator (\texttt{qiskit}) along with their $95\%$ confidence intervals (shaded
  area). The blue and orange curves are theoretical upper bounds for $n$ with
  $\alpha=0.95,\;0.99$ computed from \Eq{\ref{eq:n-theory}}.
  }
  \label{fig:n-Gc}
\end{figure*}

To compute expectation values over AGP, we need to implement the
corresponding BCS wavefunction,
correlated it, and perform post-selection at the end. To this end, it is
crucial that we use a number conserving correlator, since 
\begin{align}
    \hat{U}\rAGP = \hat{U} \left(\Proj\rBCS\right) = \Proj \hat{U}\rBCS, 
\end{align}
is true if and only if $[\Proj, \hat{U}]=0$, where $\hat{U}$ is the correlator
and $\Proj$ is the number projection operator. The disentangled uCC as
described in \Sec{\ref{subsec:uCC-agp}} satisfies this requirement. 

In general, however, $\Ebcs{\hat{N}} \neq 2N$, due to the broken $U(1)$ gauge
degree of freedom in the BCS wavefunction \cite{ring_nuclear_1980}. 
This could lead to nonoptimal
sampling during post-selection since we get samples that do not belong to the
desired particle number sector on average. Fortunately, we can fix the gauge to
$\E{\hat{N}} = 2N$ easily during the AGP optimization using {projected BCS} or 
NHFB by introducing a chemical potential.
Doing so amounts to multiplying all
$\eta_p$ by a constant number which does not change expectation values over
AGP \cite{khamoshi_efficient_2019}. Alternatively we can find this constant 
algebraically. See \Appx{\ref{subappx:fix-gauge}}. 

A major advantage of using post-selection {compared to} 
gauge integration {over the same Pauli set} is the lack of 
an additional ancilla qubit needed to perform the Hadamard test, 
and the \bigO{M} circuit depth for the controlled $R_z(\phi_i)$ rotations. 
However, the difference in the scaling
of number of measurements is more involved. For the integral, the
number of measurements scales as \bigO{M} corresponding to the total number of
grid points, and it is independent of the Hamiltonian. To derive the scaling for
post-selection, we formulate the problem as follows: \textit{What is the
minimum number of measurements needed to get at least one observation in the correct number sector with $\alpha \times 100 $ percent confidence?} We show in
\Appx{\ref{subappx:ps-scaling}} that the number of measurements, $n$, 
is given by
\begin{align}\label{eq:n-theory}
    n = \ceil*{\frac{\log (1-{\alpha})}{\log{(1-P)}}},
\end{align}
where $P = {(\prod_p u_p^2)} S_N^M$, and $S_N^M$ 
is an elementary symmetric
polynomial associated with the norm of AGP \cite{khamoshi_efficient_2019}. 
Clearly $n$ depends on the details
of the Hamiltonian through $P$, which itself is a function of $\{\eta_p\}$. The
largest value of $n$ occurs when number symmetry is strongly broken and all
$\{\eta_p\}$ approach the same value. This is the worst case scenario for
post-selection and, as we show in \Appx{\ref{subappx:ps-scaling}}, it is 
asymptotically bounded
above by \bigO{\sqrt{M}}. In regimes where number symmetry does not break, the
scaling could be as low as \bigO{1}.

\subsection{Numerical experiments}\label{subsec:ps-experiment}

\begin{figure*}[t]
  \centering
  \subfloat[]{{\includegraphics[width=\columnwidth]{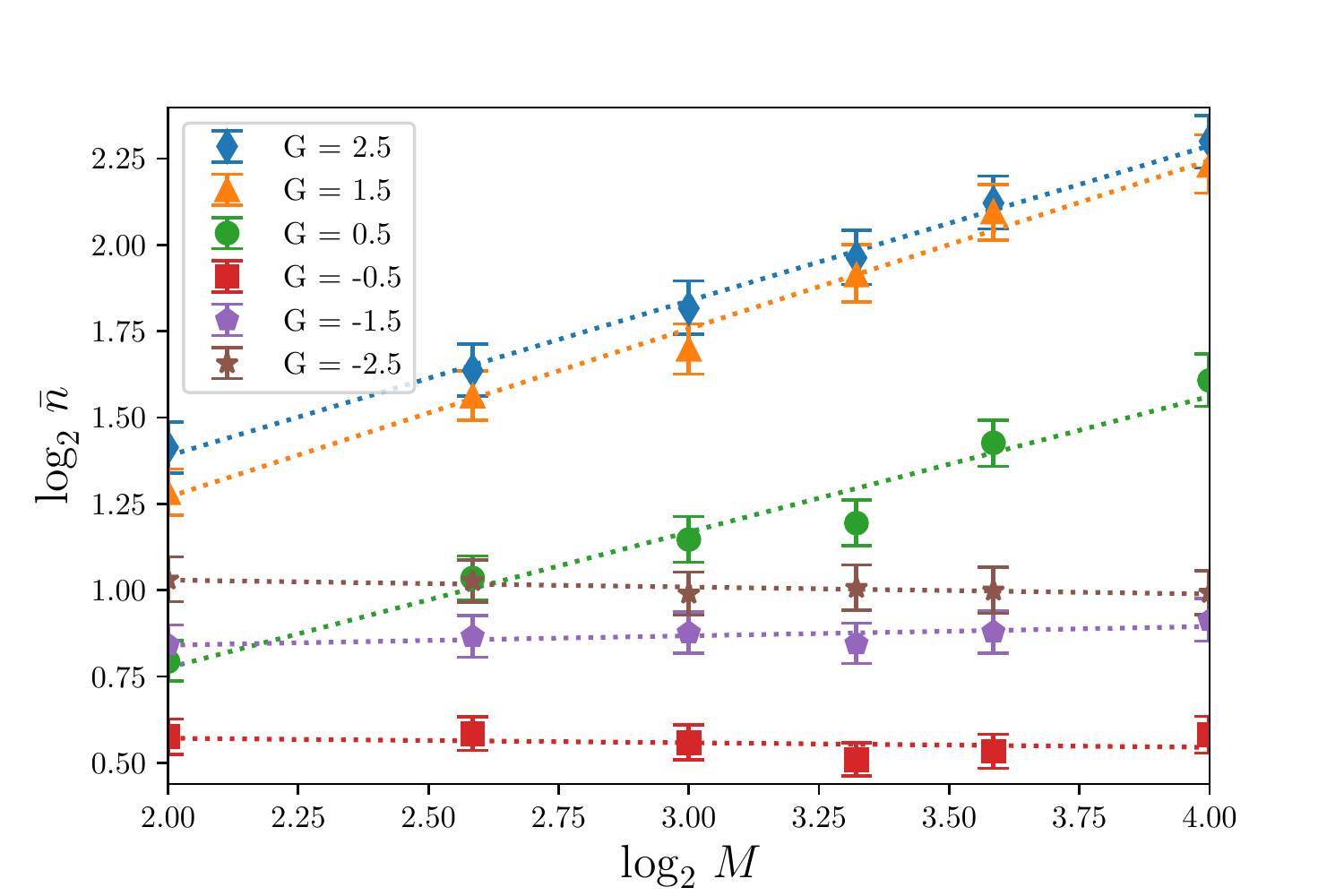}}}
  \subfloat[]{{\includegraphics[width=\columnwidth]{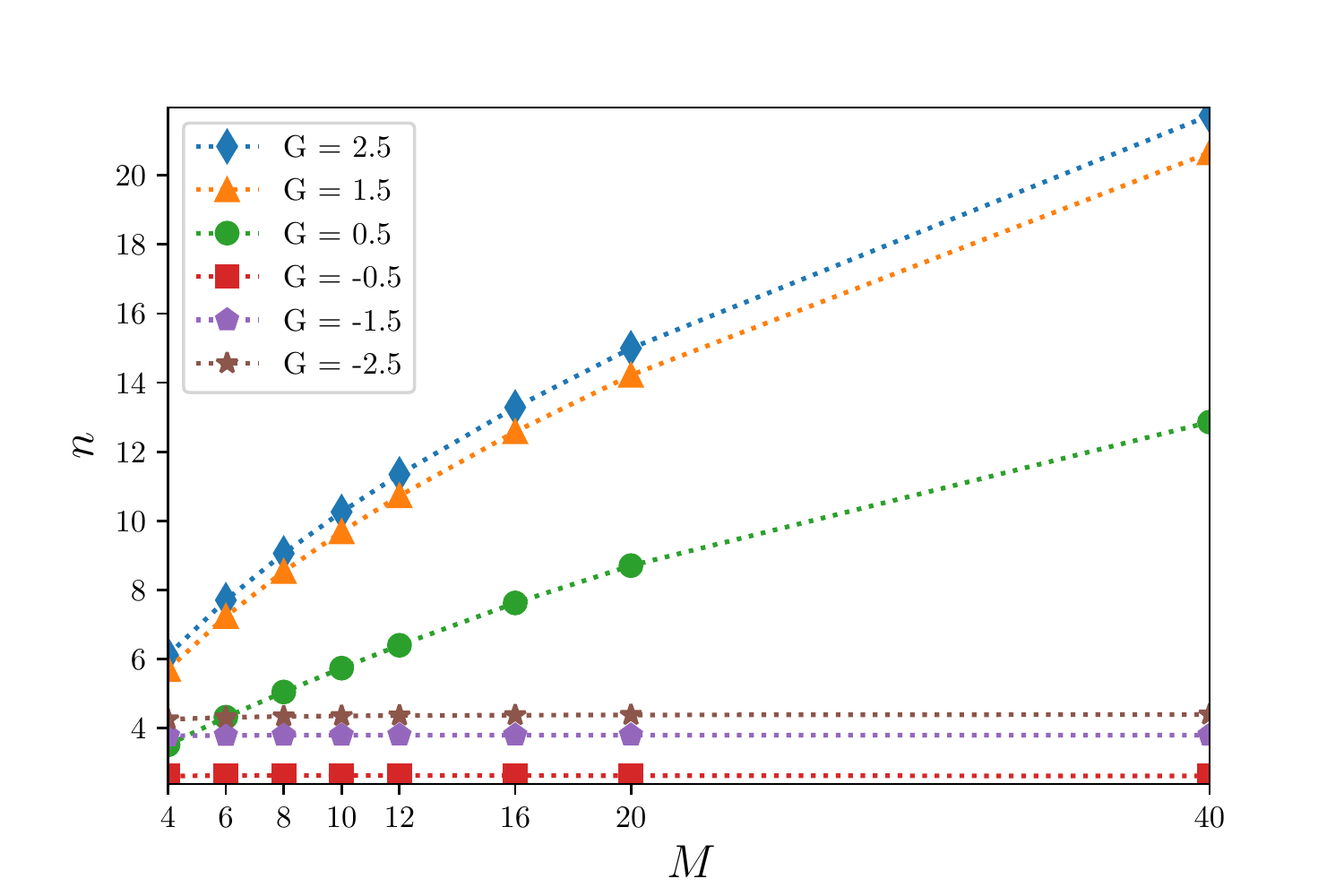}}}
  \caption{Scaling of post-selection as function of system size, $M$, for the 
  pairing Hamiltonian. (a) The average number of measurements, $\bar{n}$, 
  to get 
  one observation with the correct particle number from repeated sampling on 
  a quantum emulator (\texttt{qiskit}). The error bars correspond to the $95\%$ 
  confidence intervals. This was performed for $M=4,6,8,10,12,16$ at 
  different $G$ values. (b) The theoretical number of 
  measurements on a linear scale computed from \Eq{\ref{eq:n-theory}} with 
  $\alpha=0.95$. This 
  allows us to estimate $n$ for system much larger than those we can simulate 
  on \texttt{qiskit}.} 
  \label{fig:n-scaling}
\end{figure*}

We put the scaling to numerical test by doing post-selection for the pairing
Hamiltonian \cite{ring_nuclear_1980, dukelsky_colloquium:_2004},
\begin{align}
    H &= \sum_p \epsilon_p \N{p} - G \sum_{pq}\Pdag{p}\Pp{q},    
\end{align}
written here in the pairing algebra; $\epsilon_p=p \Delta\epsilon$ are the
single particle energy levels and $G$ tunes the strength of the pair-wise
interaction and has infinite range. The pairing model is an ideal
benchmark as the mean-field breaks number symmetry---gives rise 
to a BCS state---at $G=G_c>0$ where number fluctuations get larger as $G$ 
increases; for all $G < G_c$ (including $G<0$) the mean-field admits a single 
HF Slater determinant for which the corresponding particle number fluctuations 
are small.

In \Fig{\ref{fig:n-Gc}} we plot $n$ computed from \Eq{\ref{eq:n-theory}} 
with $\alpha=0.95$ and $0.99$ as a 
function of $G/G_c$ for $8$ and $16$ orbitals at half-filling. Half-filling is 
the ideal case for the integration as it requires the fewest number of grid 
points, whereas the opposite is true for post-selection. We also numerically 
simulated post-selection on \texttt{qiskit} \cite{noauthor_qiskit_2019}, 
using the QASM simulation 
libraries in the absence of noise. For every $G$ point, we simulated the BCS 
state and performed multiple measurements until we got an outcome with the 
correct particle number. This process was repeated $10^3$ to estimate $n$ by 
the empirical mean, $\bar{n}$. The $95\%$ confidence intervals were computed 
from bootstrapping the samples $10^4$ times and are shown with the shaded area 
in the plot. As we can see, the sample means along with their confidence 
intervals are below the number of measurements needed for the integral for all 
points, even in $G \gg G_c$ where number fluctuations are large. This suggests 
that number symmetry restoration with post-selection could be a viable 
alternative to gauge integration even in small systems with large number 
fluctuations.

In \Fig{\ref{fig:n-scaling}} we numerically investigate the scaling of 
post-selection as a function 
of system size $M$. On the left hand  side of \Fig{\ref{fig:n-scaling}} we 
plot the sample means, 
$\bar{n}$, computed at different $G/G_c$ points on a log-log scale as a 
function of the system size $M$. Linear fits to the mean resulted in slopes: 
$m=0.449, 0.488, 0.392, -0.01, 0.027, -0.02$ rounded to the closest digits in 
decreasing order of $G$. On the right side of the figure, we used \Eq{\ref
{eq:n-theory}} to 
compute $n$ beyond what we could simulate on \texttt{qiskit}, 
with system size as large 
as $M=40$. The results suggest \bigO{1} scaling in $G<0$, where number symmetry 
does not break, and a \bigO{\sqrt{M}} where number symmetry breaks. This 
is inline with the \bigO{\sqrt{M}} asymptotic scaling found analytically and 
shown in detail in \Appx{\ref{subappx:ps-scaling}}.

In summary, Table \ref{table:scaling} compares the scaling cost of phase 
estimation, exact gauge integration, and post-selection as methods to restore 
number symmetry. For phase estimation, one could follow Refs. \cite
{mcardle_error-mitigated_2019, lacroix_symmetry-assisted_2020} and use our 
technique to guarantee an \bigO{\sqrt{M}} scaling for the number of 
measurements. If mid-circuit measurement is feasible on a quantum computer, 
then phase estimation can be carried out with only one ancilla qubit i.e. \bigO{1} scaling instead of \bigO{\log(M)}.

\begin{table}[b]
\centering
  \begin{tabular}{ |c|c|c|c| } 
  \hline
  Method & Measurements & Depth & Ancilla qubits\\
  \hline\hline
  Phase estimation & \bigO{\sqrt{M}} & \bigO{M \log(M)} & \bigO{\log(M)}\\    
  Gauge integral & \bigO{M} & \bigO{M} & \bigO{1}\\
  Post-selection & \bigO{\sqrt{M}} & \bigO{1} & None\\
  \hline
  \end{tabular}
  \caption{
    Scaling of number symmetry restoration using different methods as a 
    function of system size ($M$) for a given number--conserving Pauli fragment.
  }
  \label{table:scaling}  
\end{table}

\section{Application}\label{sec:application}

\begin{figure*}[t]
  \centering
  \subfloat[]{{\includegraphics[width=9cm]{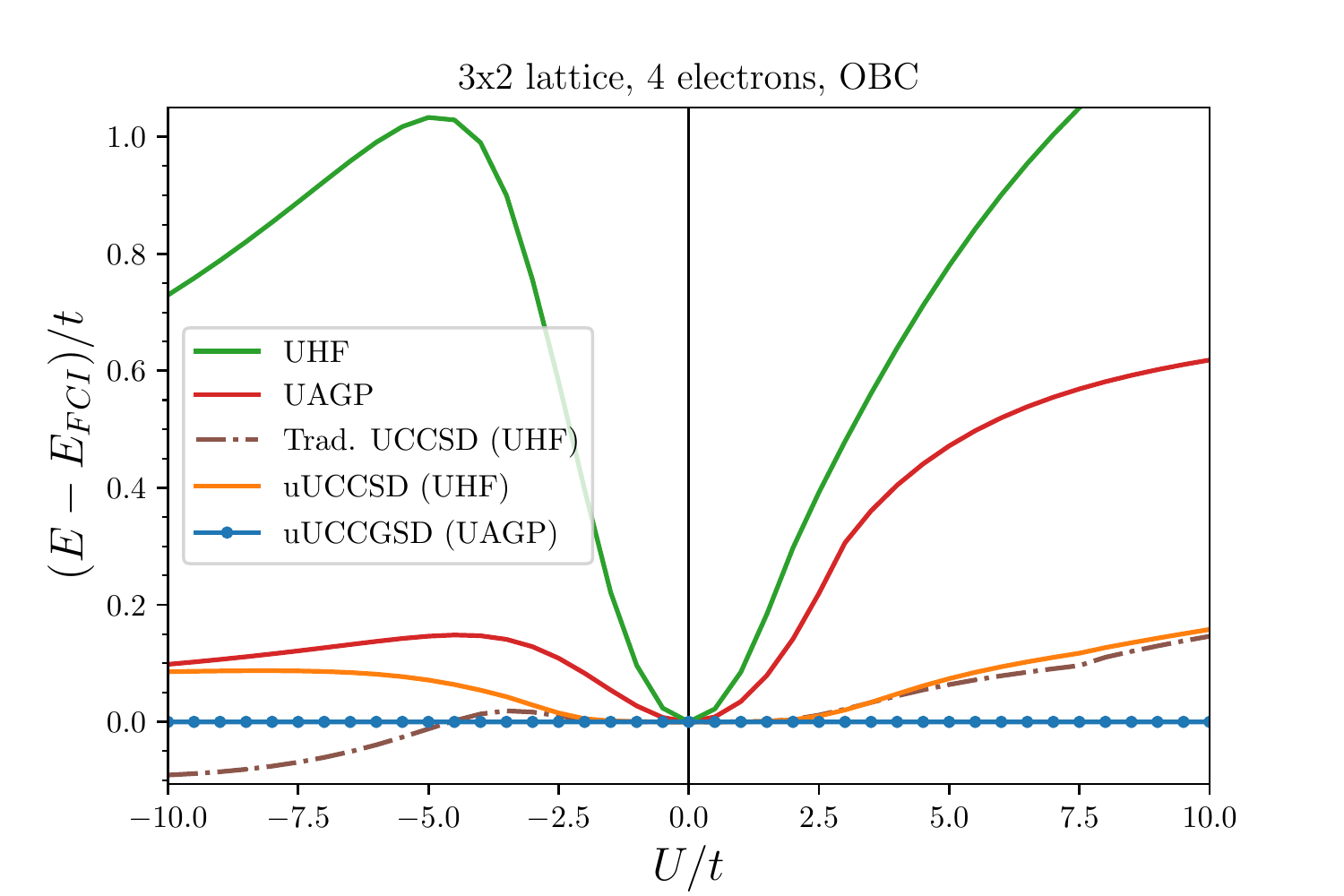}}}
  \subfloat[]{{\includegraphics[width=9cm]{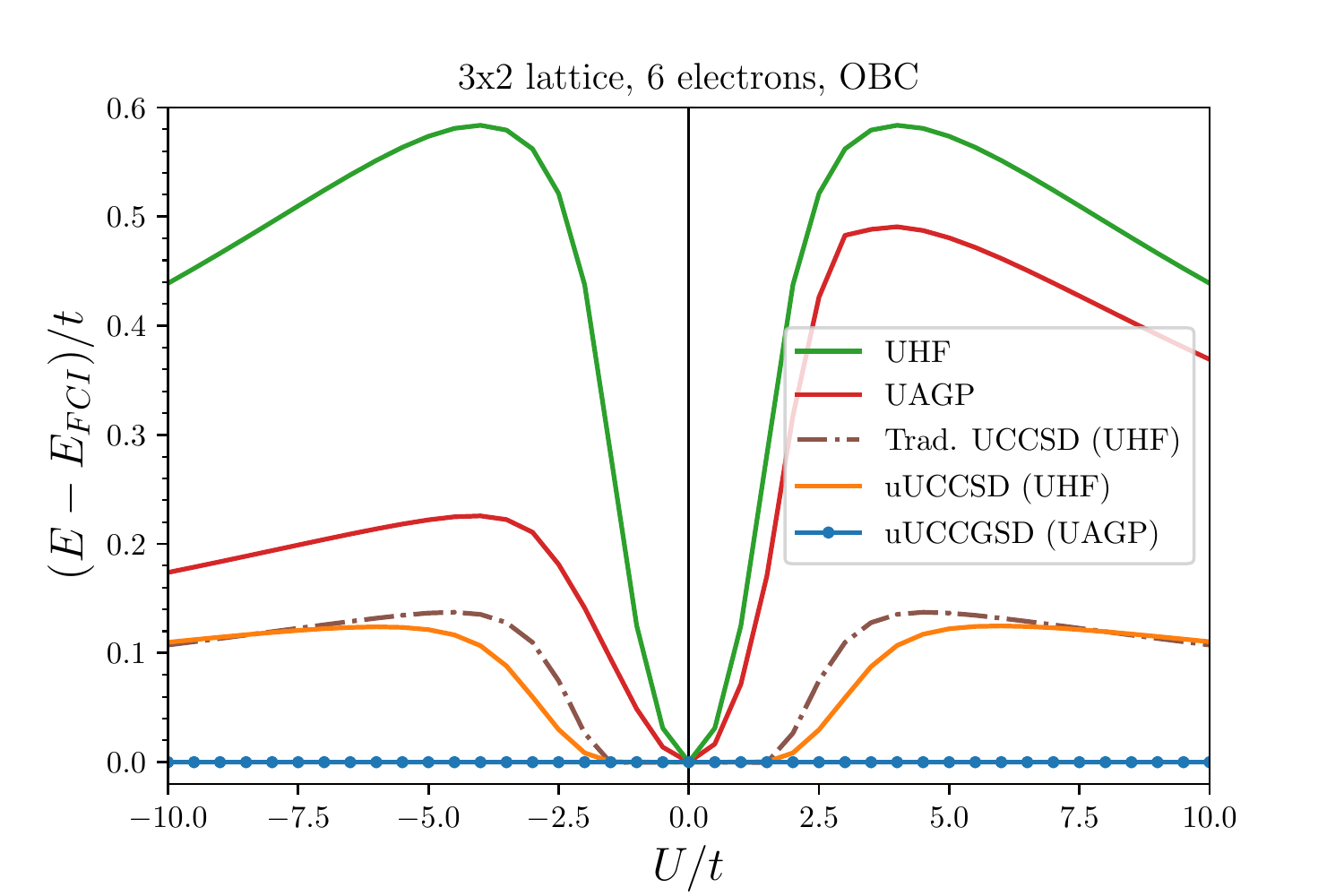}}}
  \caption{Energy error of fixed-reference calculations for 2D Hubbard
  with open boundary conditions (OBC) away
  from half-filling (a) and at half-filling (b) as a function of $U/t$.
  Traditional (Trad.) or disentangled unitary 
  (u) spin-unrestricted coupled cluster
  calculations were performed based on UHF or UAGP references with
  single and double excitations (UCCSD) or their generalized coupled cluster 
  counterpart (UCCGSD).}
  \label{fig:fixed-ref}
\end{figure*}

Having developed a method for how to optimize and implement AGP for quantum
computers, we concentrate on building correlation atop of AGP for
seniority nonzero applications in this section. Our aim is to showcase our 
method by benchmarking it against the ground state
of the single-band Fermi--Hubbard model,
\begin{align}
    H
    = -t \sum_{\langle p,q \rangle} \sum_{\sigma \in \{\uparrow,\downarrow\}}
    \left(\cdag{p\sigma}\cn{q\sigma} + h.c.\right)
    + U \sum_{p} n_{p\uparrow}n_{p\downarrow},
\end{align}
written here in its on-site basis, 
where $n_{p\sigma}=\cdag{p\sigma}\cn{p\sigma}$.
We set $t=1$ for simplicity for the rest of  
this section. For this Hamiltonian, number symmetry breaks at some finite $U<0$ 
value; and when $U>0$, the mean-filed breaks spin-symmetry at a critical $U>0$ .
While one of the main advantages of using AGP is for models wherein the 
two-body interactions are attractive, we showcase our method in both the 
repulsive and attractive regimes of this Hamiltonian. Our 
strategy is to allow the $S^2$ symmetry to break (i.e. use UAGP) and 
correlate the resulting wavefunction. This helps us illustrate the utility of 
AGP in all regimes---attractive, repulsive, weak and strong limits. Indeed, to 
improve the results even more, one can envision restoring $S_z$ and/or $S^2$ 
which tend to break at sufficient strong correlation, but we leave those for 
future work. 

We first present the results for the fixed-reference methods, 
wherein we have optimized UAGP on a classical computer and use quantum 
computers to optimize uCC atop of AGP. In the later half of this section, we 
demonstrate another powerful \ansatz based on a particle--hole unitary, which 
as we shall see, needs to be optimized simultaneously with the 
reference AGP. 

The Implementation details of this section can be found in 
\Appx{\ref{appx:ucc-imp}}.

\subsection{Fixed-reference}\label{subsec:fixed-reference}

We implement our \ansatze with the following structure,
\begin{align}\label{eq:ucc-ordering}
    \ket{\psi} = 
    \underbrace{\prod_k e^{\sigma_k}}_{\text{orbital rotation}}
    \underbrace{\left(\prod_d e^{\hat{D}_d}\right)}_{\text{doubles}}
    \underbrace{\left(\prod_s e^{\hat{S}_s}\right)}_{\text{singles}}\ket{\phi},
\end{align} 
where $\ket{\phi}$ can be AGP or HF. The orbital rotation is optional 
but it allows one to measure the Hamiltonian in the on-site basis which is
significantly sparser and simpler than the natural orbital basis (see \Sec{\ref{subsec:uCC-agp}}). The on-site Hubbard Hamiltonian is also convenient for post-selection as it maps to  
\begin{multline}
H =
\frac{t}{2} \sum_{\langle p,q \rangle} \sum_{\sigma \in \{\uparrow,\downarrow\}}
\left( \Sx{p\sigma}\Sx{q\sigma} + \Sy{p\sigma}\Sy{q\sigma} \right)\prod_{n}\Sz{n} \\
+ \frac{U}{4} \sum_p \left(1-\Sz{p\uparrow}\right) \left(1-\Sz{p\downarrow}\right),
\end{multline}
under the JW transformation. The only fragment that needs to be diagonalized is the nearest-neighbors hopping terms, for which we can apply the same circuit as the seniority-zero Hamiltonians shown in \Fig{\ref{fig:diag-xx-yy}}.

In \Fig{\ref{fig:fixed-ref}} we show the VQE calculation results for the 2D
Hubbard model at half-filling and away from half-filling {with open boundary conditions (OBC)}. For the doped
system, the number symmetry becomes strongly broken as $U<0$ get smaller. 
This is the kind of regime where we expect AGP-based methods to excel and
traditional methods to struggle. Indeed, traditional UCCSD over-correlates 
in the 4-electron case as $U$ gets smaller and becomes difficult to converge 
starting near $U\approx-8$ up to some finite regime of $U$. In the $U>0$ regime 
of the same doped system, in addition to $S^2$, $S_z$ symmetry breaks at 
mean-field near $U \sim 7.5$; only the $S^2$ symmetry breaks for the 
half-filled case. 

We plot uUCCGSD based on UAGP for all values of $U$. As a point of reference, 
we have plotted uUCCSD on UHF as well. 
The AGP-based results captured energies with
absolute errors as small as $10^{-9}$ for both systems which is where we set
the tolerance of our optimization. While this confirms that our AGP-based
method is capable of accessing all relevant parts of the Hilbert space (not
just seniority-zero for example) and get almost exact energies for these
system, the high accuracy might be in part due its high number of variational
parameters compared to the dimension of FCI. There are 870 variational
parameters in uUCCGSD while the FCI dimension of the half-filled and doped
systems are ${12 \choose 6}=924$ and ${12 \choose 4}=495$ respectively. We get
equally good results with uUCCGSD on UHF for these systems. While this in
principle can be tested by going to larger systems, our current implementation
of the code allows for maximum system size of 12 spin-orbitals. 

\begin{figure*}[t]
  \centering
  \subfloat[]{{\includegraphics[width=9cm]{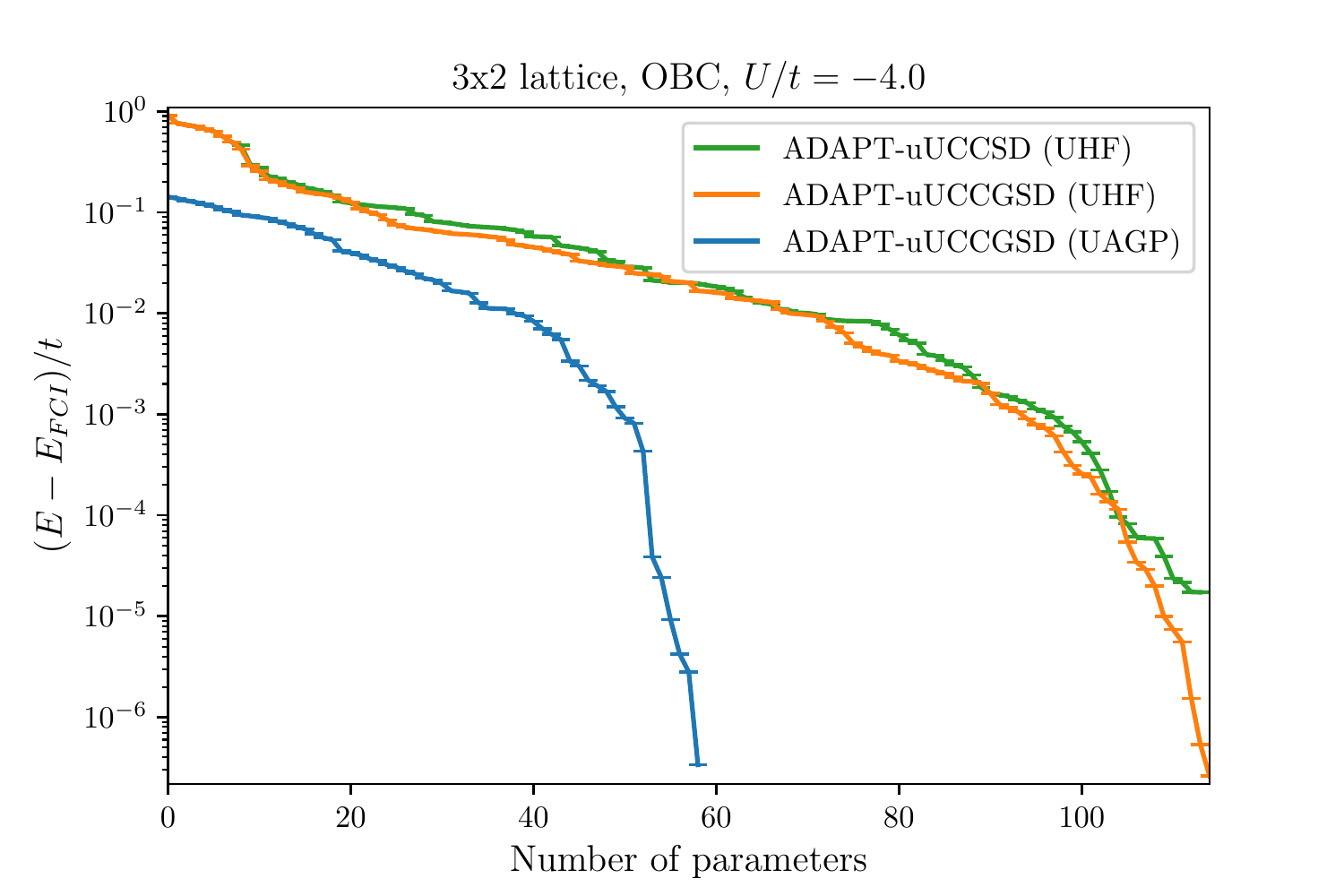}}} 
  \subfloat[]{{\includegraphics[width=9cm]{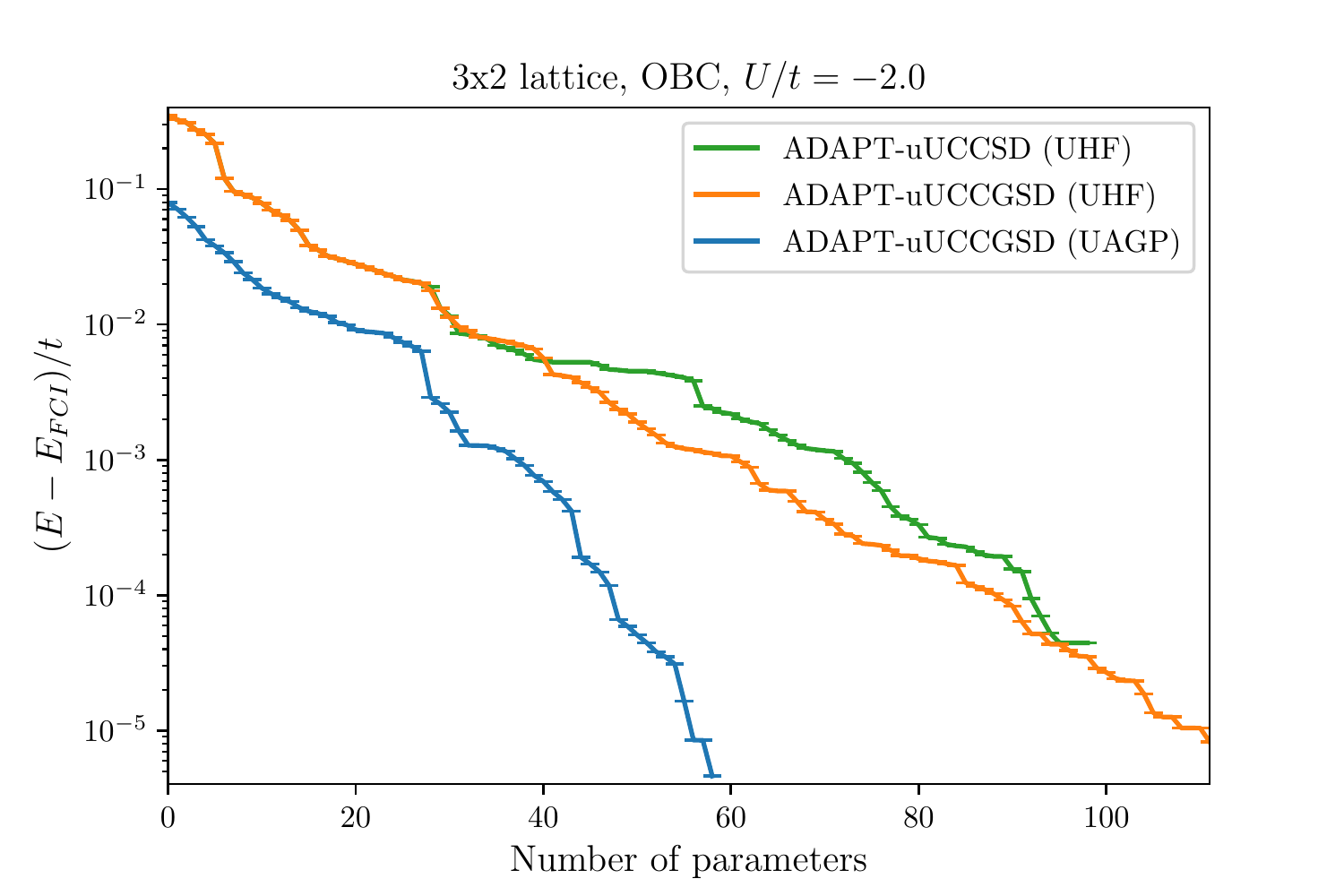}}}
  \caption{ADAPT-VQE calculations for HF- and AGP-based disentangled uCC
  (fixed-reference) for the 4 electron system at the strongly correlated point ($U/t=-4$)
  where mean-field methods struggle the most and a slightly less correlated point ($U/t=-2$)
  for comparison.}
  \label{fig:adapt}
\end{figure*}

To find a sparser representation of the disentangled uUCCGSD we performed
ADAPT-VQE \cite{grimsley_adaptive_2019} calculations with the same operator 
pool as uUCCGSD for the
4-electron system. ADAPT-VQE was designed and tested for HF-based methods, but
we follow the same idea for our implementation and use AGP as our reference. We
set ADAPT's stopping criterion to be $\epsilon=10^{-3}$. As shown in
\Fig{\ref{fig:adapt}} energy errors of $10^{-6}$ were obtained with as few as
60 parameters with AGP. For comparison we also plotted particle--hole and
general index HF-based calculations with the same computational settings. We
observe the largest gains with AGP-based ADAPT in the regimes where number
symmetry breaks and the system is strongly correlated. 
In large $U>0$, we did not see significant advantage
in using AGP compared with HF. More studies are needed to tailor ADAPT for
AGP-based methods. {
After submitting this paper, the recent work by 
Refs. \cite{tsuchimochi_adaptive_2022, bertels_symmetry_2022} has come to our 
attention in which the authors studied the affect of symmetry breaking in the 
ADAPT-VQE framework. While neither investigated the restoration of particle 
number symmetry, our results with AGP is complementary to their findings in 
that restoring symmetries that spontaneously break at the mean-field level
could greatly benefit ADAPT's convergence.
}

In the next section, we develope a particle--hole \ansatz on AGP which has 
considerably {fewer} parameters than uUCCGSD. 

\subsection{Reference-optimized uCC}\label{subsec:oo-uCC}
\begin{figure*}[t]
    \centering
    \subfloat{{\includegraphics[width=9cm]{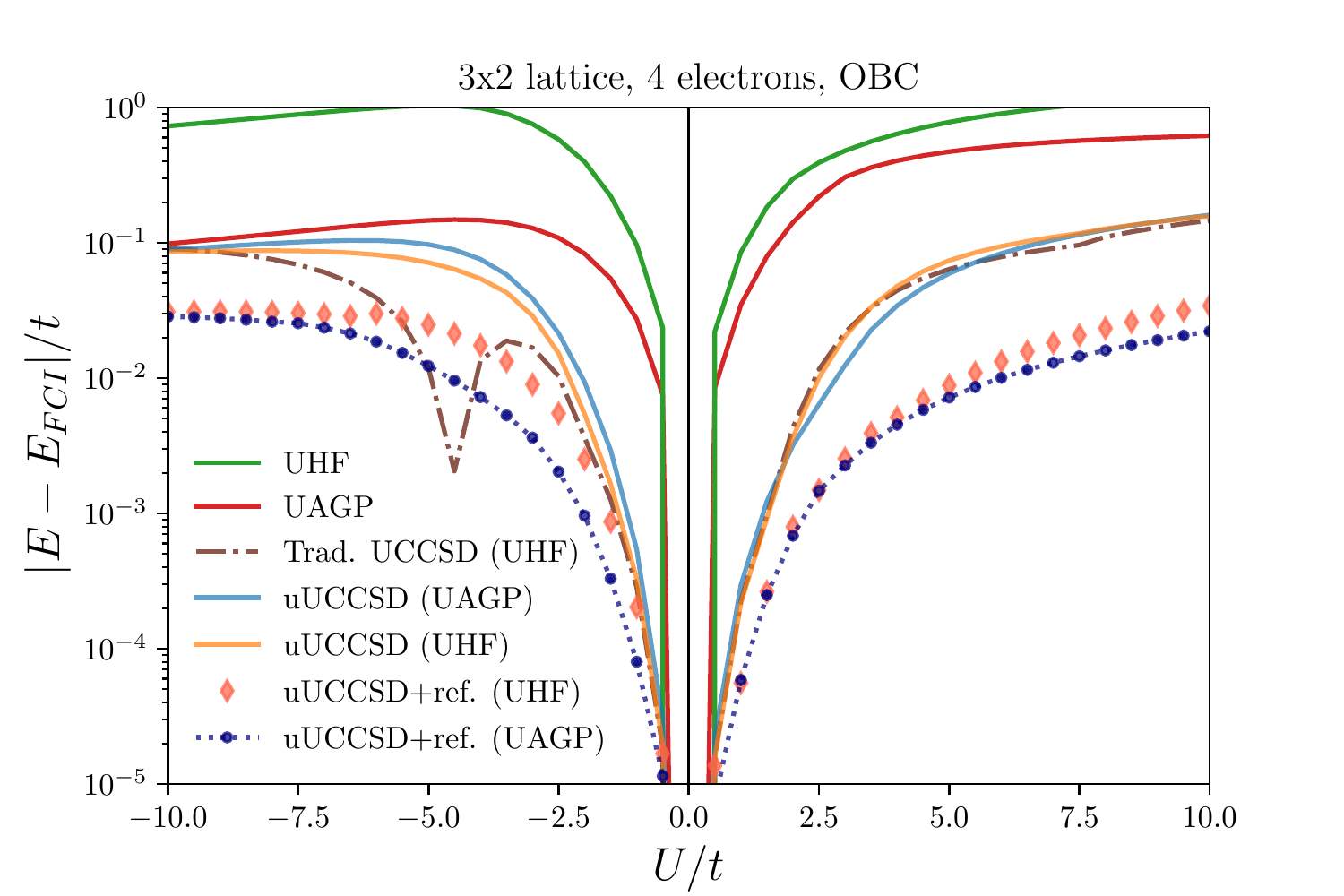}}}
    \subfloat{{\includegraphics[width=9cm]{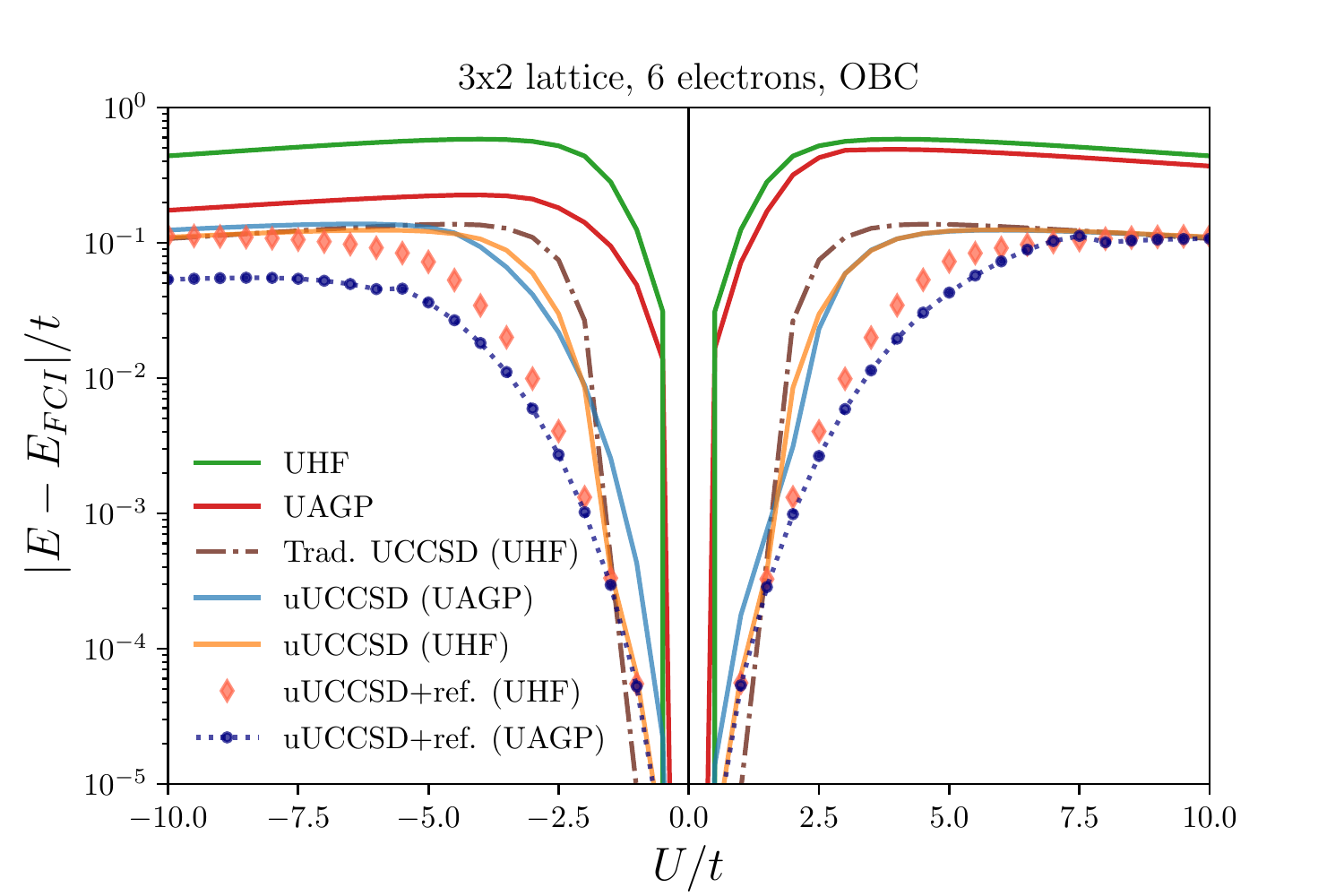}}}\\
    \subfloat{{\includegraphics[width=9cm]{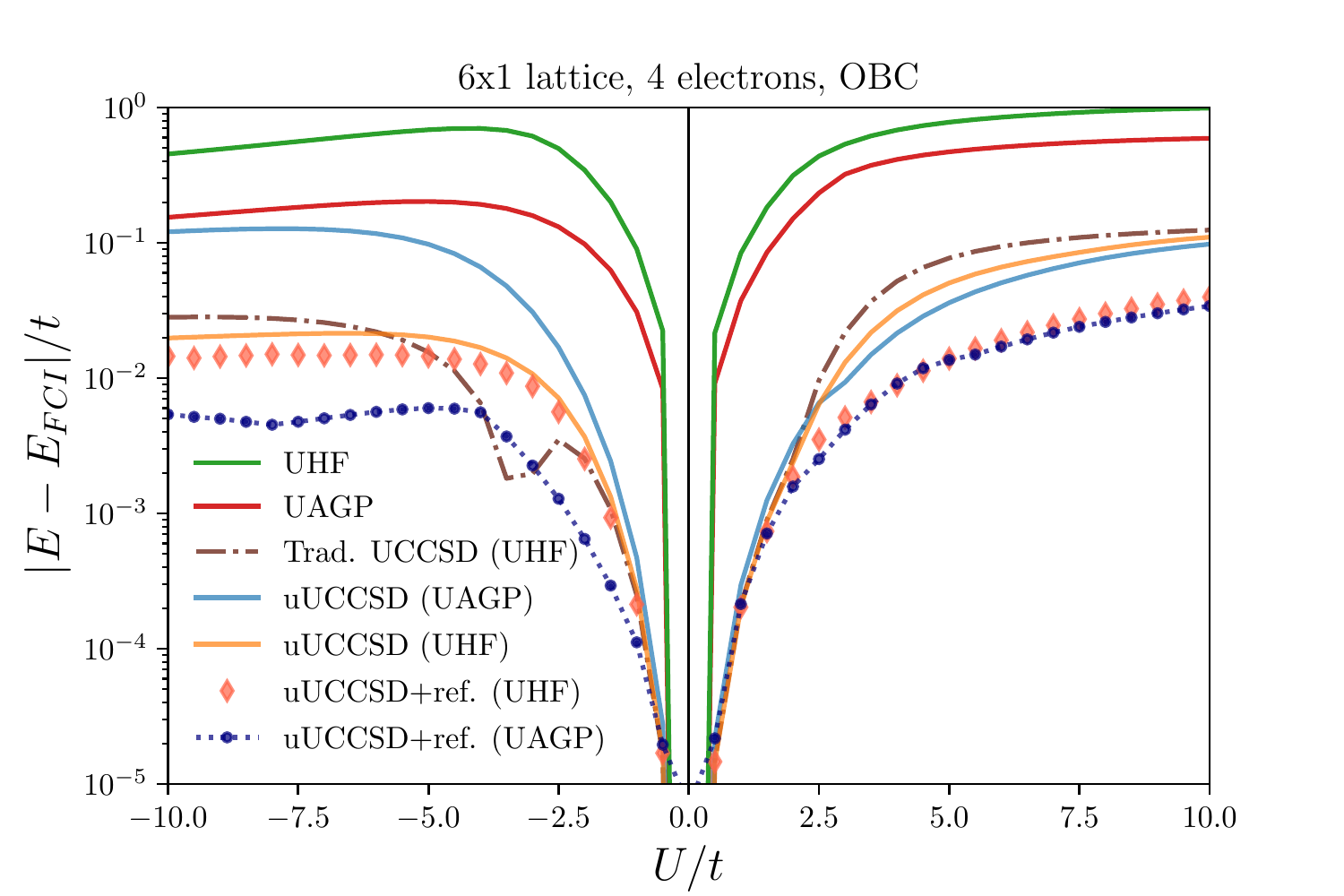}}}
    \subfloat{{\includegraphics[width=9cm]{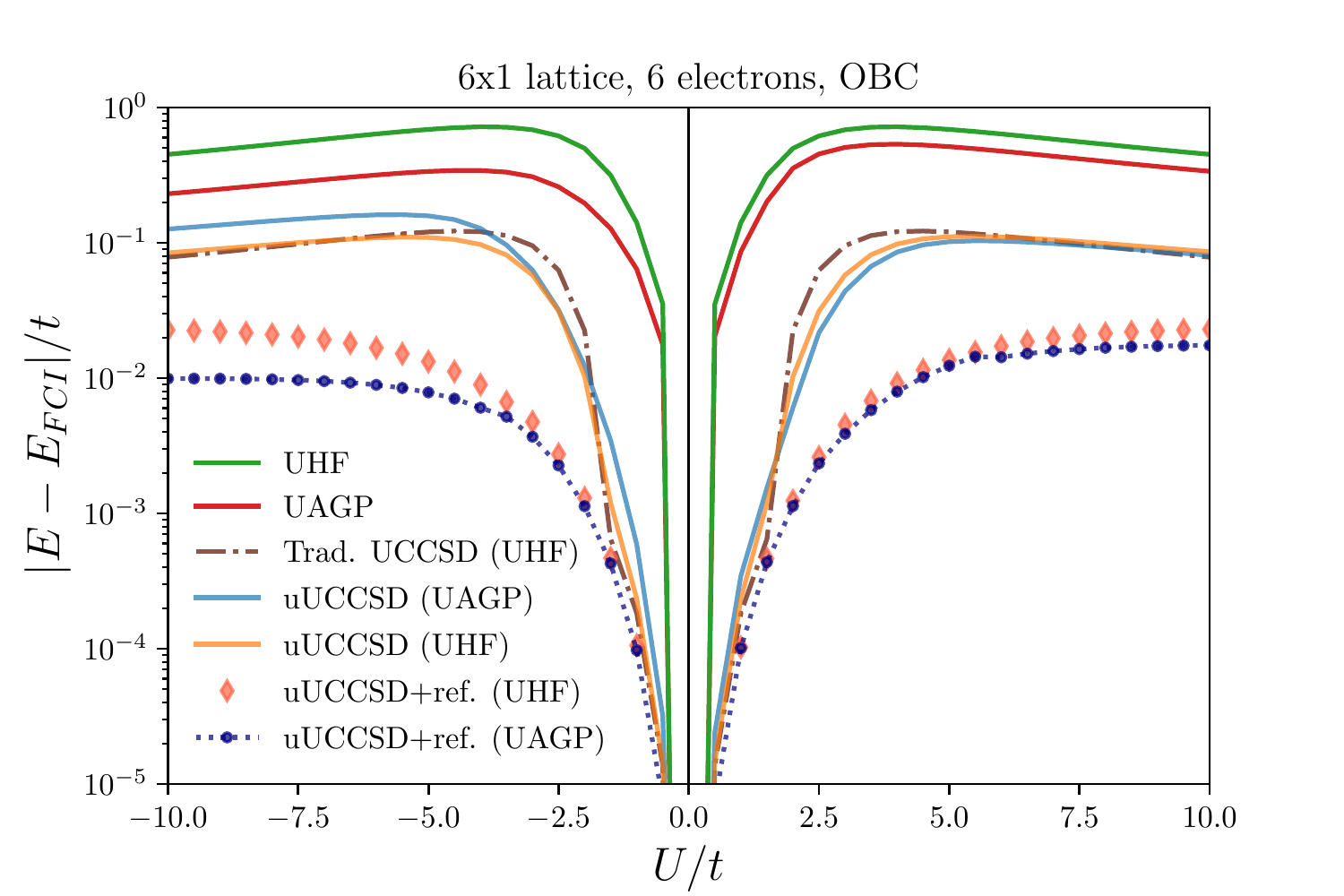}}}    
    \caption{A particle--hole uCC \ansatz based on UAGP and UHF tested for one and two
    dimensional Hubbard model at and away from half-filling. Methods involving
    reference optimization are denoted by ``+ref."}
    \label{fig:oo-uCC}
\end{figure*}
One of the main advantages of using the uCC \ansatz with particle--hole 
excitations is that it is significantly more concise compared to one 
based on general operators. Ideally, we would like to develop a particle--hole uCC 
\ansatz based on AGP. 

In general, the natural orbitals of AGP and HF are different. In the weakly 
correlated limit (e.g. $U\rightarrow0$ for the Hubbard Hamiltonian), the HF and 
AGP orbitals converge to the same values and we get $\eta_1, \cdots,
\eta_N \rightarrow 1$ and $\eta_{N+1}, \cdots,\eta_{M} \rightarrow 0$ (up to a 
global constat) so that AGP becomes just a single Slater determinant 
\cite{dutta_geminal_2020}. For finite, 
nonzero $U$, we designate the first $N$ orbitals of the largest occupations
as \textit{occupied} and the
rest \textit{virtual}, thereby defining a ``particle--hole" picture
from which we can define our cluster operators.
In our numerical experiments,
particle--hole disentangled uCC makes only a small improvement on mean-field
AGP in the attractive regime. However, we can remedy this by optimizing the AGP 
orbitals and geminal coefficients in the presence of the particle--hole 
correlator. As we shall see, we obtain significant 
improvement by so doing. Our observation is consistent with the results 
obtained in Ref. \cite{baran_variational_2021} which was carried out for 
the pairing Hamiltonian.

Again, we implemented \Eq{\ref{eq:ucc-ordering}} for all of our 
\textit{Ansatz}. However, instead of fixed
orbital-rotation parameters, we optimized them simultaneously with uCC
amplitudes. For AGP-based methods, we also optimized $\{\eta_p\}$ as discussed
in \Sec{\ref{subsec:agp-optimization}}. Indeed, for HF, reference optimization 
corresponds to optimizing the orbitals only. For all cases, the Hamiltonian was 
measured in the on-site basis. 

In \Fig{\ref{fig:oo-uCC}} we show the results for the particle--hole uCC based 
on UAGP for 6 sites, 1- and 2-dimensions, with 4 and 6 electrons. We performed
UHF-based calculations with the same optimization settings and \ansatz ordering
as those for AGP. Just as in the fixed-reference case, we observed largest
gains in the attractive regime. Nevertheless, our numerical results suggest
that AGP can be just as competitive as HF in $U>0$ as a starting point for
corrected method in all of the systems that we tried. 

\section{Discussions}\label{sec:discussions}
A desirable property for electronic structure methods is accurately  
describing the weakly and strongly correlated regimes seamlessly.
One way to achieve this goal is to restore symmetries that break at the 
mean-field level as the result of strong correlation
and capture the remaining correlation energy using CC-type correlators.

Building on our previous work that dealt with correlating AGP in the
seniority-zero subspace, we developed a coupled cluster method to correlate AGP 
for general, seniority nonzero Hamiltonians on a quantum computer. We 
benchmarked our numerical results against the ground state of 
the Fermi--Hubbard 
Hamiltonian---a model Hamiltonian that breaks number symmetry when $U<0$ and 
breaks various spin-symmetries when $U>0$. 

We proposed two techniques: First, we optimize
AGP on a classical computer then correlate it on a quantum computer. We 
showed that uUCCGSD on AGP is highly accurate and capable of accessing all 
relevant parts of the Hilbert space. 
Using a na\"ive implementation of ADAPT-VQE 
on AGP in a 2-dimensional, doped Hubbard model, we observed that 
AGP-based uUCCGSD could give a sparser,
yet highly accurate \ansatz in the regime where number 
symmetry breaks. Our second technique uses a particle--hole uCC based on 
AGP. The separation between particles and holes in AGP is not as 
well-defined as HF, and the orbitals are generally very different. 
However, by relaxing the reference in the presence of the 
particle--hole uCC correlator, we obtained excellent results in our numerical 
benchmarks.

In addition to correlating AGP, we illustrated how
post-selection can fit well into our formalism to restore particle number. 
Symmetry restoration can be realized in a variety of ways
on a quantum computer, including gauge integration and phase estimation.
Post-selection relieves the need for implementing the Hadamard tests to
evaluate the integral and is less expensive than phase estimation. Our main 
contribution is to show that the number of measurements in post-selection 
scales as \bigO{\sqrt{M}} in the worst case (where number is strongly-broken) 
in the absence of noise. We proved this analytically, and tested it numerically 
for the pairing Hamiltonian using a quantum emulator.
Our numerical results suggest scaling of as low as \bigO{1} in the regime where 
number symmetry does not break. In contrast, the measurement cost of exact 
gauge integration is \bigO{M} and independent of the strength of the 
correlation.

Post-selection on particle number is often a necessary step for noise 
mitigation on NISQ hardware. However, when performed on an \ansatz that 
deliberately breaks number symmetry, one can recover additional correlation 
energy, even in the regimes that number symmetry does not break. This 
observation together with the fact that the cost of constructing disentangled 
uCC on AGP is the same as HF on a quantum computer, could make AGP an 
attractive wavefunction for the strong correlation. Future work could benefit 
from restoring spin-symmetries, such as $S^2$ and $S_z$ to further improve the 
accuracy of AGP-based methods for molecules. {While issues related to 
the barren plateau problem \cite{mcclean_barren_2018} 
are less prominent for physically inspired \ansatze, 
future work could also explore convergence properties of AGP-based uCC.}

\section{Acknowledgments}
This work was supported by the U.S. Department of Energy under Award No. 
DE-SC0019374. G.E.S. is a Welch Foundation Chair (C-0036). 
A.K. is thankful to Garnet Kin-Lic Chan for insightful consultations 
on post-selection and thanks Jonathon P.
Misiewicz amd Nicholas H. Stair for helpful inputs regarding 
\texttt{QForte}. G.P.C thanks Carlos A. Jimenez-Hoyos for discussions 
regarding NHFB and its implementation. A.K. and G.P.C are grateful to Thomas M. 
Henderson for helpful discussions. 

\section*{Appendix}
\appendix

\section{Bloch--Messiah decomposition}\label{sec:bm}

An HFB state is defined as the vacuum of Bogoliubov quasiparticles,
\begin{equation}
  \ket{\text{HFB}} = \prod_{p=1}^{2M} \beta_p \rVac,
\end{equation}
where the \textit{quasiparticle operators} $\beta_p$, $\beta^\dag_p$
are defined by a unitary canonical transformation $\mathcal{W}$
of the physical fermionic operators
$c_p$, $c^\dag_p$,
\begin{equation}
  \begin{pmatrix}
    \beta\\ \beta^\dag
  \end{pmatrix}
  = \mathcal{W}^\dag
  \begin{pmatrix}
    c\\ c^\dag
  \end{pmatrix}
  = \begin{pmatrix}
    U^\dag & V^\dag\\
    V^T & U^T
  \end{pmatrix}
  \begin{pmatrix}
    c\\ c^\dag
  \end{pmatrix},
\end{equation}
wherein $U$ and $V$ are known as the Bogoliubov coefficients.
By the Bloch--Messiah decomposition, we can bring $U$ and $V$ into a
canonical form, which enables us to write the HFB as a
BCS wavefunction in the natural orbital basis. As a consequence,
a NHFB state optimized on a classical computer can be used
to prepare an AGP on a quantum computer.

\subsection{Bloch--Messiah theorem}\label{sub:bmthm}

The Bloch--Messiah theorem states that $\mathcal{W}$ can be decomposed
into three unitary canonical transformations of special forms,
\begin{equation}
  \mathcal{W}
  = \begin{pmatrix}
    U & V^*\\
    V & U^*
  \end{pmatrix}
  = \begin{pmatrix}
    D & 0\\
    0 & D^*
  \end{pmatrix}
  \begin{pmatrix}
    \bar{U} & \bar{V}\\
    \bar{V} & \bar{U}
  \end{pmatrix}
  \begin{pmatrix}
    C & 0\\
    0 & C^*
  \end{pmatrix},
\end{equation}
where $D$ and $C$ are unitary matrices and
\begin{subequations}
  \label{eq:ubarvbar}
  \begin{align}
    \bar{U}
    &= \bigoplus_{p=1}^{M}
    \begin{pmatrix}
      u_p & 0\\
      0 & u_p
    \end{pmatrix},\\
    \bar{V}
    &= \bigoplus_{p=1}^{M}
    \begin{pmatrix}
      0 & v_p\\
      -v_p & 0
    \end{pmatrix},
  \end{align}
\end{subequations}
are real matrices whose nonzero elements are BCS coefficients $\{u_p, v_p\}$.
$D$ is called the canonical orbital coefficients and is identical
to the natural orbital coefficients of the geminal defined in
\Eq{\ref{eq:natorb}}, hence the same symbol.
This can be readily seen by recognizing
$\eta = (V U^{-1})^*$ and $\bar{\eta} = (\bar{V} \bar{U}^{-1})^*$.
Incidentally, the most general form of the Bloch--Messiah theorem applies
to both odd and even electron systems, albeit we restrict ourselves to even
electron systems herein.

\subsection{Computing Bloch--Messiah decomposition}\label{sub:bmcomp}

Given an HFB, we denote its one-particle reduce density matrix by $\rho$
and pairing matrix by $\kappa$. $\rho$ is Hermitian and $\kappa$ antisymmetric.
They relate to the Bogoliubov coefficients through \cite{ring_nuclear_1980}
\begin{subequations}
  \begin{align}
    \rho &= V^* V^T,\\
    \kappa &= V^* U^T,
  \end{align}
\end{subequations}
and thus satisfy
\begin{subequations}
  \label{eq:rhokappa}
  \begin{align}
    \rho \kappa &= \kappa \rho^*,\\
    \rho - \rho^2 &= \kappa \kappa^\dag.
  \end{align}
\end{subequations}

Computing the Bloch--Messiah decomposition amounts to simultaneously
diagonalizing $\rho$ and canonicalizing $\kappa$ \cite{Hua_theory_1944}
in the sense of
\begin{subequations}
  \label{eq:diagrhokappa}
  \begin{align}
    \bar{\rho}
    &= \bigoplus_{p=1}^M 
    \begin{pmatrix}
      \rho_p & 0\\
      0 & \rho_p
    \end{pmatrix}
    = D^\dag \rho D,\\
    \bar{\kappa}
    &= \bigoplus_{p=1}^M
    \begin{pmatrix}
      0 & \kappa_p\\
      -\kappa_p & 0
    \end{pmatrix}
    = D^\dag \kappa D^*,
  \end{align}
\end{subequations}
where $\bar{\rho}$ and $\bar{\kappa}$ are real.
To do this, we just need to canonicalize $\kappa$ in each
degenerate eigen subspace of $\rho$.
Canonicalization of an antisymmetric matrix can be performed using
algorithms described in \cite{wimmer_algorithm_2012}.
Subsequently, $\{u_p, v_p\}$, or equivalently $\{\eta_p\}$, can
be determined from the values of $\{\rho_p, \kappa_p\}$.

Specifically for UHFB, $\rho$ and $\kappa$ have the following
spin block structures:
\begin{subequations}
    \begin{align}
        \rho &=
        \begin{pmatrix}
            \rho\saa & 0\\
            0 & \rho\sbb
        \end{pmatrix},\\
        \kappa &=
        \begin{pmatrix}
            0 & \kappa\sab\\
            -(\kappa\sab)^T & 0
        \end{pmatrix}.
    \end{align}
\end{subequations}
By \Eq{\ref{eq:rhokappa}}, we have
\begin{subequations}
  \begin{align}
    \left[\rho\saa, \: \kappa\sab \left(\kappa\sab\right)^\dag\right] &= 0,\\
    \left[\rho\sbb, \: \left(\kappa\sab\right)^T \kappa^{\alpha\beta *}\right] &= 0,
  \end{align}
\end{subequations}
which implies that we can simultaneously diagonalize the Hermitian matrices
$\rho\saa$, $\rho\sbb$, and $\kappa\sab \left(\kappa\sab\right)^\dag$.
We can show that the eigenvalues of $\rho\saa$ are identical to those
of $\rho\sbb$. Moreover, we can always find unitary
matrices $D^\alpha$ and $D^\beta$ such that the columns of $D^\alpha$ are the
eigenvectors of $\rho\saa$ and the left singular vectors of $\kappa\sab$,
while the columns of $D^\beta$ are the eigenvectors of $\rho\sbb$ and
the complex conjugate of the right singular vectors of $\kappa\sab$.
The matrix of canonical orbital coefficients is then written as
\begin{equation}
  \label{eq:uhfb_coefficients}
  D = \begin{pmatrix}
    D^\alpha & 0\\
    0 & D^\beta
  \end{pmatrix}
  P^T,
\end{equation}
where $P$ is the permutation matrix of
\begin{equation}
  \begin{pmatrix}
    1 & 2 & \cdots &    M & M+1 & M+2 & \cdots & 2M\\
    1 & 3 & \cdots & 2M-1 &   2 &   4 & \cdots & 2M  
  \end{pmatrix}.
\end{equation}
This permutation arises because we choose to put the paired spin-orbitals
$p$ and $\bar{p}$ adjacent to each other in \Eq{\ref{eq:ubarvbar}}.
We see from \Eq{\ref{eq:uhfb_coefficients}} that a $(p,\, \bar{p})$ pair
in UHFB is a spin pair. Consequently, the natural orbital pairs in UAGP
are spin pairs.

We can carry out a similar procedure to find the canonical orbital
coefficients $D$ and thereby the Bloch--Messiah decomposition for RHFB.
In this case, $D$ is in the form of \Eq{\ref{eq:uhfb_coefficients}}
with $D^\alpha = D^\beta$.

\section{Post-selection analysis}\label{appx:post-selection}

\subsection{Fixing the gauge}\label{subappx:fix-gauge}

As evident from \Eq{\ref{eq:agp}}, multiplying all $\{\eta_p\}$ by a constant, $c$, amounts
to multiplying AGP by an inconsequential global constant $c^N$. The same
however is not true for the BCS wavefunction, since a global constant cannot be
factorized. In BCS, multiplying all $\eta_p$ changes the average particle
number. Therefore, given a set of $\eta_p$, we can solve for $c$ so that the average particle number is
fixed. Recall that, 
\begin{align}
    \Ebcs{\hat{N}} = 
    \sum_{p=1}^M 2 {v_p^2} = 2 \sum_{p=1}^M \frac{{\eta_p^2}}{1+{\eta_p^2}},
\end{align}
where we used the fact that $\eta_p = v_p/u_p$ and ${u_p^2} + {v_p^2}=1$ in the
last equality. We need, $\E{\hat{N}} = 2N$, so we can solve for $c$ (it
can be taken to be real) that satisfies, 
\begin{align}
    \sum_p \frac{(c\eta_p)^2}{1+(c\eta_p)^2} - N = 0.
\end{align}
We are not aware of a simple {closed-form} solution. 
However, it is easy to
solve for $c$ numerically with standard root finding algorithms. This can be
particularly helpful when optimizing the reference AGP on a quantum computer
such as \Sec{\ref{subsec:oo-uCC}} or \Sec{\ref{subsec:agp-optimization}}, 
where for every guess of $\{\eta_p\}$, they can be scaled to
maximize sampling outcomes. 

\subsection{Scaling of post-selection}\label{subappx:ps-scaling}

We want show that the asymptotic scaling of post-selection is no worse that 
\bigO{\sqrt{M}}. To this end, we first derive an analytical expression for the 
number of measurements needed to obtain at least one sample in the correct 
particle number sector. We then find an upper bound to this 
expression which gives us the asymptotic scaling.

Typically, we are interested in computing the expectation value of some 
observable $\hat{O}$ over correlated AGP. we 
concentrate on number conserving operators, since otherwise their expectation 
values over AGP are zero. Let $\hat{U}$ define a number conserving correlator so
that $[\Proj, \hat{U}]=0$. We want to compute
\begin{align} \label{eq:sample-from-agp}
    \frac{\Eagp{\hat{U}^{\dagger}\hat{O}\hat{U}}}{\AGPnorm} = 
    \sum_{\alpha} \lambda_{\alpha} 
    \frac{|\bra{\lambda_{\alpha}} \hat{V}\hat{U}\rAGP|^2}{\AGPnorm},
\end{align}
where $\hat{V}$ diagonalizes $\hat{O}$, (i.e. $\hat{O} = 
\hat{V}^{\dagger}{\hat{\Lambda}}\hat{V}$) and $\ket{\lambda_{\alpha}}$ is a bit string 
corresponding to eigenvector of ${\hat{\Lambda}}$ with eigenvalue $\lambda_{\alpha}$. 
In practice, we need to compute \Eq{\ref{eq:sample-from-agp}} using the 
projected BCS wavefunction. 
Therefore we have
\begin{align}\label{eq:sample-from-pbcs}
    |\bra{\lambda_{\alpha}} \Proj \hat{V} \hat{U} \rBCS|^2 = 
    |\bra{\lambda_{\alpha}} \hat{V} \hat{U}\Proj \rBCS|^2 \nonumber \\
    = \underbrace{\frac{|\bra{\lambda_{\alpha}} \hat{V}\hat{U}\rAGP|^2}
    {\AGPnorm}}_{\prob{\lambda_\alpha | N}}
    \underbrace{\frac{|\langle\text{AGP}\rBCS|^2}{\AGPnorm}}_{\prob{N}},
\end{align}
where $\prob{N}$ is the probability of getting a state with $N$ pairs and $\prob
{\lambda_\alpha | N}$ is the probability of getting $\lambda_\alpha$ 
conditioned on $N$. This shows that, the scaling of 
post-selection is solely determined by $\prob{N}$ and is independent of the 
operator $\hat{O}$ and the correlator $\hat{U}$ so long as they are 
number conserving. 

From \Eq{\ref{eq:sample-from-pbcs}} we have
\begin{align}\label{eq:prob-N}
    \prob{N} = \prod_p {u_p^2} \AGPnorm,
\end{align}    
where $\AGPnorm = \sum_{1 \leq p_1< \cdots <p_{N} \leq M} 
{\eta_{p_1}^2} \cdots {\eta_{p_N}^2}
^= S_N^M $ is an elementary symmetric polynomial (ESP) 
\cite{khamoshi_efficient_2019}. If we did $n$ 
independent measurements, the probability of getting $k$ observations in the
desired particle sector (in the absence of noise) follows a Binomial
distribution,
\begin{align}
    {\prob{k; n}} = {n\choose k} P^{k}(1-P)^{n-k},
\end{align}
where $P$ is a shorthand notation for $\prob{N}$. Therefore, it follows that 
\begin{align} \label{eq:n-theory-appx}
    \prob{k \geq 1} = \alpha \implies n = \frac{\log (1-{\alpha})}{\log{(1-P)}}.
\end{align}
rounded to the closest integer (ceiling). For example, if we want to be 95\% 
confident that there is at least one observation in the desired particle 
sector, we need a sample size of $n = \lceil {-2.996}/{\log{(1-P)}} \rceil$, 
where $P$ needs to be computed numerically from \Eq{\ref{eq:prob-N}}.

Finding an upper bound for $n$ is equivalent to bounding $P$ from below. The 
precise value of $P$ depends on a given Hamiltonian and varies at different 
correlation regimes; except for special cases, an analytic expressions for $P$ 
is not known. However, since $0<P<1$ we can look at its limiting cases to find 
the lower and upper bounds. 

As $P\rightarrow 1$, we get $n\rightarrow1$, which corresponds to the HF limit 
of AGP where fluctuations in particle number are minimal. Indeed this is the 
upper bound to $P$. The lower limit corresponds to the opposite case where 
number symmetry is strongly broken and the number fluctuations are at their 
peak. Physically, this is associated with the superfluid phase where the 
occupation number of all orbitals become equal. In this case---assuming 
$\{\eta_p\}$ normalized so that $\Ebcs{\hat{N}}=2N$ (\textit{vide supra})---we 
get
\begin{subequations}
    \begin{align}
        & v_p^2 = \frac{N}{M}, \quad u_p^2 = \frac{M - N}{M},  \\
        & \eta_p^2 = \frac{N}{M-N}.
    \end{align}
\end{subequations}
The lower-bound of $P$ can be obtained by plugging these expressions into 
\Eq{\ref{eq:prob-N}},
so we arrive at
\begin{align}
    P = \frac{N^N(M - N)^{M-N}}{M^M} {M \choose N}.
\end{align}
Using Stirling's approximation,
\begin{align}
    {M \choose N} \sim 
    {M^M \over N^N (M-N)^{M-N}} \sqrt{M \over 2\pi N (M-N)},
\end{align} we get
\begin{align}
    P \approx \sqrt{M \over 2\pi N (M-N)}.
\end{align}
Let $\xi = N/M$ define an intensive quantity. We may assume $\xi \leq 1/2$ for
all practical purposes. Simplifying $P$ gives
  \begin{align}\label{eq:approx-P}
      P & \approx \frac{1}{\sqrt{ 2\pi M \xi(1-\xi)}} \geq \frac{1}{\sqrt{ \pi M / 2}}.
  \end{align}
Finally, to get the asymptotic scaling for $n$, we Taylor expand near $P\rightarrow 0$,
\begin{align}
    n \propto  \frac{1}{-\log{(1-P)}}
    = \frac{1}{P} - \frac{1}{2} - \frac{P}{12} + \mathcal{O}(P^2), 
\end{align}
and plug in \Eq{\ref{eq:approx-P}} to obtain
\begin{align}
    n &\propto \sqrt{\frac{ \pi M}{2}} + \mathcal{O}(P).
\end{align}
This finishes the proof as it shows $n \sim \mathcal{O}(\sqrt{M})$

\section{Implementation details}\label{appx:ucc-imp}
In \Sec{\ref{sec:application}}, we used \texttt{PySCF} \cite{sun_p_2018} to 
carry out all UHF and traditional 
UCCSD calculations. Our quantum algorithms were 
implemented using the \texttt{QForte} \cite{stair_qforte_2022} 
software with the ``state-vector" emulator in the absence of noise. We
made slight modifications to \texttt{QForte} 
libraries to implement AGP and orbital 
rotation/optimization in an in-house code. 
For the fixed-reference UHF-based calculations, we obtained the 
molecular orbital coefficients from \texttt{PySCF}; for AGP-based
calculations, we used an in-house NHFB code.

For computational feasibility of our quantum algorithms, 
we implemented AGP by brute force; we looped over 
all states in the Hilbert space and assigned appropriate coefficients to each 
state. The procedure can be more easily seen if we expand \Eq{\ref{eq:agp}} 
to get 
\begin{align}
  \rAGP = \sum_{ 1 \leq p_1< \cdots <p_N \leq M} \eta _{p_1} \cdots \eta _{p_N}
  \Pdag{p_1} \cdots \Pdag{p_{N}} \rVac.
\end{align}
For example, for the $M=4$, $N=2$ state, we have 
\begin{align} 
  \rAGP &= \nonumber \\
  &{}\eta_1\eta_2\ket{00001111} + \eta_1\eta_3\ket{00110011} \nonumber \\
+ &{}\eta_1\eta_4\ket{11000011} + \eta_2\eta_3\ket{00111100} \nonumber \\
+ &{}\eta_2\eta_4\ket{11001100} + \eta_3\eta_4\ket{11110000},
\end{align}

Note that this state is not normalized, i.e. $\AGPnorm \neq 1$; it is 
necessary to divide all energy and gradient expressions with appropriate 
normalization. The gradient with respect to $\eta_p$ can be computed from 
\cite{khamoshi_efficient_2019}
\begin{align}
  \frac{d}{d \eta_p} \rAGP = \frac{1}{2 \eta_p} \N{p} \rAGP.
\end{align}
On a quantum computer, however, we can efficiently compute the gradient by 
applying the shift-rule to the parameters of the unitary that implements the 
BCS wavefunction and number projecting the resulting term. This follows from 
the fact that $\left[\frac{d}{d\eta_p}, \Proj\right]=0$.

We relied on the existing implementation for ADAPT and VQE calculations which 
use the \texttt{SciPy} libraries \cite{virtanen_scipy_2020} to do the 
intermediate optimizations. We set the
optimizer to be L-BFGS-B \cite{byrd_limited_1995}; 
the gradients threshold as well as the energy 
difference tolerance were set to $10^{-9}$. In ADAPT, there is an additional 
tolerance, $\epsilon$, that determines when it terminates; 
we set $\epsilon = 10^{-3}$ for the results shown in 
\Fig{\ref{fig:adapt}}. In all calculations, the 
gradients were computed analytically. 

To implement orbital rotation, we used an in-house code to perform a QR 
decomposition in terms of Givens rotations; we followed the work of 
Ref. \cite{kivlichan_quantum_2018} closely and implemented the rotations 
in \texttt{QForte} using 1-body rotations acting 
on nearest neighbors. For the orbital optimization of \Sec{\ref{subsec:oo-uCC}}
we treated each of these 1-body rotations as independent parameters. Note 
that care must be given in making sure spin symmetries are retained. In all of 
our calculation with orbital optimization, we made certain that $S_z$ does not 
break---hence UAGP and UHF---by construction.

\bibliography{main}

\begin{thebibliography}{138}%
\makeatletter
\providecommand \@ifxundefined [1]{%
 \@ifx{#1\undefined}
}%
\providecommand \@ifnum [1]{%
 \ifnum #1\expandafter \@firstoftwo
 \else \expandafter \@secondoftwo
 \fi
}%
\providecommand \@ifx [1]{%
 \ifx #1\expandafter \@firstoftwo
 \else \expandafter \@secondoftwo
 \fi
}%
\providecommand \natexlab [1]{#1}%
\providecommand \enquote  [1]{``#1''}%
\providecommand \bibnamefont  [1]{#1}%
\providecommand \bibfnamefont [1]{#1}%
\providecommand \citenamefont [1]{#1}%
\providecommand \href@noop [0]{\@secondoftwo}%
\providecommand \href [0]{\begingroup \@sanitize@url \@href}%
\providecommand \@href[1]{\@@startlink{#1}\@@href}%
\providecommand \@@href[1]{\endgroup#1\@@endlink}%
\providecommand \@sanitize@url [0]{\catcode `\\12\catcode `\$12\catcode
  `\&12\catcode `\#12\catcode `\^12\catcode `\_12\catcode `\%12\relax}%
\providecommand \@@startlink[1]{}%
\providecommand \@@endlink[0]{}%
\providecommand \url  [0]{\begingroup\@sanitize@url \@url }%
\providecommand \@url [1]{\endgroup\@href {#1}{\urlprefix }}%
\providecommand \urlprefix  [0]{URL }%
\providecommand \Eprint [0]{\href }%
\providecommand \doibase [0]{http://dx.doi.org/}%
\providecommand \selectlanguage [0]{\@gobble}%
\providecommand \bibinfo  [0]{\@secondoftwo}%
\providecommand \bibfield  [0]{\@secondoftwo}%
\providecommand \translation [1]{[#1]}%
\providecommand \BibitemOpen [0]{}%
\providecommand \bibitemStop [0]{}%
\providecommand \bibitemNoStop [0]{.\EOS\space}%
\providecommand \EOS [0]{\spacefactor3000\relax}%
\providecommand \BibitemShut  [1]{\csname bibitem#1\endcsname}%
\let\auto@bib@innerbib\@empty
\bibitem [{\citenamefont {Cao}\ \emph {et~al.}(2019)\citenamefont {Cao},
  \citenamefont {Romero}, \citenamefont {Olson}, \citenamefont {Degroote},
  \citenamefont {Johnson}, \citenamefont {Kieferov{\'a}}, \citenamefont
  {Kivlichan}, \citenamefont {Menke}, \citenamefont {Peropadre}, \citenamefont
  {Sawaya}, \citenamefont {Sim}, \citenamefont {Veis},\ and\ \citenamefont
  {Aspuru-Guzik}}]{cao_quantum_2019}%
  \BibitemOpen
  \bibfield  {author} {\bibinfo {author} {\bibfnamefont {Y.}~\bibnamefont
  {Cao}}, \bibinfo {author} {\bibfnamefont {J.}~\bibnamefont {Romero}},
  \bibinfo {author} {\bibfnamefont {J.~P.}\ \bibnamefont {Olson}}, \bibinfo
  {author} {\bibfnamefont {M.}~\bibnamefont {Degroote}}, \bibinfo {author}
  {\bibfnamefont {P.~D.}\ \bibnamefont {Johnson}}, \bibinfo {author}
  {\bibfnamefont {M.}~\bibnamefont {Kieferov{\'a}}}, \bibinfo {author}
  {\bibfnamefont {I.~D.}\ \bibnamefont {Kivlichan}}, \bibinfo {author}
  {\bibfnamefont {T.}~\bibnamefont {Menke}}, \bibinfo {author} {\bibfnamefont
  {B.}~\bibnamefont {Peropadre}}, \bibinfo {author} {\bibfnamefont {N.~P.~D.}\
  \bibnamefont {Sawaya}}, \bibinfo {author} {\bibfnamefont {S.}~\bibnamefont
  {Sim}}, \bibinfo {author} {\bibfnamefont {L.}~\bibnamefont {Veis}}, \ and\
  \bibinfo {author} {\bibfnamefont {A.}~\bibnamefont {Aspuru-Guzik}},\ }\href
  {\doibase 10.1021/acs.chemrev.8b00803} {\bibfield  {journal} {\bibinfo
  {journal} {Chem. Rev.}\ }\textbf {\bibinfo {volume} {119}},\ \bibinfo {pages}
  {10856} (\bibinfo {year} {2019})}\BibitemShut {NoStop}%
\bibitem [{\citenamefont {McArdle}\ \emph {et~al.}(2020)\citenamefont
  {McArdle}, \citenamefont {Endo}, \citenamefont {Aspuru-Guzik}, \citenamefont
  {Benjamin},\ and\ \citenamefont {Yuan}}]{mcardle_quantum_2020}%
  \BibitemOpen
  \bibfield  {author} {\bibinfo {author} {\bibfnamefont {S.}~\bibnamefont
  {McArdle}}, \bibinfo {author} {\bibfnamefont {S.}~\bibnamefont {Endo}},
  \bibinfo {author} {\bibfnamefont {A.}~\bibnamefont {Aspuru-Guzik}}, \bibinfo
  {author} {\bibfnamefont {S.~C.}\ \bibnamefont {Benjamin}}, \ and\ \bibinfo
  {author} {\bibfnamefont {X.}~\bibnamefont {Yuan}},\ }\href {\doibase
  10.1103/RevModPhys.92.015003} {\bibfield  {journal} {\bibinfo  {journal}
  {Rev. Mod. Phys.}\ }\textbf {\bibinfo {volume} {92}},\ \bibinfo {pages}
  {015003} (\bibinfo {year} {2020})}\BibitemShut {NoStop}%
\bibitem [{\citenamefont {Helgaker}\ \emph {et~al.}(2000)\citenamefont
  {Helgaker}, \citenamefont {J{\o}rgensen},\ and\ \citenamefont
  {Olsen}}]{helgaker_molecular_2000}%
  \BibitemOpen
  \bibfield  {author} {\bibinfo {author} {\bibfnamefont {T.}~\bibnamefont
  {Helgaker}}, \bibinfo {author} {\bibfnamefont {P.}~\bibnamefont
  {J{\o}rgensen}}, \ and\ \bibinfo {author} {\bibfnamefont {J.}~\bibnamefont
  {Olsen}},\ }\href@noop {} {\emph {\bibinfo {title} {Molecular
  electronic-structure theory}}}\ (\bibinfo  {publisher} {Wiley},\ \bibinfo
  {address} {Chichester; New York},\ \bibinfo {year} {2000})\BibitemShut
  {NoStop}%
\bibitem [{\citenamefont {Scuseria}\ \emph {et~al.}(1987)\citenamefont
  {Scuseria}, \citenamefont {Scheiner}, \citenamefont {Lee}, \citenamefont
  {Rice},\ and\ \citenamefont {Schaefer}}]{scuseria_closedshell_1987}%
  \BibitemOpen
  \bibfield  {author} {\bibinfo {author} {\bibfnamefont {G.~E.}\ \bibnamefont
  {Scuseria}}, \bibinfo {author} {\bibfnamefont {A.~C.}\ \bibnamefont
  {Scheiner}}, \bibinfo {author} {\bibfnamefont {T.~J.}\ \bibnamefont {Lee}},
  \bibinfo {author} {\bibfnamefont {J.~E.}\ \bibnamefont {Rice}}, \ and\
  \bibinfo {author} {\bibfnamefont {H.~F.}\ \bibnamefont {Schaefer}},\ }\href
  {\doibase 10.1063/1.452039} {\bibfield  {journal} {\bibinfo  {journal} {J.
  Chem. Phys.}\ }\textbf {\bibinfo {volume} {86}},\ \bibinfo {pages} {2881}
  (\bibinfo {year} {1987})}\BibitemShut {NoStop}%
\bibitem [{\citenamefont {Bartlett}\ and\ \citenamefont {Musia{\l
  }}(2007)}]{bartlett_coupled-cluster_2007}%
  \BibitemOpen
  \bibfield  {author} {\bibinfo {author} {\bibfnamefont {R.~J.}\ \bibnamefont
  {Bartlett}}\ and\ \bibinfo {author} {\bibfnamefont {M.}~\bibnamefont
  {Musia{\l }}},\ }\href {\doibase 10.1103/RevModPhys.79.291} {\bibfield
  {journal} {\bibinfo  {journal} {Rev. Mod. Phys.}\ }\textbf {\bibinfo {volume}
  {79}},\ \bibinfo {pages} {291} (\bibinfo {year} {2007})}\BibitemShut
  {NoStop}%
\bibitem [{\citenamefont {Bulik}\ \emph {et~al.}(2015)\citenamefont {Bulik},
  \citenamefont {Henderson},\ and\ \citenamefont {Scuseria}}]{bulik_can_2015}%
  \BibitemOpen
  \bibfield  {author} {\bibinfo {author} {\bibfnamefont {I.~W.}\ \bibnamefont
  {Bulik}}, \bibinfo {author} {\bibfnamefont {T.~M.}\ \bibnamefont
  {Henderson}}, \ and\ \bibinfo {author} {\bibfnamefont {G.~E.}\ \bibnamefont
  {Scuseria}},\ }\href {\doibase 10.1021/acs.jctc.5b00422} {\bibfield
  {journal} {\bibinfo  {journal} {J. Chem. Theory Comput.}\ }\textbf {\bibinfo
  {volume} {11}},\ \bibinfo {pages} {3171} (\bibinfo {year}
  {2015})}\BibitemShut {NoStop}%
\bibitem [{\citenamefont {Ring}\ and\ \citenamefont
  {Schuck}(1980)}]{ring_nuclear_1980}%
  \BibitemOpen
  \bibfield  {author} {\bibinfo {author} {\bibfnamefont {P.}~\bibnamefont
  {Ring}}\ and\ \bibinfo {author} {\bibfnamefont {P.}~\bibnamefont {Schuck}},\
  }\href {https://www.springer.com/gp/book/9783540212065} {\emph {\bibinfo
  {title} {The {Nuclear} {Many}-{Body} {Problem}}}},\ Theoretical and
  {Mathematical} {Physics}, {The} {Nuclear} {Many}-{Body} {Problem}\ (\bibinfo
  {publisher} {Springer-Verlag},\ \bibinfo {address} {Berlin Heidelberg},\
  \bibinfo {year} {1980})\BibitemShut {NoStop}%
\bibitem [{\citenamefont {Blaizot}\ and\ \citenamefont
  {Ripka}(1986)}]{blaizot_quantum_1986}%
  \BibitemOpen
  \bibfield  {author} {\bibinfo {author} {\bibfnamefont {J.-P.}\ \bibnamefont
  {Blaizot}}\ and\ \bibinfo {author} {\bibfnamefont {G.}~\bibnamefont
  {Ripka}},\ }\href@noop {} {\emph {\bibinfo {title} {Quantum theory of finite
  systems}}}\ (\bibinfo  {publisher} {MIT Press},\ \bibinfo {address}
  {Cambridge, Mass.},\ \bibinfo {year} {1986})\BibitemShut {NoStop}%
\bibitem [{\citenamefont {L{\"o}wdin}(1955)}]{lowdin_quantum_1955}%
  \BibitemOpen
  \bibfield  {author} {\bibinfo {author} {\bibfnamefont {P.-O.}\ \bibnamefont
  {L{\"o}wdin}},\ }\href {\doibase 10.1103/PhysRev.97.1509} {\bibfield
  {journal} {\bibinfo  {journal} {Phys. Rev.}\ }\textbf {\bibinfo {volume}
  {97}},\ \bibinfo {pages} {1509} (\bibinfo {year} {1955})}\BibitemShut
  {NoStop}%
\bibitem [{\citenamefont {Lykos}\ and\ \citenamefont
  {Pratt}(1963)}]{lykos_discussion_1963}%
  \BibitemOpen
  \bibfield  {author} {\bibinfo {author} {\bibfnamefont {P.}~\bibnamefont
  {Lykos}}\ and\ \bibinfo {author} {\bibfnamefont {G.~W.}\ \bibnamefont
  {Pratt}},\ }\href {\doibase 10.1103/RevModPhys.35.496} {\bibfield  {journal}
  {\bibinfo  {journal} {Rev. Mod. Phys.}\ }\textbf {\bibinfo {volume} {35}},\
  \bibinfo {pages} {496} (\bibinfo {year} {1963})}\BibitemShut {NoStop}%
\bibitem [{\citenamefont {Scuseria}\ \emph {et~al.}(2011)\citenamefont
  {Scuseria}, \citenamefont {Jim{\'e}nez-Hoyos}, \citenamefont {Henderson},
  \citenamefont {Samanta},\ and\ \citenamefont
  {Ellis}}]{scuseria_projected_2011}%
  \BibitemOpen
  \bibfield  {author} {\bibinfo {author} {\bibfnamefont {G.~E.}\ \bibnamefont
  {Scuseria}}, \bibinfo {author} {\bibfnamefont {C.~A.}\ \bibnamefont
  {Jim{\'e}nez-Hoyos}}, \bibinfo {author} {\bibfnamefont {T.~M.}\ \bibnamefont
  {Henderson}}, \bibinfo {author} {\bibfnamefont {K.}~\bibnamefont {Samanta}},
  \ and\ \bibinfo {author} {\bibfnamefont {J.~K.}\ \bibnamefont {Ellis}},\
  }\href {\doibase 10.1063/1.3643338} {\bibfield  {journal} {\bibinfo
  {journal} {J. Chem. Phys.}\ }\textbf {\bibinfo {volume} {135}},\ \bibinfo
  {pages} {124108} (\bibinfo {year} {2011})}\BibitemShut {NoStop}%
\bibitem [{\citenamefont {Jim{\'e}nez-Hoyos}\ \emph {et~al.}(2012)\citenamefont
  {Jim{\'e}nez-Hoyos}, \citenamefont {Henderson}, \citenamefont {Tsuchimochi},\
  and\ \citenamefont {Scuseria}}]{jimenez-hoyos_projected_2012}%
  \BibitemOpen
  \bibfield  {author} {\bibinfo {author} {\bibfnamefont {C.~A.}\ \bibnamefont
  {Jim{\'e}nez-Hoyos}}, \bibinfo {author} {\bibfnamefont {T.~M.}\ \bibnamefont
  {Henderson}}, \bibinfo {author} {\bibfnamefont {T.}~\bibnamefont
  {Tsuchimochi}}, \ and\ \bibinfo {author} {\bibfnamefont {G.~E.}\ \bibnamefont
  {Scuseria}},\ }\href {\doibase 10.1063/1.4705280} {\bibfield  {journal}
  {\bibinfo  {journal} {J. Chem. Phys.}\ }\textbf {\bibinfo {volume} {136}},\
  \bibinfo {pages} {164109} (\bibinfo {year} {2012})}\BibitemShut {NoStop}%
\bibitem [{\citenamefont {Evangelista}(2011)}]{evangelista_alternative_2011}%
  \BibitemOpen
  \bibfield  {author} {\bibinfo {author} {\bibfnamefont {F.~A.}\ \bibnamefont
  {Evangelista}},\ }\href {\doibase 10.1063/1.3598471} {\bibfield  {journal}
  {\bibinfo  {journal} {J. Chem. Phys.}\ }\textbf {\bibinfo {volume} {134}},\
  \bibinfo {pages} {224102} (\bibinfo {year} {2011})}\BibitemShut {NoStop}%
\bibitem [{\citenamefont {Duguet}(2015)}]{duguet_symmetry_2015}%
  \BibitemOpen
  \bibfield  {author} {\bibinfo {author} {\bibfnamefont {T.}~\bibnamefont
  {Duguet}},\ }\href {\doibase 10.1088/0954-3899/42/2/025107} {\bibfield
  {journal} {\bibinfo  {journal} {J. Phys. G: Nucl. Part. Phys.}\ }\textbf
  {\bibinfo {volume} {42}},\ \bibinfo {pages} {025107} (\bibinfo {year}
  {2015})}\BibitemShut {NoStop}%
\bibitem [{\citenamefont {Qiu}\ \emph {et~al.}(2017)\citenamefont {Qiu},
  \citenamefont {Henderson}, \citenamefont {Zhao},\ and\ \citenamefont
  {Scuseria}}]{qiu_projected_2017}%
  \BibitemOpen
  \bibfield  {author} {\bibinfo {author} {\bibfnamefont {Y.}~\bibnamefont
  {Qiu}}, \bibinfo {author} {\bibfnamefont {T.~M.}\ \bibnamefont {Henderson}},
  \bibinfo {author} {\bibfnamefont {J.}~\bibnamefont {Zhao}}, \ and\ \bibinfo
  {author} {\bibfnamefont {G.~E.}\ \bibnamefont {Scuseria}},\ }\href {\doibase
  10.1063/1.4991020} {\bibfield  {journal} {\bibinfo  {journal} {J. Chem.
  Phys.}\ }\textbf {\bibinfo {volume} {147}},\ \bibinfo {pages} {064111}
  (\bibinfo {year} {2017})}\BibitemShut {NoStop}%
\bibitem [{\citenamefont {Cerezo}\ \emph {et~al.}(2021)\citenamefont {Cerezo},
  \citenamefont {Arrasmith}, \citenamefont {Babbush}, \citenamefont {Benjamin},
  \citenamefont {Endo}, \citenamefont {Fujii}, \citenamefont {McClean},
  \citenamefont {Mitarai}, \citenamefont {Yuan}, \citenamefont {Cincio},\ and\
  \citenamefont {Coles}}]{cerezo_variational_2021}%
  \BibitemOpen
  \bibfield  {author} {\bibinfo {author} {\bibfnamefont {M.}~\bibnamefont
  {Cerezo}}, \bibinfo {author} {\bibfnamefont {A.}~\bibnamefont {Arrasmith}},
  \bibinfo {author} {\bibfnamefont {R.}~\bibnamefont {Babbush}}, \bibinfo
  {author} {\bibfnamefont {S.~C.}\ \bibnamefont {Benjamin}}, \bibinfo {author}
  {\bibfnamefont {S.}~\bibnamefont {Endo}}, \bibinfo {author} {\bibfnamefont
  {K.}~\bibnamefont {Fujii}}, \bibinfo {author} {\bibfnamefont {J.~R.}\
  \bibnamefont {McClean}}, \bibinfo {author} {\bibfnamefont {K.}~\bibnamefont
  {Mitarai}}, \bibinfo {author} {\bibfnamefont {X.}~\bibnamefont {Yuan}},
  \bibinfo {author} {\bibfnamefont {L.}~\bibnamefont {Cincio}}, \ and\ \bibinfo
  {author} {\bibfnamefont {P.~J.}\ \bibnamefont {Coles}},\ }\href {\doibase
  10.1038/s42254-021-00348-9} {\bibfield  {journal} {\bibinfo  {journal} {Nat.
  Rev. Phys.}\ }\textbf {\bibinfo {volume} {3}},\ \bibinfo {pages} {625}
  (\bibinfo {year} {2021})}\BibitemShut {NoStop}%
\bibitem [{\citenamefont {Shen}\ \emph {et~al.}(2017)\citenamefont {Shen},
  \citenamefont {Zhang}, \citenamefont {Zhang}, \citenamefont {Zhang},
  \citenamefont {Yung},\ and\ \citenamefont {Kim}}]{shen_quantum_2017}%
  \BibitemOpen
  \bibfield  {author} {\bibinfo {author} {\bibfnamefont {Y.}~\bibnamefont
  {Shen}}, \bibinfo {author} {\bibfnamefont {X.}~\bibnamefont {Zhang}},
  \bibinfo {author} {\bibfnamefont {S.}~\bibnamefont {Zhang}}, \bibinfo
  {author} {\bibfnamefont {J.-N.}\ \bibnamefont {Zhang}}, \bibinfo {author}
  {\bibfnamefont {M.-H.}\ \bibnamefont {Yung}}, \ and\ \bibinfo {author}
  {\bibfnamefont {K.}~\bibnamefont {Kim}},\ }\href {\doibase
  10.1103/PhysRevA.95.020501} {\bibfield  {journal} {\bibinfo  {journal} {Phys.
  Rev. A}\ }\textbf {\bibinfo {volume} {95}},\ \bibinfo {pages} {020501}
  (\bibinfo {year} {2017})}\BibitemShut {NoStop}%
\bibitem [{\citenamefont {Tilly}\ \emph {et~al.}(2021)\citenamefont {Tilly},
  \citenamefont {Chen}, \citenamefont {Cao}, \citenamefont {Picozzi},
  \citenamefont {Setia}, \citenamefont {Li}, \citenamefont {Grant},
  \citenamefont {Wossnig}, \citenamefont {Rungger}, \citenamefont {Booth},\
  and\ \citenamefont {Tennyson}}]{tilly_variational_2021}%
  \BibitemOpen
  \bibfield  {author} {\bibinfo {author} {\bibfnamefont {J.}~\bibnamefont
  {Tilly}}, \bibinfo {author} {\bibfnamefont {H.}~\bibnamefont {Chen}},
  \bibinfo {author} {\bibfnamefont {S.}~\bibnamefont {Cao}}, \bibinfo {author}
  {\bibfnamefont {D.}~\bibnamefont {Picozzi}}, \bibinfo {author} {\bibfnamefont
  {K.}~\bibnamefont {Setia}}, \bibinfo {author} {\bibfnamefont
  {Y.}~\bibnamefont {Li}}, \bibinfo {author} {\bibfnamefont {E.}~\bibnamefont
  {Grant}}, \bibinfo {author} {\bibfnamefont {L.}~\bibnamefont {Wossnig}},
  \bibinfo {author} {\bibfnamefont {I.}~\bibnamefont {Rungger}}, \bibinfo
  {author} {\bibfnamefont {G.~H.}\ \bibnamefont {Booth}}, \ and\ \bibinfo
  {author} {\bibfnamefont {J.}~\bibnamefont {Tennyson}},\ }\href
  {http://arxiv.org/abs/2111.05176} {\bibfield  {journal} {\bibinfo  {journal}
  {arXiv:2111.05176 [quant-ph]}\ } (\bibinfo {year} {2021})}\BibitemShut
  {NoStop}%
\bibitem [{\citenamefont {Anand}\ \emph {et~al.}(2022)\citenamefont {Anand},
  \citenamefont {Schleich}, \citenamefont {Alperin-Lea}, \citenamefont
  {Jensen}, \citenamefont {Sim}, \citenamefont {D{\'i}az-Tinoco}, \citenamefont
  {Kottmann}, \citenamefont {Degroote}, \citenamefont {Izmaylov},\ and\
  \citenamefont {Aspuru-Guzik}}]{anand_quantum_2022}%
  \BibitemOpen
  \bibfield  {author} {\bibinfo {author} {\bibfnamefont {A.}~\bibnamefont
  {Anand}}, \bibinfo {author} {\bibfnamefont {P.}~\bibnamefont {Schleich}},
  \bibinfo {author} {\bibfnamefont {S.}~\bibnamefont {Alperin-Lea}}, \bibinfo
  {author} {\bibfnamefont {P.~W.~K.}\ \bibnamefont {Jensen}}, \bibinfo {author}
  {\bibfnamefont {S.}~\bibnamefont {Sim}}, \bibinfo {author} {\bibfnamefont
  {M.}~\bibnamefont {D{\'i}az-Tinoco}}, \bibinfo {author} {\bibfnamefont
  {J.~S.}\ \bibnamefont {Kottmann}}, \bibinfo {author} {\bibfnamefont
  {M.}~\bibnamefont {Degroote}}, \bibinfo {author} {\bibfnamefont {A.~F.}\
  \bibnamefont {Izmaylov}}, \ and\ \bibinfo {author} {\bibfnamefont
  {A.}~\bibnamefont {Aspuru-Guzik}},\ }\href {\doibase 10.1039/D1CS00932J}
  {\bibfield  {journal} {\bibinfo  {journal} {Chem. Soc. Rev.}\ }\textbf
  {\bibinfo {volume} {51}},\ \bibinfo {pages} {1659} (\bibinfo {year}
  {2022})}\BibitemShut {NoStop}%
\bibitem [{\citenamefont {Evangelista}\ \emph {et~al.}(2019)\citenamefont
  {Evangelista}, \citenamefont {Chan},\ and\ \citenamefont
  {Scuseria}}]{evangelista_exact_2019}%
  \BibitemOpen
  \bibfield  {author} {\bibinfo {author} {\bibfnamefont {F.~A.}\ \bibnamefont
  {Evangelista}}, \bibinfo {author} {\bibfnamefont {G.~K.-L.}\ \bibnamefont
  {Chan}}, \ and\ \bibinfo {author} {\bibfnamefont {G.~E.}\ \bibnamefont
  {Scuseria}},\ }\href {\doibase 10.1063/1.5133059} {\bibfield  {journal}
  {\bibinfo  {journal} {J. Chem. Phys.}\ }\textbf {\bibinfo {volume} {151}},\
  \bibinfo {pages} {244112} (\bibinfo {year} {2019})}\BibitemShut {NoStop}%
\bibitem [{\citenamefont {Grimsley}\ \emph {et~al.}(2020)\citenamefont
  {Grimsley}, \citenamefont {Claudino}, \citenamefont {Economou}, \citenamefont
  {Barnes},\ and\ \citenamefont {Mayhall}}]{grimsley_is_2020}%
  \BibitemOpen
  \bibfield  {author} {\bibinfo {author} {\bibfnamefont {H.~R.}\ \bibnamefont
  {Grimsley}}, \bibinfo {author} {\bibfnamefont {D.}~\bibnamefont {Claudino}},
  \bibinfo {author} {\bibfnamefont {S.~E.}\ \bibnamefont {Economou}}, \bibinfo
  {author} {\bibfnamefont {E.}~\bibnamefont {Barnes}}, \ and\ \bibinfo {author}
  {\bibfnamefont {N.~J.}\ \bibnamefont {Mayhall}},\ }\href {\doibase
  10.1021/acs.jctc.9b01083} {\bibfield  {journal} {\bibinfo  {journal} {J.
  Chem. Theory Comput.}\ }\textbf {\bibinfo {volume} {16}},\ \bibinfo {pages}
  {1} (\bibinfo {year} {2020})}\BibitemShut {NoStop}%
\bibitem [{\citenamefont {Izmaylov}\ \emph
  {et~al.}(2020{\natexlab{a}})\citenamefont {Izmaylov}, \citenamefont
  {D{\'i}az-Tinoco},\ and\ \citenamefont {Lang}}]{izmaylov_order_2020}%
  \BibitemOpen
  \bibfield  {author} {\bibinfo {author} {\bibfnamefont {A.~F.}\ \bibnamefont
  {Izmaylov}}, \bibinfo {author} {\bibfnamefont {M.}~\bibnamefont
  {D{\'i}az-Tinoco}}, \ and\ \bibinfo {author} {\bibfnamefont {R.~A.}\
  \bibnamefont {Lang}},\ }\href {\doibase 10.1039/D0CP01707H} {\bibfield
  {journal} {\bibinfo  {journal} {Phys. Chem. Chem. Phys.}\ }\textbf {\bibinfo
  {volume} {22}},\ \bibinfo {pages} {12980} (\bibinfo {year}
  {2020}{\natexlab{a}})}\BibitemShut {NoStop}%
\bibitem [{\citenamefont {Peruzzo}\ \emph {et~al.}(2014)\citenamefont
  {Peruzzo}, \citenamefont {McClean}, \citenamefont {Shadbolt}, \citenamefont
  {Yung}, \citenamefont {Zhou}, \citenamefont {Love}, \citenamefont
  {Aspuru-Guzik},\ and\ \citenamefont
  {O{\textquoteright}Brien}}]{peruzzo_variational_2014}%
  \BibitemOpen
  \bibfield  {author} {\bibinfo {author} {\bibfnamefont {A.}~\bibnamefont
  {Peruzzo}}, \bibinfo {author} {\bibfnamefont {J.}~\bibnamefont {McClean}},
  \bibinfo {author} {\bibfnamefont {P.}~\bibnamefont {Shadbolt}}, \bibinfo
  {author} {\bibfnamefont {M.-H.}\ \bibnamefont {Yung}}, \bibinfo {author}
  {\bibfnamefont {X.-Q.}\ \bibnamefont {Zhou}}, \bibinfo {author}
  {\bibfnamefont {P.~J.}\ \bibnamefont {Love}}, \bibinfo {author}
  {\bibfnamefont {A.}~\bibnamefont {Aspuru-Guzik}}, \ and\ \bibinfo {author}
  {\bibfnamefont {J.~L.}\ \bibnamefont {O{\textquoteright}Brien}},\ }\href
  {\doibase 10.1038/ncomms5213} {\bibfield  {journal} {\bibinfo  {journal}
  {Nat. Commun.}\ }\textbf {\bibinfo {volume} {5}},\ \bibinfo {pages} {1}
  (\bibinfo {year} {2014})}\BibitemShut {NoStop}%
\bibitem [{\citenamefont {McClean}\ \emph {et~al.}(2016)\citenamefont
  {McClean}, \citenamefont {Romero}, \citenamefont {Babbush},\ and\
  \citenamefont {Aspuru-Guzik}}]{mcclean_theory_2016}%
  \BibitemOpen
  \bibfield  {author} {\bibinfo {author} {\bibfnamefont {J.~R.}\ \bibnamefont
  {McClean}}, \bibinfo {author} {\bibfnamefont {J.}~\bibnamefont {Romero}},
  \bibinfo {author} {\bibfnamefont {R.}~\bibnamefont {Babbush}}, \ and\
  \bibinfo {author} {\bibfnamefont {A.}~\bibnamefont {Aspuru-Guzik}},\ }\href
  {\doibase 10.1088/1367-2630/18/2/023023} {\bibfield  {journal} {\bibinfo
  {journal} {New J. Phys.}\ }\textbf {\bibinfo {volume} {18}},\ \bibinfo
  {pages} {023023} (\bibinfo {year} {2016})}\BibitemShut {NoStop}%
\bibitem [{\citenamefont {Grimsley}\ \emph {et~al.}(2019)\citenamefont
  {Grimsley}, \citenamefont {Economou}, \citenamefont {Barnes},\ and\
  \citenamefont {Mayhall}}]{grimsley_adaptive_2019}%
  \BibitemOpen
  \bibfield  {author} {\bibinfo {author} {\bibfnamefont {H.~R.}\ \bibnamefont
  {Grimsley}}, \bibinfo {author} {\bibfnamefont {S.~E.}\ \bibnamefont
  {Economou}}, \bibinfo {author} {\bibfnamefont {E.}~\bibnamefont {Barnes}}, \
  and\ \bibinfo {author} {\bibfnamefont {N.~J.}\ \bibnamefont {Mayhall}},\
  }\href {\doibase 10.1038/s41467-019-10988-2} {\bibfield  {journal} {\bibinfo
  {journal} {Nat. Commun.}\ }\textbf {\bibinfo {volume} {10}},\ \bibinfo
  {pages} {3007} (\bibinfo {year} {2019})}\BibitemShut {NoStop}%
\bibitem [{\citenamefont {Tang}\ \emph {et~al.}(2021)\citenamefont {Tang},
  \citenamefont {Shkolnikov}, \citenamefont {Barron}, \citenamefont {Grimsley},
  \citenamefont {Mayhall}, \citenamefont {Barnes},\ and\ \citenamefont
  {Economou}}]{tang_qubit-adapt-vqe_2021}%
  \BibitemOpen
  \bibfield  {author} {\bibinfo {author} {\bibfnamefont {H.~L.}\ \bibnamefont
  {Tang}}, \bibinfo {author} {\bibfnamefont {V.}~\bibnamefont {Shkolnikov}},
  \bibinfo {author} {\bibfnamefont {G.~S.}\ \bibnamefont {Barron}}, \bibinfo
  {author} {\bibfnamefont {H.~R.}\ \bibnamefont {Grimsley}}, \bibinfo {author}
  {\bibfnamefont {N.~J.}\ \bibnamefont {Mayhall}}, \bibinfo {author}
  {\bibfnamefont {E.}~\bibnamefont {Barnes}}, \ and\ \bibinfo {author}
  {\bibfnamefont {S.~E.}\ \bibnamefont {Economou}},\ }\href {\doibase
  10.1103/PRXQuantum.2.020310} {\bibfield  {journal} {\bibinfo  {journal} {PRX
  Quantum}\ }\textbf {\bibinfo {volume} {2}},\ \bibinfo {pages} {020310}
  (\bibinfo {year} {2021})}\BibitemShut {NoStop}%
\bibitem [{\citenamefont {Grimsley}\ \emph {et~al.}(2022)\citenamefont
  {Grimsley}, \citenamefont {Barron}, \citenamefont {Barnes}, \citenamefont
  {Economou},\ and\ \citenamefont {Mayhall}}]{grimsley_adapt-vqe_2022}%
  \BibitemOpen
  \bibfield  {author} {\bibinfo {author} {\bibfnamefont {H.~R.}\ \bibnamefont
  {Grimsley}}, \bibinfo {author} {\bibfnamefont {G.~S.}\ \bibnamefont
  {Barron}}, \bibinfo {author} {\bibfnamefont {E.}~\bibnamefont {Barnes}},
  \bibinfo {author} {\bibfnamefont {S.~E.}\ \bibnamefont {Economou}}, \ and\
  \bibinfo {author} {\bibfnamefont {N.~J.}\ \bibnamefont {Mayhall}},\ }\href
  {\doibase 10.48550/ARXIV.2204.07179} {\  (\bibinfo {year} {2022}),\
  10.48550/ARXIV.2204.07179}\BibitemShut {NoStop}%
\bibitem [{\citenamefont {Romero}\ \emph {et~al.}(2022)\citenamefont {Romero},
  \citenamefont {Engel}, \citenamefont {Tang},\ and\ \citenamefont
  {Economou}}]{romero_solving_2022}%
  \BibitemOpen
  \bibfield  {author} {\bibinfo {author} {\bibfnamefont {A.~M.}\ \bibnamefont
  {Romero}}, \bibinfo {author} {\bibfnamefont {J.}~\bibnamefont {Engel}},
  \bibinfo {author} {\bibfnamefont {H.~L.}\ \bibnamefont {Tang}}, \ and\
  \bibinfo {author} {\bibfnamefont {S.~E.}\ \bibnamefont {Economou}},\ }\href
  {\doibase 10.1103/PhysRevC.105.064317} {\bibfield  {journal} {\bibinfo
  {journal} {Phys. Rev. C}\ }\textbf {\bibinfo {volume} {105}},\ \bibinfo
  {pages} {064317} (\bibinfo {year} {2022})}\BibitemShut {NoStop}%
\bibitem [{\citenamefont {Ryabinkin}\ \emph {et~al.}(2018)\citenamefont
  {Ryabinkin}, \citenamefont {Yen}, \citenamefont {Genin},\ and\ \citenamefont
  {Izmaylov}}]{ryabinkin_qubit_2018}%
  \BibitemOpen
  \bibfield  {author} {\bibinfo {author} {\bibfnamefont {I.~G.}\ \bibnamefont
  {Ryabinkin}}, \bibinfo {author} {\bibfnamefont {T.-C.}\ \bibnamefont {Yen}},
  \bibinfo {author} {\bibfnamefont {S.~N.}\ \bibnamefont {Genin}}, \ and\
  \bibinfo {author} {\bibfnamefont {A.~F.}\ \bibnamefont {Izmaylov}},\ }\href
  {\doibase 10.1021/acs.jctc.8b00932} {\bibfield  {journal} {\bibinfo
  {journal} {J. Chem. Theory Comput.}\ }\textbf {\bibinfo {volume} {14}},\
  \bibinfo {pages} {6317} (\bibinfo {year} {2018})}\BibitemShut {NoStop}%
\bibitem [{\citenamefont {Ryabinkin}\ \emph {et~al.}(2020)\citenamefont
  {Ryabinkin}, \citenamefont {Lang}, \citenamefont {Genin},\ and\ \citenamefont
  {Izmaylov}}]{ryabinkin_iterative_2020}%
  \BibitemOpen
  \bibfield  {author} {\bibinfo {author} {\bibfnamefont {I.~G.}\ \bibnamefont
  {Ryabinkin}}, \bibinfo {author} {\bibfnamefont {R.~A.}\ \bibnamefont {Lang}},
  \bibinfo {author} {\bibfnamefont {S.~N.}\ \bibnamefont {Genin}}, \ and\
  \bibinfo {author} {\bibfnamefont {A.~F.}\ \bibnamefont {Izmaylov}},\ }\href
  {\doibase 10.1021/acs.jctc.9b01084} {\bibfield  {journal} {\bibinfo
  {journal} {J. Chem. Theory Comput.}\ }\textbf {\bibinfo {volume} {16}},\
  \bibinfo {pages} {1055} (\bibinfo {year} {2020})}\BibitemShut {NoStop}%
\bibitem [{\citenamefont {Matsuzawa}\ and\ \citenamefont
  {Kurashige}(2020)}]{matsuzawa_jastrow-type_2020}%
  \BibitemOpen
  \bibfield  {author} {\bibinfo {author} {\bibfnamefont {Y.}~\bibnamefont
  {Matsuzawa}}\ and\ \bibinfo {author} {\bibfnamefont {Y.}~\bibnamefont
  {Kurashige}},\ }\href {\doibase 10.1021/acs.jctc.9b00963} {\bibfield
  {journal} {\bibinfo  {journal} {J. Chem. Theory Comput.}\ }\textbf {\bibinfo
  {volume} {16}},\ \bibinfo {pages} {944} (\bibinfo {year} {2020})}\BibitemShut
  {NoStop}%
\bibitem [{\citenamefont {Lee}\ \emph {et~al.}(2019)\citenamefont {Lee},
  \citenamefont {Huggins}, \citenamefont {Head-Gordon},\ and\ \citenamefont
  {Whaley}}]{lee_generalized_2019}%
  \BibitemOpen
  \bibfield  {author} {\bibinfo {author} {\bibfnamefont {J.}~\bibnamefont
  {Lee}}, \bibinfo {author} {\bibfnamefont {W.~J.}\ \bibnamefont {Huggins}},
  \bibinfo {author} {\bibfnamefont {M.}~\bibnamefont {Head-Gordon}}, \ and\
  \bibinfo {author} {\bibfnamefont {K.~B.}\ \bibnamefont {Whaley}},\ }\href
  {\doibase 10.1021/acs.jctc.8b01004} {\bibfield  {journal} {\bibinfo
  {journal} {J. Chem. Theory Comput.}\ }\textbf {\bibinfo {volume} {15}},\
  \bibinfo {pages} {311} (\bibinfo {year} {2019})}\BibitemShut {NoStop}%
\bibitem [{\citenamefont {Stair}\ and\ \citenamefont
  {Evangelista}(2021)}]{stair_simulating_2021}%
  \BibitemOpen
  \bibfield  {author} {\bibinfo {author} {\bibfnamefont {N.~H.}\ \bibnamefont
  {Stair}}\ and\ \bibinfo {author} {\bibfnamefont {F.~A.}\ \bibnamefont
  {Evangelista}},\ }\href {\doibase 10.1103/PRXQuantum.2.030301} {\bibfield
  {journal} {\bibinfo  {journal} {PRX Quantum}\ }\textbf {\bibinfo {volume}
  {2}},\ \bibinfo {pages} {030301} (\bibinfo {year} {2021})}\BibitemShut
  {NoStop}%
\bibitem [{\citenamefont {Kandala}\ \emph {et~al.}(2017)\citenamefont
  {Kandala}, \citenamefont {Mezzacapo}, \citenamefont {Temme}, \citenamefont
  {Takita}, \citenamefont {Brink}, \citenamefont {Chow},\ and\ \citenamefont
  {Gambetta}}]{kandala_hardware-efficient_2017}%
  \BibitemOpen
  \bibfield  {author} {\bibinfo {author} {\bibfnamefont {A.}~\bibnamefont
  {Kandala}}, \bibinfo {author} {\bibfnamefont {A.}~\bibnamefont {Mezzacapo}},
  \bibinfo {author} {\bibfnamefont {K.}~\bibnamefont {Temme}}, \bibinfo
  {author} {\bibfnamefont {M.}~\bibnamefont {Takita}}, \bibinfo {author}
  {\bibfnamefont {M.}~\bibnamefont {Brink}}, \bibinfo {author} {\bibfnamefont
  {J.~M.}\ \bibnamefont {Chow}}, \ and\ \bibinfo {author} {\bibfnamefont
  {J.~M.}\ \bibnamefont {Gambetta}},\ }\href {\doibase 10.1038/nature23879}
  {\bibfield  {journal} {\bibinfo  {journal} {Nature}\ }\textbf {\bibinfo
  {volume} {549}},\ \bibinfo {pages} {242} (\bibinfo {year}
  {2017})}\BibitemShut {NoStop}%
\bibitem [{\citenamefont {Dallaire-Demers}\ \emph {et~al.}(2019)\citenamefont
  {Dallaire-Demers}, \citenamefont {Romero}, \citenamefont {Veis},
  \citenamefont {Sim},\ and\ \citenamefont
  {Aspuru-Guzik}}]{dallaire-demers_low-depth_2019}%
  \BibitemOpen
  \bibfield  {author} {\bibinfo {author} {\bibfnamefont {P.-L.}\ \bibnamefont
  {Dallaire-Demers}}, \bibinfo {author} {\bibfnamefont {J.}~\bibnamefont
  {Romero}}, \bibinfo {author} {\bibfnamefont {L.}~\bibnamefont {Veis}},
  \bibinfo {author} {\bibfnamefont {S.}~\bibnamefont {Sim}}, \ and\ \bibinfo
  {author} {\bibfnamefont {A.}~\bibnamefont {Aspuru-Guzik}},\ }\href {\doibase
  10.1088/2058-9565/ab3951} {\bibfield  {journal} {\bibinfo  {journal} {Quantum
  Sci. Technol.}\ }\textbf {\bibinfo {volume} {4}},\ \bibinfo {pages} {045005}
  (\bibinfo {year} {2019})}\BibitemShut {NoStop}%
\bibitem [{\citenamefont {Xia}\ and\ \citenamefont
  {Kais}(2020)}]{xia_qubit_2020}%
  \BibitemOpen
  \bibfield  {author} {\bibinfo {author} {\bibfnamefont {R.}~\bibnamefont
  {Xia}}\ and\ \bibinfo {author} {\bibfnamefont {S.}~\bibnamefont {Kais}},\
  }\href {\doibase 10.1088/2058-9565/abbc74} {\bibfield  {journal} {\bibinfo
  {journal} {Quantum Sci. Technol.}\ }\textbf {\bibinfo {volume} {6}},\
  \bibinfo {pages} {015001} (\bibinfo {year} {2020})}\BibitemShut {NoStop}%
\bibitem [{\citenamefont {Anselmetti}\ \emph {et~al.}(2021)\citenamefont
  {Anselmetti}, \citenamefont {Wierichs}, \citenamefont {Gogolin},\ and\
  \citenamefont {Parrish}}]{anselmetti_local_2021}%
  \BibitemOpen
  \bibfield  {author} {\bibinfo {author} {\bibfnamefont {G.-L.~R.}\
  \bibnamefont {Anselmetti}}, \bibinfo {author} {\bibfnamefont
  {D.}~\bibnamefont {Wierichs}}, \bibinfo {author} {\bibfnamefont
  {C.}~\bibnamefont {Gogolin}}, \ and\ \bibinfo {author} {\bibfnamefont
  {R.~M.}\ \bibnamefont {Parrish}},\ }\href {\doibase 10.1088/1367-2630/ac2cb3}
  {\bibfield  {journal} {\bibinfo  {journal} {New J. Phys.}\ }\textbf {\bibinfo
  {volume} {23}},\ \bibinfo {pages} {113010} (\bibinfo {year}
  {2021})}\BibitemShut {NoStop}%
\bibitem [{\citenamefont {Huggins}\ \emph {et~al.}(2022)\citenamefont
  {Huggins}, \citenamefont {O{\textquoteright}Gorman}, \citenamefont {Rubin},
  \citenamefont {Reichman}, \citenamefont {Babbush},\ and\ \citenamefont
  {Lee}}]{huggins_unbiasing_2022}%
  \BibitemOpen
  \bibfield  {author} {\bibinfo {author} {\bibfnamefont {W.~J.}\ \bibnamefont
  {Huggins}}, \bibinfo {author} {\bibfnamefont {B.~A.}\ \bibnamefont
  {O{\textquoteright}Gorman}}, \bibinfo {author} {\bibfnamefont {N.~C.}\
  \bibnamefont {Rubin}}, \bibinfo {author} {\bibfnamefont {D.~R.}\ \bibnamefont
  {Reichman}}, \bibinfo {author} {\bibfnamefont {R.}~\bibnamefont {Babbush}}, \
  and\ \bibinfo {author} {\bibfnamefont {J.}~\bibnamefont {Lee}},\ }\href
  {\doibase 10.1038/s41586-021-04351-z} {\bibfield  {journal} {\bibinfo
  {journal} {Nature}\ }\textbf {\bibinfo {volume} {603}},\ \bibinfo {pages}
  {416} (\bibinfo {year} {2022})}\BibitemShut {NoStop}%
\bibitem [{\citenamefont {Preskill}(2018)}]{preskill_quantum_2018}%
  \BibitemOpen
  \bibfield  {author} {\bibinfo {author} {\bibfnamefont {J.}~\bibnamefont
  {Preskill}},\ }\href {\doibase 10.22331/q-2018-08-06-79} {\bibfield
  {journal} {\bibinfo  {journal} {Quantum}\ }\textbf {\bibinfo {volume} {2}},\
  \bibinfo {pages} {79} (\bibinfo {year} {2018})}\BibitemShut {NoStop}%
\bibitem [{\citenamefont {Leymann}\ and\ \citenamefont
  {Barzen}(2020)}]{leymann_bitter_2020}%
  \BibitemOpen
  \bibfield  {author} {\bibinfo {author} {\bibfnamefont {F.}~\bibnamefont
  {Leymann}}\ and\ \bibinfo {author} {\bibfnamefont {J.}~\bibnamefont
  {Barzen}},\ }\href {\doibase 10.1088/2058-9565/abae7d} {\bibfield  {journal}
  {\bibinfo  {journal} {Quantum Sci. Technol.}\ }\textbf {\bibinfo {volume}
  {5}},\ \bibinfo {pages} {044007} (\bibinfo {year} {2020})}\BibitemShut
  {NoStop}%
\bibitem [{\citenamefont {Duguet}\ and\ \citenamefont
  {Signoracci}(2017)}]{duguet_symmetry_2017}%
  \BibitemOpen
  \bibfield  {author} {\bibinfo {author} {\bibfnamefont {T.}~\bibnamefont
  {Duguet}}\ and\ \bibinfo {author} {\bibfnamefont {A.}~\bibnamefont
  {Signoracci}},\ }\href {\doibase 10.1088/0954-3899/44/1/015103} {\bibfield
  {journal} {\bibinfo  {journal} {J. Phys. G: Nucl. Part. Phys.}\ }\textbf
  {\bibinfo {volume} {44}},\ \bibinfo {pages} {015103} (\bibinfo {year}
  {2017})}\BibitemShut {NoStop}%
\bibitem [{\citenamefont {Wahlen-Strothman}\ \emph {et~al.}(2017)\citenamefont
  {Wahlen-Strothman}, \citenamefont {Henderson}, \citenamefont {Hermes},
  \citenamefont {Degroote}, \citenamefont {Qiu}, \citenamefont {Zhao},
  \citenamefont {Dukelsky},\ and\ \citenamefont
  {Scuseria}}]{wahlen-strothman_merging_2017}%
  \BibitemOpen
  \bibfield  {author} {\bibinfo {author} {\bibfnamefont {J.~M.}\ \bibnamefont
  {Wahlen-Strothman}}, \bibinfo {author} {\bibfnamefont {T.~M.}\ \bibnamefont
  {Henderson}}, \bibinfo {author} {\bibfnamefont {M.~R.}\ \bibnamefont
  {Hermes}}, \bibinfo {author} {\bibfnamefont {M.}~\bibnamefont {Degroote}},
  \bibinfo {author} {\bibfnamefont {Y.}~\bibnamefont {Qiu}}, \bibinfo {author}
  {\bibfnamefont {J.}~\bibnamefont {Zhao}}, \bibinfo {author} {\bibfnamefont
  {J.}~\bibnamefont {Dukelsky}}, \ and\ \bibinfo {author} {\bibfnamefont
  {G.~E.}\ \bibnamefont {Scuseria}},\ }\href {\doibase 10.1063/1.4974989}
  {\bibfield  {journal} {\bibinfo  {journal} {The Journal of Chemical Physics}\
  }\textbf {\bibinfo {volume} {146}},\ \bibinfo {pages} {054110} (\bibinfo
  {year} {2017})}\BibitemShut {NoStop}%
\bibitem [{\citenamefont {Qiu}\ \emph {et~al.}(2019)\citenamefont {Qiu},
  \citenamefont {Henderson}, \citenamefont {Duguet},\ and\ \citenamefont
  {Scuseria}}]{qiu_particle-number_2019}%
  \BibitemOpen
  \bibfield  {author} {\bibinfo {author} {\bibfnamefont {Y.}~\bibnamefont
  {Qiu}}, \bibinfo {author} {\bibfnamefont {T.~M.}\ \bibnamefont {Henderson}},
  \bibinfo {author} {\bibfnamefont {T.}~\bibnamefont {Duguet}}, \ and\ \bibinfo
  {author} {\bibfnamefont {G.~E.}\ \bibnamefont {Scuseria}},\ }\href {\doibase
  10.1103/PhysRevC.99.044301} {\bibfield  {journal} {\bibinfo  {journal} {Phys.
  Rev. C}\ }\textbf {\bibinfo {volume} {99}},\ \bibinfo {pages} {044301}
  (\bibinfo {year} {2019})}\BibitemShut {NoStop}%
\bibitem [{\citenamefont {Song}\ \emph {et~al.}(2022)\citenamefont {Song},
  \citenamefont {Henderson},\ and\ \citenamefont {Scuseria}}]{song_power_2022}%
  \BibitemOpen
  \bibfield  {author} {\bibinfo {author} {\bibfnamefont {R.}~\bibnamefont
  {Song}}, \bibinfo {author} {\bibfnamefont {T.~M.}\ \bibnamefont {Henderson}},
  \ and\ \bibinfo {author} {\bibfnamefont {G.~E.}\ \bibnamefont {Scuseria}},\
  }\href {\doibase 10.1063/5.0080165} {\bibfield  {journal} {\bibinfo
  {journal} {J. Chem. Phys.}\ }\textbf {\bibinfo {volume} {156}},\ \bibinfo
  {pages} {104105} (\bibinfo {year} {2022})}\BibitemShut {NoStop}%
\bibitem [{\citenamefont {Sheikh}\ \emph {et~al.}(2021)\citenamefont {Sheikh},
  \citenamefont {Dobaczewski}, \citenamefont {Ring}, \citenamefont {Robledo},\
  and\ \citenamefont {Yannouleas}}]{sheikh_symmetry_2021}%
  \BibitemOpen
  \bibfield  {author} {\bibinfo {author} {\bibfnamefont {J.~A.}\ \bibnamefont
  {Sheikh}}, \bibinfo {author} {\bibfnamefont {J.}~\bibnamefont {Dobaczewski}},
  \bibinfo {author} {\bibfnamefont {P.}~\bibnamefont {Ring}}, \bibinfo {author}
  {\bibfnamefont {L.~M.}\ \bibnamefont {Robledo}}, \ and\ \bibinfo {author}
  {\bibfnamefont {C.}~\bibnamefont {Yannouleas}},\ }\href {\doibase
  10.1088/1361-6471/ac288a} {\bibfield  {journal} {\bibinfo  {journal} {J.
  Phys. G: Nucl. Part. Phys.}\ }\textbf {\bibinfo {volume} {48}},\ \bibinfo
  {pages} {123001} (\bibinfo {year} {2021})}\BibitemShut {NoStop}%
\bibitem [{\citenamefont {Izmaylov}(2019)}]{izmaylov_construction_2019}%
  \BibitemOpen
  \bibfield  {author} {\bibinfo {author} {\bibfnamefont {A.~F.}\ \bibnamefont
  {Izmaylov}},\ }\href {\doibase 10.1021/acs.jpca.9b01103} {\bibfield
  {journal} {\bibinfo  {journal} {J. Phys. Chem. A}\ }\textbf {\bibinfo
  {volume} {123}},\ \bibinfo {pages} {3429} (\bibinfo {year}
  {2019})}\BibitemShut {NoStop}%
\bibitem [{\citenamefont {Yen}\ \emph {et~al.}(2019)\citenamefont {Yen},
  \citenamefont {Lang},\ and\ \citenamefont {Izmaylov}}]{yen_exact_2019}%
  \BibitemOpen
  \bibfield  {author} {\bibinfo {author} {\bibfnamefont {T.-C.}\ \bibnamefont
  {Yen}}, \bibinfo {author} {\bibfnamefont {R.~A.}\ \bibnamefont {Lang}}, \
  and\ \bibinfo {author} {\bibfnamefont {A.~F.}\ \bibnamefont {Izmaylov}},\
  }\href {\doibase 10.1063/1.5110682} {\bibfield  {journal} {\bibinfo
  {journal} {J. Chem. Phys.}\ }\textbf {\bibinfo {volume} {151}},\ \bibinfo
  {pages} {164111} (\bibinfo {year} {2019})}\BibitemShut {NoStop}%
\bibitem [{\citenamefont {Tsuchimochi}\ \emph {et~al.}(2020)\citenamefont
  {Tsuchimochi}, \citenamefont {Mori},\ and\ \citenamefont
  {Ten-no}}]{tsuchimochi_spin-projection_2020}%
  \BibitemOpen
  \bibfield  {author} {\bibinfo {author} {\bibfnamefont {T.}~\bibnamefont
  {Tsuchimochi}}, \bibinfo {author} {\bibfnamefont {Y.}~\bibnamefont {Mori}}, \
  and\ \bibinfo {author} {\bibfnamefont {S.~L.}\ \bibnamefont {Ten-no}},\
  }\href {\doibase 10.1103/PhysRevResearch.2.043142} {\bibfield  {journal}
  {\bibinfo  {journal} {Phys. Rev. Research}\ }\textbf {\bibinfo {volume}
  {2}},\ \bibinfo {pages} {043142} (\bibinfo {year} {2020})}\BibitemShut
  {NoStop}%
\bibitem [{\citenamefont {Seki}\ \emph {et~al.}(2020)\citenamefont {Seki},
  \citenamefont {Shirakawa},\ and\ \citenamefont
  {Yunoki}}]{seki_symmetry-adapted_2020}%
  \BibitemOpen
  \bibfield  {author} {\bibinfo {author} {\bibfnamefont {K.}~\bibnamefont
  {Seki}}, \bibinfo {author} {\bibfnamefont {T.}~\bibnamefont {Shirakawa}}, \
  and\ \bibinfo {author} {\bibfnamefont {S.}~\bibnamefont {Yunoki}},\ }\href
  {\doibase 10.1103/PhysRevA.101.052340} {\bibfield  {journal} {\bibinfo
  {journal} {Phys. Rev. A}\ }\textbf {\bibinfo {volume} {101}},\ \bibinfo
  {pages} {052340} (\bibinfo {year} {2020})}\BibitemShut {NoStop}%
\bibitem [{\citenamefont {Lacroix}(2020)}]{lacroix_symmetry-assisted_2020}%
  \BibitemOpen
  \bibfield  {author} {\bibinfo {author} {\bibfnamefont {D.}~\bibnamefont
  {Lacroix}},\ }\href {\doibase 10.1103/PhysRevLett.125.230502} {\bibfield
  {journal} {\bibinfo  {journal} {Phys. Rev. Lett.}\ }\textbf {\bibinfo
  {volume} {125}},\ \bibinfo {pages} {230502} (\bibinfo {year}
  {2020})}\BibitemShut {NoStop}%
\bibitem [{\citenamefont {Siwach}\ and\ \citenamefont
  {Lacroix}(2021)}]{siwach_filtering_2021}%
  \BibitemOpen
  \bibfield  {author} {\bibinfo {author} {\bibfnamefont {P.}~\bibnamefont
  {Siwach}}\ and\ \bibinfo {author} {\bibfnamefont {D.}~\bibnamefont
  {Lacroix}},\ }\href {\doibase 10.1103/PhysRevA.104.062435} {\bibfield
  {journal} {\bibinfo  {journal} {Phys. Rev. A}\ }\textbf {\bibinfo {volume}
  {104}},\ \bibinfo {pages} {062435} (\bibinfo {year} {2021})}\BibitemShut
  {NoStop}%
\bibitem [{\citenamefont {Seki}\ and\ \citenamefont
  {Yunoki}(2022)}]{seki_spatial_2022}%
  \BibitemOpen
  \bibfield  {author} {\bibinfo {author} {\bibfnamefont {K.}~\bibnamefont
  {Seki}}\ and\ \bibinfo {author} {\bibfnamefont {S.}~\bibnamefont {Yunoki}},\
  }\href {\doibase 10.1103/PhysRevA.105.032419} {\bibfield  {journal} {\bibinfo
   {journal} {Phys. Rev. A}\ }\textbf {\bibinfo {volume} {105}},\ \bibinfo
  {pages} {032419} (\bibinfo {year} {2022})}\BibitemShut {NoStop}%
\bibitem [{\citenamefont {Ruiz~Guzman}\ and\ \citenamefont
  {Lacroix}(2022)}]{ruiz_guzman_accessing_2022}%
  \BibitemOpen
  \bibfield  {author} {\bibinfo {author} {\bibfnamefont {E.~A.}\ \bibnamefont
  {Ruiz~Guzman}}\ and\ \bibinfo {author} {\bibfnamefont {D.}~\bibnamefont
  {Lacroix}},\ }\href {\doibase 10.1103/PhysRevC.105.024324} {\bibfield
  {journal} {\bibinfo  {journal} {Phys. Rev. C}\ }\textbf {\bibinfo {volume}
  {105}},\ \bibinfo {pages} {024324} (\bibinfo {year} {2022})}\BibitemShut
  {NoStop}%
\bibitem [{\citenamefont {Khamoshi}\ \emph {et~al.}(2020)\citenamefont
  {Khamoshi}, \citenamefont {Evangelista},\ and\ \citenamefont
  {Scuseria}}]{khamoshi_correlating_2020}%
  \BibitemOpen
  \bibfield  {author} {\bibinfo {author} {\bibfnamefont {A.}~\bibnamefont
  {Khamoshi}}, \bibinfo {author} {\bibfnamefont {F.~A.}\ \bibnamefont
  {Evangelista}}, \ and\ \bibinfo {author} {\bibfnamefont {G.~E.}\ \bibnamefont
  {Scuseria}},\ }\href {\doibase 10.1088/2058-9565/abc1bb} {\bibfield
  {journal} {\bibinfo  {journal} {Quantum Sci. Technol.}\ }\textbf {\bibinfo
  {volume} {6}},\ \bibinfo {pages} {014004} (\bibinfo {year}
  {2020})}\BibitemShut {NoStop}%
\bibitem [{\citenamefont {Yang}(1962)}]{yang_concept_1962}%
  \BibitemOpen
  \bibfield  {author} {\bibinfo {author} {\bibfnamefont {C.~N.}\ \bibnamefont
  {Yang}},\ }\href {\doibase 10.1103/RevModPhys.34.694} {\bibfield  {journal}
  {\bibinfo  {journal} {Rev. Mod. Phys.}\ }\textbf {\bibinfo {volume} {34}},\
  \bibinfo {pages} {694} (\bibinfo {year} {1962})}\BibitemShut {NoStop}%
\bibitem [{\citenamefont {Coleman}(1965)}]{coleman_structure_1965}%
  \BibitemOpen
  \bibfield  {author} {\bibinfo {author} {\bibfnamefont {A.~J.}\ \bibnamefont
  {Coleman}},\ }\href {\doibase 10.1063/1.1704794} {\bibfield  {journal}
  {\bibinfo  {journal} {J. Math. Phys.}\ }\textbf {\bibinfo {volume} {6}},\
  \bibinfo {pages} {1425} (\bibinfo {year} {1965})}\BibitemShut {NoStop}%
\bibitem [{\citenamefont {Sager}\ and\ \citenamefont
  {Mazziotti}(2022)}]{sager_cooper-pair_2022}%
  \BibitemOpen
  \bibfield  {author} {\bibinfo {author} {\bibfnamefont {L.~M.}\ \bibnamefont
  {Sager}}\ and\ \bibinfo {author} {\bibfnamefont {D.~A.}\ \bibnamefont
  {Mazziotti}},\ }\href {\doibase 10.1103/PhysRevResearch.4.013003} {\bibfield
  {journal} {\bibinfo  {journal} {Phys. Rev. Research}\ }\textbf {\bibinfo
  {volume} {4}},\ \bibinfo {pages} {013003} (\bibinfo {year}
  {2022})}\BibitemShut {NoStop}%
\bibitem [{\citenamefont {Bardeen}\ \emph {et~al.}(1957)\citenamefont
  {Bardeen}, \citenamefont {Cooper},\ and\ \citenamefont
  {Schrieffer}}]{bardeen_theory_1957}%
  \BibitemOpen
  \bibfield  {author} {\bibinfo {author} {\bibfnamefont {J.}~\bibnamefont
  {Bardeen}}, \bibinfo {author} {\bibfnamefont {L.~N.}\ \bibnamefont {Cooper}},
  \ and\ \bibinfo {author} {\bibfnamefont {J.~R.}\ \bibnamefont {Schrieffer}},\
  }\href {\doibase 10.1103/PhysRev.108.1175} {\bibfield  {journal} {\bibinfo
  {journal} {Phys. Rev.}\ }\textbf {\bibinfo {volume} {108}},\ \bibinfo {pages}
  {1175} (\bibinfo {year} {1957})}\BibitemShut {NoStop}%
\bibitem [{\citenamefont {Brink}\ and\ \citenamefont
  {Broglia}(2005)}]{brink_nuclear_2005}%
  \BibitemOpen
  \bibfield  {author} {\bibinfo {author} {\bibfnamefont {D.~M.}\ \bibnamefont
  {Brink}}\ and\ \bibinfo {author} {\bibfnamefont {R.~A.}\ \bibnamefont
  {Broglia}},\ }\href {\doibase 10.1017/CBO9780511534911} {\emph {\bibinfo
  {title} {Nuclear {Superfluidity}: {Pairing} in {Finite} {Systems}}}},\
  \bibinfo {edition} {1st}\ ed.\ (\bibinfo  {publisher} {Cambridge University
  Press},\ \bibinfo {year} {2005})\BibitemShut {NoStop}%
\bibitem [{\citenamefont {Sedrakian}\ and\ \citenamefont
  {Clark}(2019)}]{sedrakian_superfluidity_2019}%
  \BibitemOpen
  \bibfield  {author} {\bibinfo {author} {\bibfnamefont {A.}~\bibnamefont
  {Sedrakian}}\ and\ \bibinfo {author} {\bibfnamefont {J.~W.}\ \bibnamefont
  {Clark}},\ }\href {\doibase 10.1140/epja/i2019-12863-6} {\bibfield  {journal}
  {\bibinfo  {journal} {Eur. Phys. J. A}\ }\textbf {\bibinfo {volume} {55}},\
  \bibinfo {pages} {167} (\bibinfo {year} {2019})}\BibitemShut {NoStop}%
\bibitem [{\citenamefont {Bach}\ \emph {et~al.}(1994)\citenamefont {Bach},
  \citenamefont {Lieb},\ and\ \citenamefont {Solovej}}]{bach_generalized_1994}%
  \BibitemOpen
  \bibfield  {author} {\bibinfo {author} {\bibfnamefont {V.}~\bibnamefont
  {Bach}}, \bibinfo {author} {\bibfnamefont {E.~H.}\ \bibnamefont {Lieb}}, \
  and\ \bibinfo {author} {\bibfnamefont {J.~P.}\ \bibnamefont {Solovej}},\
  }\href {\doibase 10.1007/BF02188656} {\bibfield  {journal} {\bibinfo
  {journal} {J Stat Phys}\ }\textbf {\bibinfo {volume} {76}},\ \bibinfo {pages}
  {3} (\bibinfo {year} {1994})}\BibitemShut {NoStop}%
\bibitem [{\citenamefont {Surj{\'a}n}(1999)}]{surjan_introduction_1999}%
  \BibitemOpen
  \bibfield  {author} {\bibinfo {author} {\bibfnamefont {P.~R.}\ \bibnamefont
  {Surj{\'a}n}},\ }\href {https://doi.org/10.1007/3-540-48972-X_4} {\bibfield
  {journal} {\bibinfo  {journal} {Top. Curr. Chem.}\ }\textbf {\bibinfo
  {volume} {203}},\ \bibinfo {pages} {63} (\bibinfo {year} {1999})}\BibitemShut
  {NoStop}%
\bibitem [{\citenamefont {Henderson}\ and\ \citenamefont
  {Scuseria}(2019)}]{henderson_geminal-based_2019}%
  \BibitemOpen
  \bibfield  {author} {\bibinfo {author} {\bibfnamefont {T.~M.}\ \bibnamefont
  {Henderson}}\ and\ \bibinfo {author} {\bibfnamefont {G.~E.}\ \bibnamefont
  {Scuseria}},\ }\href {\doibase 10.1063/1.5116715} {\bibfield  {journal}
  {\bibinfo  {journal} {J. Chem. Phys.}\ }\textbf {\bibinfo {volume} {151}},\
  \bibinfo {pages} {051101} (\bibinfo {year} {2019})}\BibitemShut {NoStop}%
\bibitem [{\citenamefont {Khamoshi}\ \emph {et~al.}(2019)\citenamefont
  {Khamoshi}, \citenamefont {Henderson},\ and\ \citenamefont
  {Scuseria}}]{khamoshi_efficient_2019}%
  \BibitemOpen
  \bibfield  {author} {\bibinfo {author} {\bibfnamefont {A.}~\bibnamefont
  {Khamoshi}}, \bibinfo {author} {\bibfnamefont {T.~M.}\ \bibnamefont
  {Henderson}}, \ and\ \bibinfo {author} {\bibfnamefont {G.~E.}\ \bibnamefont
  {Scuseria}},\ }\href {\doibase 10.1063/1.5127850} {\bibfield  {journal}
  {\bibinfo  {journal} {J. Chem. Phys.}\ }\textbf {\bibinfo {volume} {151}},\
  \bibinfo {pages} {184103} (\bibinfo {year} {2019})}\BibitemShut {NoStop}%
\bibitem [{\citenamefont {Dutta}\ \emph {et~al.}(2020)\citenamefont {Dutta},
  \citenamefont {Henderson},\ and\ \citenamefont
  {Scuseria}}]{dutta_geminal_2020}%
  \BibitemOpen
  \bibfield  {author} {\bibinfo {author} {\bibfnamefont {R.}~\bibnamefont
  {Dutta}}, \bibinfo {author} {\bibfnamefont {T.~M.}\ \bibnamefont
  {Henderson}}, \ and\ \bibinfo {author} {\bibfnamefont {G.~E.}\ \bibnamefont
  {Scuseria}},\ }\href {\doibase 10.1021/acs.jctc.0c00807} {\bibfield
  {journal} {\bibinfo  {journal} {J. Chem. Theory Comput.}\ }\textbf {\bibinfo
  {volume} {16}},\ \bibinfo {pages} {6358} (\bibinfo {year}
  {2020})}\BibitemShut {NoStop}%
\bibitem [{\citenamefont {Henderson}\ and\ \citenamefont
  {Scuseria}(2020)}]{henderson_correlating_2020}%
  \BibitemOpen
  \bibfield  {author} {\bibinfo {author} {\bibfnamefont {T.~M.}\ \bibnamefont
  {Henderson}}\ and\ \bibinfo {author} {\bibfnamefont {G.~E.}\ \bibnamefont
  {Scuseria}},\ }\href {\doibase 10.1063/5.0021144} {\bibfield  {journal}
  {\bibinfo  {journal} {J. Chem. Phys.}\ }\textbf {\bibinfo {volume} {153}},\
  \bibinfo {pages} {084111} (\bibinfo {year} {2020})}\BibitemShut {NoStop}%
\bibitem [{\citenamefont {Dutta}\ \emph {et~al.}(2021)\citenamefont {Dutta},
  \citenamefont {Chen}, \citenamefont {Henderson},\ and\ \citenamefont
  {Scuseria}}]{dutta_construction_2021}%
  \BibitemOpen
  \bibfield  {author} {\bibinfo {author} {\bibfnamefont {R.}~\bibnamefont
  {Dutta}}, \bibinfo {author} {\bibfnamefont {G.~P.}\ \bibnamefont {Chen}},
  \bibinfo {author} {\bibfnamefont {T.~M.}\ \bibnamefont {Henderson}}, \ and\
  \bibinfo {author} {\bibfnamefont {G.~E.}\ \bibnamefont {Scuseria}},\ }\href
  {\doibase 10.1063/5.0045006} {\bibfield  {journal} {\bibinfo  {journal} {J.
  Chem. Phys.}\ }\textbf {\bibinfo {volume} {154}},\ \bibinfo {pages} {114112}
  (\bibinfo {year} {2021})}\BibitemShut {NoStop}%
\bibitem [{\citenamefont {Khamoshi}\ \emph {et~al.}(2021)\citenamefont
  {Khamoshi}, \citenamefont {Chen}, \citenamefont {Henderson},\ and\
  \citenamefont {Scuseria}}]{khamoshi_exploring_2021}%
  \BibitemOpen
  \bibfield  {author} {\bibinfo {author} {\bibfnamefont {A.}~\bibnamefont
  {Khamoshi}}, \bibinfo {author} {\bibfnamefont {G.~P.}\ \bibnamefont {Chen}},
  \bibinfo {author} {\bibfnamefont {T.~M.}\ \bibnamefont {Henderson}}, \ and\
  \bibinfo {author} {\bibfnamefont {G.~E.}\ \bibnamefont {Scuseria}},\ }\href
  {\doibase 10.1063/5.0039618} {\bibfield  {journal} {\bibinfo  {journal} {J.
  Chem. Phys.}\ }\textbf {\bibinfo {volume} {154}},\ \bibinfo {pages} {074113}
  (\bibinfo {year} {2021})}\BibitemShut {NoStop}%
\bibitem [{\citenamefont {Casula}\ and\ \citenamefont
  {Sorella}(2003)}]{casula_geminal_2003}%
  \BibitemOpen
  \bibfield  {author} {\bibinfo {author} {\bibfnamefont {M.}~\bibnamefont
  {Casula}}\ and\ \bibinfo {author} {\bibfnamefont {S.}~\bibnamefont
  {Sorella}},\ }\href {\doibase 10.1063/1.1604379} {\bibfield  {journal}
  {\bibinfo  {journal} {J. Chem. Phys.}\ }\textbf {\bibinfo {volume} {119}},\
  \bibinfo {pages} {6500} (\bibinfo {year} {2003})}\BibitemShut {NoStop}%
\bibitem [{\citenamefont {Casula}\ \emph {et~al.}(2004)\citenamefont {Casula},
  \citenamefont {Attaccalite},\ and\ \citenamefont
  {Sorella}}]{casula_correlated_2004}%
  \BibitemOpen
  \bibfield  {author} {\bibinfo {author} {\bibfnamefont {M.}~\bibnamefont
  {Casula}}, \bibinfo {author} {\bibfnamefont {C.}~\bibnamefont {Attaccalite}},
  \ and\ \bibinfo {author} {\bibfnamefont {S.}~\bibnamefont {Sorella}},\ }\href
  {\doibase 10.1063/1.1794632} {\bibfield  {journal} {\bibinfo  {journal} {J.
  Chem. Phys.}\ }\textbf {\bibinfo {volume} {121}},\ \bibinfo {pages} {7110}
  (\bibinfo {year} {2004})}\BibitemShut {NoStop}%
\bibitem [{\citenamefont {Wei}\ and\ \citenamefont
  {Neuscamman}(2018)}]{wei_reduced_2018}%
  \BibitemOpen
  \bibfield  {author} {\bibinfo {author} {\bibfnamefont {H.}~\bibnamefont
  {Wei}}\ and\ \bibinfo {author} {\bibfnamefont {E.}~\bibnamefont
  {Neuscamman}},\ }\href {\doibase 10.1063/1.5047207} {\bibfield  {journal}
  {\bibinfo  {journal} {J. Chem. Phys.}\ }\textbf {\bibinfo {volume} {149}},\
  \bibinfo {pages} {184106} (\bibinfo {year} {2018})}\BibitemShut {NoStop}%
\bibitem [{\citenamefont {Genovese}\ \emph {et~al.}(2019)\citenamefont
  {Genovese}, \citenamefont {Meninno},\ and\ \citenamefont
  {Sorella}}]{genovese_assessing_2019}%
  \BibitemOpen
  \bibfield  {author} {\bibinfo {author} {\bibfnamefont {C.}~\bibnamefont
  {Genovese}}, \bibinfo {author} {\bibfnamefont {A.}~\bibnamefont {Meninno}}, \
  and\ \bibinfo {author} {\bibfnamefont {S.}~\bibnamefont {Sorella}},\ }\href
  {\doibase 10.1063/1.5081933} {\bibfield  {journal} {\bibinfo  {journal} {J.
  Chem. Phys.}\ }\textbf {\bibinfo {volume} {150}},\ \bibinfo {pages} {084102}
  (\bibinfo {year} {2019})}\BibitemShut {NoStop}%
\bibitem [{\citenamefont {Genovese}\ and\ \citenamefont
  {Sorella}(2020)}]{genovese_nature_2020}%
  \BibitemOpen
  \bibfield  {author} {\bibinfo {author} {\bibfnamefont {C.}~\bibnamefont
  {Genovese}}\ and\ \bibinfo {author} {\bibfnamefont {S.}~\bibnamefont
  {Sorella}},\ }\href {\doibase 10.1063/5.0023067} {\bibfield  {journal}
  {\bibinfo  {journal} {J. Chem. Phys.}\ }\textbf {\bibinfo {volume} {153}},\
  \bibinfo {pages} {164301} (\bibinfo {year} {2020})}\BibitemShut {NoStop}%
\bibitem [{\citenamefont
  {Neuscamman}(2013{\natexlab{a}})}]{neuscamman_jastrow_2013}%
  \BibitemOpen
  \bibfield  {author} {\bibinfo {author} {\bibfnamefont {E.}~\bibnamefont
  {Neuscamman}},\ }\href {\doibase 10.1063/1.4829835} {\bibfield  {journal}
  {\bibinfo  {journal} {J. Chem. Phys.}\ }\textbf {\bibinfo {volume} {139}},\
  \bibinfo {pages} {194105} (\bibinfo {year} {2013}{\natexlab{a}})}\BibitemShut
  {NoStop}%
\bibitem [{\citenamefont
  {Neuscamman}(2013{\natexlab{b}})}]{neuscamman_communication_2013}%
  \BibitemOpen
  \bibfield  {author} {\bibinfo {author} {\bibfnamefont {E.}~\bibnamefont
  {Neuscamman}},\ }\href {\doibase 10.1063/1.4829536} {\bibfield  {journal}
  {\bibinfo  {journal} {J. Chem. Phys.}\ }\textbf {\bibinfo {volume} {139}},\
  \bibinfo {pages} {181101} (\bibinfo {year} {2013}{\natexlab{b}})}\BibitemShut
  {NoStop}%
\bibitem [{\citenamefont {Fecteau}\ \emph
  {et~al.}(2020{\natexlab{a}})\citenamefont {Fecteau}, \citenamefont
  {Berthiaume}, \citenamefont {Khalfoun},\ and\ \citenamefont
  {Johnson}}]{fecteau_richardson-gaudin_2020}%
  \BibitemOpen
  \bibfield  {author} {\bibinfo {author} {\bibfnamefont {C.-E.}\ \bibnamefont
  {Fecteau}}, \bibinfo {author} {\bibfnamefont {F.}~\bibnamefont {Berthiaume}},
  \bibinfo {author} {\bibfnamefont {M.}~\bibnamefont {Khalfoun}}, \ and\
  \bibinfo {author} {\bibfnamefont {P.~A.}\ \bibnamefont {Johnson}},\ }\href
  {https://doi.org/10.1007/s10910-020-01197-0} {\bibfield  {journal} {\bibinfo
  {journal} {J. Math. Chem.}\ } (\bibinfo {year}
  {2020}{\natexlab{a}})}\BibitemShut {NoStop}%
\bibitem [{\citenamefont {Johnson}\ \emph {et~al.}(2020)\citenamefont
  {Johnson}, \citenamefont {Fecteau}, \citenamefont {Berthiaume}, \citenamefont
  {Cloutier}, \citenamefont {Carrier}, \citenamefont {Gratton}, \citenamefont
  {Bultinck}, \citenamefont {De~Baerdemacker}, \citenamefont {Van~Neck},
  \citenamefont {Limacher},\ and\ \citenamefont
  {Ayers}}]{johnson_richardsongaudin_2020}%
  \BibitemOpen
  \bibfield  {author} {\bibinfo {author} {\bibfnamefont {P.~A.}\ \bibnamefont
  {Johnson}}, \bibinfo {author} {\bibfnamefont {C.-E.}\ \bibnamefont
  {Fecteau}}, \bibinfo {author} {\bibfnamefont {F.}~\bibnamefont {Berthiaume}},
  \bibinfo {author} {\bibfnamefont {S.}~\bibnamefont {Cloutier}}, \bibinfo
  {author} {\bibfnamefont {L.}~\bibnamefont {Carrier}}, \bibinfo {author}
  {\bibfnamefont {M.}~\bibnamefont {Gratton}}, \bibinfo {author} {\bibfnamefont
  {P.}~\bibnamefont {Bultinck}}, \bibinfo {author} {\bibfnamefont
  {S.}~\bibnamefont {De~Baerdemacker}}, \bibinfo {author} {\bibfnamefont
  {D.}~\bibnamefont {Van~Neck}}, \bibinfo {author} {\bibfnamefont
  {P.}~\bibnamefont {Limacher}}, \ and\ \bibinfo {author} {\bibfnamefont
  {P.~W.}\ \bibnamefont {Ayers}},\ }\href {\doibase 10.1063/5.0022189}
  {\bibfield  {journal} {\bibinfo  {journal} {J. Chem. Phys.}\ }\textbf
  {\bibinfo {volume} {153}},\ \bibinfo {pages} {104110} (\bibinfo {year}
  {2020})}\BibitemShut {NoStop}%
\bibitem [{\citenamefont {Johnson}\ \emph {et~al.}(2021)\citenamefont
  {Johnson}, \citenamefont {Fortin}, \citenamefont {Cloutier},\ and\
  \citenamefont {Fecteau}}]{johnson_transition_2021}%
  \BibitemOpen
  \bibfield  {author} {\bibinfo {author} {\bibfnamefont {P.~A.}\ \bibnamefont
  {Johnson}}, \bibinfo {author} {\bibfnamefont {H.}~\bibnamefont {Fortin}},
  \bibinfo {author} {\bibfnamefont {S.}~\bibnamefont {Cloutier}}, \ and\
  \bibinfo {author} {\bibfnamefont {C.-{\'E}.}\ \bibnamefont {Fecteau}},\
  }\href {\doibase 10.1063/5.0041051} {\bibfield  {journal} {\bibinfo
  {journal} {J. Chem. Phys.}\ }\textbf {\bibinfo {volume} {154}},\ \bibinfo
  {pages} {124125} (\bibinfo {year} {2021})}\BibitemShut {NoStop}%
\bibitem [{\citenamefont {Fecteau}\ \emph
  {et~al.}(2020{\natexlab{b}})\citenamefont {Fecteau}, \citenamefont {Fortin},
  \citenamefont {Cloutier},\ and\ \citenamefont
  {Johnson}}]{fecteau_reduced_2020}%
  \BibitemOpen
  \bibfield  {author} {\bibinfo {author} {\bibfnamefont {C.-E.}\ \bibnamefont
  {Fecteau}}, \bibinfo {author} {\bibfnamefont {H.}~\bibnamefont {Fortin}},
  \bibinfo {author} {\bibfnamefont {S.}~\bibnamefont {Cloutier}}, \ and\
  \bibinfo {author} {\bibfnamefont {P.~A.}\ \bibnamefont {Johnson}},\ }\href
  {\doibase 10.1063/5.0027393} {\bibfield  {journal} {\bibinfo  {journal} {J.
  Chem. Phys.}\ }\textbf {\bibinfo {volume} {153}},\ \bibinfo {pages} {164117}
  (\bibinfo {year} {2020}{\natexlab{b}})}\BibitemShut {NoStop}%
\bibitem [{\citenamefont {Fecteau}\ \emph {et~al.}(2022)\citenamefont
  {Fecteau}, \citenamefont {Cloutier}, \citenamefont {Moisset}, \citenamefont
  {Boulay}, \citenamefont {Bultinck}, \citenamefont {Faribault},\ and\
  \citenamefont {Johnson}}]{fecteau_near-exact_2022}%
  \BibitemOpen
  \bibfield  {author} {\bibinfo {author} {\bibfnamefont {C.-{\'E}.}\
  \bibnamefont {Fecteau}}, \bibinfo {author} {\bibfnamefont {S.}~\bibnamefont
  {Cloutier}}, \bibinfo {author} {\bibfnamefont {J.-D.}\ \bibnamefont
  {Moisset}}, \bibinfo {author} {\bibfnamefont {J.}~\bibnamefont {Boulay}},
  \bibinfo {author} {\bibfnamefont {P.}~\bibnamefont {Bultinck}}, \bibinfo
  {author} {\bibfnamefont {A.}~\bibnamefont {Faribault}}, \ and\ \bibinfo
  {author} {\bibfnamefont {P.~A.}\ \bibnamefont {Johnson}},\ }\href {\doibase
  10.1063/5.0091338} {\bibfield  {journal} {\bibinfo  {journal} {J. Chem.
  Phys.}\ }\textbf {\bibinfo {volume} {156}},\ \bibinfo {pages} {194103}
  (\bibinfo {year} {2022})}\BibitemShut {NoStop}%
\bibitem [{\citenamefont {Moisset}\ \emph {et~al.}(2022)\citenamefont
  {Moisset}, \citenamefont {Fecteau},\ and\ \citenamefont
  {Johnson}}]{moisset_density_2022}%
  \BibitemOpen
  \bibfield  {author} {\bibinfo {author} {\bibfnamefont {J.-D.}\ \bibnamefont
  {Moisset}}, \bibinfo {author} {\bibfnamefont {C.-{\'E}.}\ \bibnamefont
  {Fecteau}}, \ and\ \bibinfo {author} {\bibfnamefont {P.~A.}\ \bibnamefont
  {Johnson}},\ }\href {\doibase 10.1063/5.0088602} {\bibfield  {journal}
  {\bibinfo  {journal} {J. Chem. Phys.}\ }\textbf {\bibinfo {volume} {156}},\
  \bibinfo {pages} {214110} (\bibinfo {year} {2022})}\BibitemShut {NoStop}%
\bibitem [{\citenamefont {Endo}\ \emph {et~al.}(2018)\citenamefont {Endo},
  \citenamefont {Benjamin},\ and\ \citenamefont {Li}}]{endo_practical_2018}%
  \BibitemOpen
  \bibfield  {author} {\bibinfo {author} {\bibfnamefont {S.}~\bibnamefont
  {Endo}}, \bibinfo {author} {\bibfnamefont {S.~C.}\ \bibnamefont {Benjamin}},
  \ and\ \bibinfo {author} {\bibfnamefont {Y.}~\bibnamefont {Li}},\ }\href
  {\doibase 10.1103/PhysRevX.8.031027} {\bibfield  {journal} {\bibinfo
  {journal} {Phys. Rev. X}\ }\textbf {\bibinfo {volume} {8}},\ \bibinfo {pages}
  {031027} (\bibinfo {year} {2018})}\BibitemShut {NoStop}%
\bibitem [{\citenamefont {Bonet-Monroig}\ \emph {et~al.}(2018)\citenamefont
  {Bonet-Monroig}, \citenamefont {Sagastizabal}, \citenamefont {Singh},\ and\
  \citenamefont {O'Brien}}]{bonet-monroig_low-cost_2018}%
  \BibitemOpen
  \bibfield  {author} {\bibinfo {author} {\bibfnamefont {X.}~\bibnamefont
  {Bonet-Monroig}}, \bibinfo {author} {\bibfnamefont {R.}~\bibnamefont
  {Sagastizabal}}, \bibinfo {author} {\bibfnamefont {M.}~\bibnamefont {Singh}},
  \ and\ \bibinfo {author} {\bibfnamefont {T.~E.}\ \bibnamefont {O'Brien}},\
  }\href {\doibase 10.1103/PhysRevA.98.062339} {\bibfield  {journal} {\bibinfo
  {journal} {Phys. Rev. A}\ }\textbf {\bibinfo {volume} {98}},\ \bibinfo
  {pages} {062339} (\bibinfo {year} {2018})}\BibitemShut {NoStop}%
\bibitem [{\citenamefont {Kandala}\ \emph {et~al.}(2019)\citenamefont
  {Kandala}, \citenamefont {Temme}, \citenamefont {C{\'o}rcoles}, \citenamefont
  {Mezzacapo}, \citenamefont {Chow},\ and\ \citenamefont
  {Gambetta}}]{kandala_error_2019}%
  \BibitemOpen
  \bibfield  {author} {\bibinfo {author} {\bibfnamefont {A.}~\bibnamefont
  {Kandala}}, \bibinfo {author} {\bibfnamefont {K.}~\bibnamefont {Temme}},
  \bibinfo {author} {\bibfnamefont {A.~D.}\ \bibnamefont {C{\'o}rcoles}},
  \bibinfo {author} {\bibfnamefont {A.}~\bibnamefont {Mezzacapo}}, \bibinfo
  {author} {\bibfnamefont {J.~M.}\ \bibnamefont {Chow}}, \ and\ \bibinfo
  {author} {\bibfnamefont {J.~M.}\ \bibnamefont {Gambetta}},\ }\href {\doibase
  10.1038/s41586-019-1040-7} {\bibfield  {journal} {\bibinfo  {journal}
  {Nature}\ }\textbf {\bibinfo {volume} {567}},\ \bibinfo {pages} {491}
  (\bibinfo {year} {2019})}\BibitemShut {NoStop}%
\bibitem [{\citenamefont {Endo}\ \emph {et~al.}(2021)\citenamefont {Endo},
  \citenamefont {Cai}, \citenamefont {Benjamin},\ and\ \citenamefont
  {Yuan}}]{endo_hybrid_2021}%
  \BibitemOpen
  \bibfield  {author} {\bibinfo {author} {\bibfnamefont {S.}~\bibnamefont
  {Endo}}, \bibinfo {author} {\bibfnamefont {Z.}~\bibnamefont {Cai}}, \bibinfo
  {author} {\bibfnamefont {S.~C.}\ \bibnamefont {Benjamin}}, \ and\ \bibinfo
  {author} {\bibfnamefont {X.}~\bibnamefont {Yuan}},\ }\href {\doibase
  10.7566/JPSJ.90.032001} {\bibfield  {journal} {\bibinfo  {journal} {J. Phys.
  Soc. Jpn.}\ }\textbf {\bibinfo {volume} {90}},\ \bibinfo {pages} {032001}
  (\bibinfo {year} {2021})}\BibitemShut {NoStop}%
\bibitem [{\citenamefont {Huggins}\ \emph {et~al.}(2021)\citenamefont
  {Huggins}, \citenamefont {McClean}, \citenamefont {Rubin}, \citenamefont
  {Jiang}, \citenamefont {Wiebe}, \citenamefont {Whaley},\ and\ \citenamefont
  {Babbush}}]{huggins_efficient_2021}%
  \BibitemOpen
  \bibfield  {author} {\bibinfo {author} {\bibfnamefont {W.~J.}\ \bibnamefont
  {Huggins}}, \bibinfo {author} {\bibfnamefont {J.~R.}\ \bibnamefont
  {McClean}}, \bibinfo {author} {\bibfnamefont {N.~C.}\ \bibnamefont {Rubin}},
  \bibinfo {author} {\bibfnamefont {Z.}~\bibnamefont {Jiang}}, \bibinfo
  {author} {\bibfnamefont {N.}~\bibnamefont {Wiebe}}, \bibinfo {author}
  {\bibfnamefont {K.~B.}\ \bibnamefont {Whaley}}, \ and\ \bibinfo {author}
  {\bibfnamefont {R.}~\bibnamefont {Babbush}},\ }\href {\doibase
  10.1038/s41534-020-00341-7} {\bibfield  {journal} {\bibinfo  {journal} {npj
  Quantum Inf}\ }\textbf {\bibinfo {volume} {7}},\ \bibinfo {pages} {23}
  (\bibinfo {year} {2021})}\BibitemShut {NoStop}%
\bibitem [{\citenamefont {Hua}(1944)}]{Hua_theory_1944}%
  \BibitemOpen
  \bibfield  {author} {\bibinfo {author} {\bibfnamefont {L.-K.}\ \bibnamefont
  {Hua}},\ }\href {\doibase 10.2307/2371910} {\bibfield  {journal} {\bibinfo
  {journal} {Am. J. Math.}\ }\textbf {\bibinfo {volume} {66}},\ \bibinfo
  {pages} {470} (\bibinfo {year} {1944})}\BibitemShut {NoStop}%
\bibitem [{\citenamefont {Henderson}\ \emph {et~al.}(2015)\citenamefont
  {Henderson}, \citenamefont {Bulik},\ and\ \citenamefont
  {Scuseria}}]{henderson_pair_2015}%
  \BibitemOpen
  \bibfield  {author} {\bibinfo {author} {\bibfnamefont {T.~M.}\ \bibnamefont
  {Henderson}}, \bibinfo {author} {\bibfnamefont {I.~W.}\ \bibnamefont
  {Bulik}}, \ and\ \bibinfo {author} {\bibfnamefont {G.~E.}\ \bibnamefont
  {Scuseria}},\ }\href {\doibase 10.1063/1.4921986} {\bibfield  {journal}
  {\bibinfo  {journal} {J. Chem. Phys.}\ }\textbf {\bibinfo {volume} {142}},\
  \bibinfo {pages} {214116} (\bibinfo {year} {2015})}\BibitemShut {NoStop}%
\bibitem [{\citenamefont {Bytautas}\ \emph {et~al.}(2011)\citenamefont
  {Bytautas}, \citenamefont {Henderson}, \citenamefont {Jim{\'e}nez-Hoyos},
  \citenamefont {Ellis},\ and\ \citenamefont
  {Scuseria}}]{bytautas_seniority_2011}%
  \BibitemOpen
  \bibfield  {author} {\bibinfo {author} {\bibfnamefont {L.}~\bibnamefont
  {Bytautas}}, \bibinfo {author} {\bibfnamefont {T.~M.}\ \bibnamefont
  {Henderson}}, \bibinfo {author} {\bibfnamefont {C.~A.}\ \bibnamefont
  {Jim{\'e}nez-Hoyos}}, \bibinfo {author} {\bibfnamefont {J.~K.}\ \bibnamefont
  {Ellis}}, \ and\ \bibinfo {author} {\bibfnamefont {G.~E.}\ \bibnamefont
  {Scuseria}},\ }\href {\doibase 10.1063/1.3613706} {\bibfield  {journal}
  {\bibinfo  {journal} {J. Chem. Phys.}\ }\textbf {\bibinfo {volume} {135}},\
  \bibinfo {pages} {044119} (\bibinfo {year} {2011})}\BibitemShut {NoStop}%
\bibitem [{\citenamefont {Stein}\ \emph {et~al.}(2014)\citenamefont {Stein},
  \citenamefont {Henderson},\ and\ \citenamefont
  {Scuseria}}]{stein_seniority_2014}%
  \BibitemOpen
  \bibfield  {author} {\bibinfo {author} {\bibfnamefont {T.}~\bibnamefont
  {Stein}}, \bibinfo {author} {\bibfnamefont {T.~M.}\ \bibnamefont
  {Henderson}}, \ and\ \bibinfo {author} {\bibfnamefont {G.~E.}\ \bibnamefont
  {Scuseria}},\ }\href {\doibase 10.1063/1.4880819} {\bibfield  {journal}
  {\bibinfo  {journal} {J. Chem. Phys.}\ }\textbf {\bibinfo {volume} {140}},\
  \bibinfo {pages} {214113} (\bibinfo {year} {2014})}\BibitemShut {NoStop}%
\bibitem [{\citenamefont {Henderson}\ \emph {et~al.}(2014)\citenamefont
  {Henderson}, \citenamefont {Bulik}, \citenamefont {Stein},\ and\
  \citenamefont {Scuseria}}]{henderson_seniority-based_2014}%
  \BibitemOpen
  \bibfield  {author} {\bibinfo {author} {\bibfnamefont {T.~M.}\ \bibnamefont
  {Henderson}}, \bibinfo {author} {\bibfnamefont {I.~W.}\ \bibnamefont
  {Bulik}}, \bibinfo {author} {\bibfnamefont {T.}~\bibnamefont {Stein}}, \ and\
  \bibinfo {author} {\bibfnamefont {G.~E.}\ \bibnamefont {Scuseria}},\ }\href
  {\doibase 10.1063/1.4904384} {\bibfield  {journal} {\bibinfo  {journal} {J.
  Chem. Phys.}\ }\textbf {\bibinfo {volume} {141}},\ \bibinfo {pages} {244104}
  (\bibinfo {year} {2014})}\BibitemShut {NoStop}%
\bibitem [{\citenamefont {Bytautas}\ \emph {et~al.}(2015)\citenamefont
  {Bytautas}, \citenamefont {Scuseria},\ and\ \citenamefont
  {Ruedenberg}}]{bytautas_seniority_2015}%
  \BibitemOpen
  \bibfield  {author} {\bibinfo {author} {\bibfnamefont {L.}~\bibnamefont
  {Bytautas}}, \bibinfo {author} {\bibfnamefont {G.~E.}\ \bibnamefont
  {Scuseria}}, \ and\ \bibinfo {author} {\bibfnamefont {K.}~\bibnamefont
  {Ruedenberg}},\ }\href {\doibase 10.1063/1.4929904} {\bibfield  {journal}
  {\bibinfo  {journal} {J. Chem. Phys.}\ }\textbf {\bibinfo {volume} {143}},\
  \bibinfo {pages} {094105} (\bibinfo {year} {2015})}\BibitemShut {NoStop}%
\bibitem [{\citenamefont {Shepherd}\ \emph {et~al.}(2016)\citenamefont
  {Shepherd}, \citenamefont {Henderson},\ and\ \citenamefont
  {Scuseria}}]{shepherd_using_2016}%
  \BibitemOpen
  \bibfield  {author} {\bibinfo {author} {\bibfnamefont {J.~J.}\ \bibnamefont
  {Shepherd}}, \bibinfo {author} {\bibfnamefont {T.~M.}\ \bibnamefont
  {Henderson}}, \ and\ \bibinfo {author} {\bibfnamefont {G.~E.}\ \bibnamefont
  {Scuseria}},\ }\href {\doibase 10.1063/1.4942770} {\bibfield  {journal}
  {\bibinfo  {journal} {J. Chem. Phys.}\ }\textbf {\bibinfo {volume} {144}},\
  \bibinfo {pages} {094112} (\bibinfo {year} {2016})}\BibitemShut {NoStop}%
\bibitem [{\citenamefont {Elfving}\ \emph {et~al.}(2021)\citenamefont
  {Elfving}, \citenamefont {Millaruelo}, \citenamefont {G{\'a}mez},\ and\
  \citenamefont {Gogolin}}]{elfving_simulating_2021}%
  \BibitemOpen
  \bibfield  {author} {\bibinfo {author} {\bibfnamefont {V.~E.}\ \bibnamefont
  {Elfving}}, \bibinfo {author} {\bibfnamefont {M.}~\bibnamefont {Millaruelo}},
  \bibinfo {author} {\bibfnamefont {J.~A.}\ \bibnamefont {G{\'a}mez}}, \ and\
  \bibinfo {author} {\bibfnamefont {C.}~\bibnamefont {Gogolin}},\ }\href
  {\doibase 10.1103/PhysRevA.103.032605} {\bibfield  {journal} {\bibinfo
  {journal} {Phys. Rev. A}\ }\textbf {\bibinfo {volume} {103}},\ \bibinfo
  {pages} {032605} (\bibinfo {year} {2021})}\BibitemShut {NoStop}%
\bibitem [{\citenamefont {Veillard}\ and\ \citenamefont
  {Clementi}(1967)}]{veillard_complete_1967}%
  \BibitemOpen
  \bibfield  {author} {\bibinfo {author} {\bibfnamefont {A.}~\bibnamefont
  {Veillard}}\ and\ \bibinfo {author} {\bibfnamefont {E.}~\bibnamefont
  {Clementi}},\ }\href {\doibase 10.1007/BF01151915} {\bibfield  {journal}
  {\bibinfo  {journal} {Theoret. Chim. Acta}\ }\textbf {\bibinfo {volume}
  {7}},\ \bibinfo {pages} {133} (\bibinfo {year} {1967})}\BibitemShut {NoStop}%
\bibitem [{\citenamefont {Couty}\ and\ \citenamefont
  {Hall}(1997)}]{couty_generalized_1997}%
  \BibitemOpen
  \bibfield  {author} {\bibinfo {author} {\bibfnamefont {M.}~\bibnamefont
  {Couty}}\ and\ \bibinfo {author} {\bibfnamefont {M.~B.}\ \bibnamefont
  {Hall}},\ }\href {\doibase 10.1021/jp963953l} {\bibfield  {journal} {\bibinfo
   {journal} {J. Phys. Chem. A}\ }\textbf {\bibinfo {volume} {101}},\ \bibinfo
  {pages} {6936} (\bibinfo {year} {1997})}\BibitemShut {NoStop}%
\bibitem [{\citenamefont {Kollmar}\ and\ \citenamefont
  {He{\ss}}(2003)}]{kollmar_new_2003}%
  \BibitemOpen
  \bibfield  {author} {\bibinfo {author} {\bibfnamefont {C.}~\bibnamefont
  {Kollmar}}\ and\ \bibinfo {author} {\bibfnamefont {B.~A.}\ \bibnamefont
  {He{\ss}}},\ }\href {\doibase 10.1063/1.1590635} {\bibfield  {journal}
  {\bibinfo  {journal} {J. Chem. Phys.}\ }\textbf {\bibinfo {volume} {119}},\
  \bibinfo {pages} {4655} (\bibinfo {year} {2003})}\BibitemShut {NoStop}%
\bibitem [{\citenamefont {Jordan}\ and\ \citenamefont
  {Wigner}(1928)}]{jordan_uber_1928}%
  \BibitemOpen
  \bibfield  {author} {\bibinfo {author} {\bibfnamefont {P.}~\bibnamefont
  {Jordan}}\ and\ \bibinfo {author} {\bibfnamefont {E.}~\bibnamefont
  {Wigner}},\ }\href {\doibase 10.1007/BF01331938} {\bibfield  {journal}
  {\bibinfo  {journal} {Z. Physik}\ }\textbf {\bibinfo {volume} {47}},\
  \bibinfo {pages} {631} (\bibinfo {year} {1928})}\BibitemShut {NoStop}%
\bibitem [{\citenamefont {Jiang}\ \emph {et~al.}(2018)\citenamefont {Jiang},
  \citenamefont {Sung}, \citenamefont {Kechedzhi}, \citenamefont
  {Smelyanskiy},\ and\ \citenamefont {Boixo}}]{jiang_quantum_2018}%
  \BibitemOpen
  \bibfield  {author} {\bibinfo {author} {\bibfnamefont {Z.}~\bibnamefont
  {Jiang}}, \bibinfo {author} {\bibfnamefont {K.~J.}\ \bibnamefont {Sung}},
  \bibinfo {author} {\bibfnamefont {K.}~\bibnamefont {Kechedzhi}}, \bibinfo
  {author} {\bibfnamefont {V.~N.}\ \bibnamefont {Smelyanskiy}}, \ and\ \bibinfo
  {author} {\bibfnamefont {S.}~\bibnamefont {Boixo}},\ }\href {\doibase
  10.1103/PhysRevApplied.9.044036} {\bibfield  {journal} {\bibinfo  {journal}
  {Phys. Rev. Applied}\ }\textbf {\bibinfo {volume} {9}},\ \bibinfo {pages}
  {044036} (\bibinfo {year} {2018})}\BibitemShut {NoStop}%
\bibitem [{\citenamefont {Thouless}(1960)}]{thouless_stability_1960}%
  \BibitemOpen
  \bibfield  {author} {\bibinfo {author} {\bibfnamefont {D.}~\bibnamefont
  {Thouless}},\ }\href {\doibase 10.1016/0029-5582(60)90048-1} {\bibfield
  {journal} {\bibinfo  {journal} {Nucl. Phys.}\ }\textbf {\bibinfo {volume}
  {21}},\ \bibinfo {pages} {225} (\bibinfo {year} {1960})}\BibitemShut
  {NoStop}%
\bibitem [{\citenamefont {Balian}\ and\ \citenamefont
  {Brezin}(1969)}]{balian_nonunitary_1969}%
  \BibitemOpen
  \bibfield  {author} {\bibinfo {author} {\bibfnamefont {R.}~\bibnamefont
  {Balian}}\ and\ \bibinfo {author} {\bibfnamefont {E.}~\bibnamefont
  {Brezin}},\ }\href {\doibase 10.1007/BF02710281} {\bibfield  {journal}
  {\bibinfo  {journal} {Nuov. Cim. B}\ }\textbf {\bibinfo {volume} {64}},\
  \bibinfo {pages} {37} (\bibinfo {year} {1969})}\BibitemShut {NoStop}%
\bibitem [{\citenamefont {Kivlichan}\ \emph {et~al.}(2018)\citenamefont
  {Kivlichan}, \citenamefont {McClean}, \citenamefont {Wiebe}, \citenamefont
  {Gidney}, \citenamefont {Aspuru-Guzik}, \citenamefont {Chan},\ and\
  \citenamefont {Babbush}}]{kivlichan_quantum_2018}%
  \BibitemOpen
  \bibfield  {author} {\bibinfo {author} {\bibfnamefont {I.~D.}\ \bibnamefont
  {Kivlichan}}, \bibinfo {author} {\bibfnamefont {J.}~\bibnamefont {McClean}},
  \bibinfo {author} {\bibfnamefont {N.}~\bibnamefont {Wiebe}}, \bibinfo
  {author} {\bibfnamefont {C.}~\bibnamefont {Gidney}}, \bibinfo {author}
  {\bibfnamefont {A.}~\bibnamefont {Aspuru-Guzik}}, \bibinfo {author}
  {\bibfnamefont {G.~K.-L.}\ \bibnamefont {Chan}}, \ and\ \bibinfo {author}
  {\bibfnamefont {R.}~\bibnamefont {Babbush}},\ }\href {\doibase
  10.1103/PhysRevLett.120.110501} {\bibfield  {journal} {\bibinfo  {journal}
  {Phys. Rev. Lett.}\ }\textbf {\bibinfo {volume} {120}},\ \bibinfo {pages}
  {110501} (\bibinfo {year} {2018})}\BibitemShut {NoStop}%
\bibitem [{\citenamefont {Schuld}\ \emph {et~al.}(2019)\citenamefont {Schuld},
  \citenamefont {Bergholm}, \citenamefont {Gogolin}, \citenamefont {Izaac},\
  and\ \citenamefont {Killoran}}]{schuld_evaluating_2019}%
  \BibitemOpen
  \bibfield  {author} {\bibinfo {author} {\bibfnamefont {M.}~\bibnamefont
  {Schuld}}, \bibinfo {author} {\bibfnamefont {V.}~\bibnamefont {Bergholm}},
  \bibinfo {author} {\bibfnamefont {C.}~\bibnamefont {Gogolin}}, \bibinfo
  {author} {\bibfnamefont {J.}~\bibnamefont {Izaac}}, \ and\ \bibinfo {author}
  {\bibfnamefont {N.}~\bibnamefont {Killoran}},\ }\href {\doibase
  10.1103/PhysRevA.99.032331} {\bibfield  {journal} {\bibinfo  {journal} {Phys.
  Rev. A}\ }\textbf {\bibinfo {volume} {99}},\ \bibinfo {pages} {032331}
  (\bibinfo {year} {2019})}\BibitemShut {NoStop}%
\bibitem [{\citenamefont {Kottmann}\ \emph {et~al.}(2021)\citenamefont
  {Kottmann}, \citenamefont {Anand},\ and\ \citenamefont
  {Aspuru-Guzik}}]{kottmann_feasible_2021}%
  \BibitemOpen
  \bibfield  {author} {\bibinfo {author} {\bibfnamefont {J.~S.}\ \bibnamefont
  {Kottmann}}, \bibinfo {author} {\bibfnamefont {A.}~\bibnamefont {Anand}}, \
  and\ \bibinfo {author} {\bibfnamefont {A.}~\bibnamefont {Aspuru-Guzik}},\
  }\href {\doibase 10.1039/D0SC06627C} {\bibfield  {journal} {\bibinfo
  {journal} {Chem. Sci.}\ }\textbf {\bibinfo {volume} {12}},\ \bibinfo {pages}
  {3497} (\bibinfo {year} {2021})}\BibitemShut {NoStop}%
\bibitem [{\citenamefont {Izmaylov}\ \emph {et~al.}(2021)\citenamefont
  {Izmaylov}, \citenamefont {Lang},\ and\ \citenamefont
  {Yen}}]{izmaylov_analytic_2021}%
  \BibitemOpen
  \bibfield  {author} {\bibinfo {author} {\bibfnamefont {A.~F.}\ \bibnamefont
  {Izmaylov}}, \bibinfo {author} {\bibfnamefont {R.~A.}\ \bibnamefont {Lang}},
  \ and\ \bibinfo {author} {\bibfnamefont {T.-C.}\ \bibnamefont {Yen}},\ }\href
  {\doibase 10.1103/PhysRevA.104.062443} {\bibfield  {journal} {\bibinfo
  {journal} {Phys. Rev. A}\ }\textbf {\bibinfo {volume} {104}},\ \bibinfo
  {pages} {062443} (\bibinfo {year} {2021})}\BibitemShut {NoStop}%
\bibitem [{\citenamefont {Mizukami}\ \emph {et~al.}(2020)\citenamefont
  {Mizukami}, \citenamefont {Mitarai}, \citenamefont {Nakagawa}, \citenamefont
  {Yamamoto}, \citenamefont {Yan},\ and\ \citenamefont
  {Ohnishi}}]{mizukami_orbital_2020}%
  \BibitemOpen
  \bibfield  {author} {\bibinfo {author} {\bibfnamefont {W.}~\bibnamefont
  {Mizukami}}, \bibinfo {author} {\bibfnamefont {K.}~\bibnamefont {Mitarai}},
  \bibinfo {author} {\bibfnamefont {Y.~O.}\ \bibnamefont {Nakagawa}}, \bibinfo
  {author} {\bibfnamefont {T.}~\bibnamefont {Yamamoto}}, \bibinfo {author}
  {\bibfnamefont {T.}~\bibnamefont {Yan}}, \ and\ \bibinfo {author}
  {\bibfnamefont {Y.-y.}\ \bibnamefont {Ohnishi}},\ }\href {\doibase
  10.1103/PhysRevResearch.2.033421} {\bibfield  {journal} {\bibinfo  {journal}
  {Phys. Rev. Research}\ }\textbf {\bibinfo {volume} {2}},\ \bibinfo {pages}
  {033421} (\bibinfo {year} {2020})}\BibitemShut {NoStop}%
\bibitem [{\citenamefont {Sokolov}\ \emph {et~al.}(2020)\citenamefont
  {Sokolov}, \citenamefont {Barkoutsos}, \citenamefont {Ollitrault},
  \citenamefont {Greenberg}, \citenamefont {Rice}, \citenamefont {Pistoia},\
  and\ \citenamefont {Tavernelli}}]{sokolov_quantum_2020}%
  \BibitemOpen
  \bibfield  {author} {\bibinfo {author} {\bibfnamefont {I.~O.}\ \bibnamefont
  {Sokolov}}, \bibinfo {author} {\bibfnamefont {P.~K.}\ \bibnamefont
  {Barkoutsos}}, \bibinfo {author} {\bibfnamefont {P.~J.}\ \bibnamefont
  {Ollitrault}}, \bibinfo {author} {\bibfnamefont {D.}~\bibnamefont
  {Greenberg}}, \bibinfo {author} {\bibfnamefont {J.}~\bibnamefont {Rice}},
  \bibinfo {author} {\bibfnamefont {M.}~\bibnamefont {Pistoia}}, \ and\
  \bibinfo {author} {\bibfnamefont {I.}~\bibnamefont {Tavernelli}},\ }\href
  {\doibase 10.1063/1.5141835} {\bibfield  {journal} {\bibinfo  {journal} {J.
  Chem. Phys.}\ }\textbf {\bibinfo {volume} {152}},\ \bibinfo {pages} {124107}
  (\bibinfo {year} {2020})}\BibitemShut {NoStop}%
\bibitem [{\citenamefont {Bloch}\ and\ \citenamefont
  {Messiah}(1962)}]{bloch_canonical_1962}%
  \BibitemOpen
  \bibfield  {author} {\bibinfo {author} {\bibfnamefont {C.}~\bibnamefont
  {Bloch}}\ and\ \bibinfo {author} {\bibfnamefont {A.}~\bibnamefont
  {Messiah}},\ }\href {\doibase 10.1016/0029-5582(62)90377-2} {\bibfield
  {journal} {\bibinfo  {journal} {Nucl. Phys.}\ }\textbf {\bibinfo {volume}
  {39}},\ \bibinfo {pages} {95} (\bibinfo {year} {1962})}\BibitemShut {NoStop}%
\bibitem [{\citenamefont {Egido}\ and\ \citenamefont
  {Ring}(1982)}]{egido_symmetry_1982}%
  \BibitemOpen
  \bibfield  {author} {\bibinfo {author} {\bibfnamefont {J.}~\bibnamefont
  {Egido}}\ and\ \bibinfo {author} {\bibfnamefont {P.}~\bibnamefont {Ring}},\
  }\href {\doibase 10.1016/0375-9474(82)90447-X} {\bibfield  {journal}
  {\bibinfo  {journal} {Nucl. Phys. A}\ }\textbf {\bibinfo {volume} {383}},\
  \bibinfo {pages} {189} (\bibinfo {year} {1982})}\BibitemShut {NoStop}%
\bibitem [{\citenamefont {Sheikh}\ and\ \citenamefont
  {Ring}(2000)}]{sheikh_symmetry-projected_2000}%
  \BibitemOpen
  \bibfield  {author} {\bibinfo {author} {\bibfnamefont {J.~A.}\ \bibnamefont
  {Sheikh}}\ and\ \bibinfo {author} {\bibfnamefont {P.}~\bibnamefont {Ring}},\
  }\href {\doibase 10.1016/S0375-9474(99)00424-8} {\bibfield  {journal}
  {\bibinfo  {journal} {Nucl. Phys. A}\ }\textbf {\bibinfo {volume} {665}},\
  \bibinfo {pages} {71} (\bibinfo {year} {2000})}\BibitemShut {NoStop}%
\bibitem [{\citenamefont {Nooijen}(2000)}]{nooijen_can_2000}%
  \BibitemOpen
  \bibfield  {author} {\bibinfo {author} {\bibfnamefont {M.}~\bibnamefont
  {Nooijen}},\ }\href {\doibase 10.1103/PhysRevLett.84.2108} {\bibfield
  {journal} {\bibinfo  {journal} {Phys. Rev. Lett.}\ }\textbf {\bibinfo
  {volume} {84}},\ \bibinfo {pages} {2108} (\bibinfo {year}
  {2000})}\BibitemShut {NoStop}%
\bibitem [{\citenamefont {Taube}\ and\ \citenamefont
  {Bartlett}(2006)}]{taube_new_2006}%
  \BibitemOpen
  \bibfield  {author} {\bibinfo {author} {\bibfnamefont {A.~G.}\ \bibnamefont
  {Taube}}\ and\ \bibinfo {author} {\bibfnamefont {R.~J.}\ \bibnamefont
  {Bartlett}},\ }\href {\doibase https://doi.org/10.1002/qua.21198} {\bibfield
  {journal} {\bibinfo  {journal} {Int. J. Quantum Chem.}\ }\textbf {\bibinfo
  {volume} {106}},\ \bibinfo {pages} {3393} (\bibinfo {year}
  {2006})}\BibitemShut {NoStop}%
\bibitem [{\citenamefont {Romero}\ \emph {et~al.}(2018)\citenamefont {Romero},
  \citenamefont {Babbush}, \citenamefont {McClean}, \citenamefont {Hempel},
  \citenamefont {Love},\ and\ \citenamefont
  {Aspuru-Guzik}}]{romero_strategies_2018}%
  \BibitemOpen
  \bibfield  {author} {\bibinfo {author} {\bibfnamefont {J.}~\bibnamefont
  {Romero}}, \bibinfo {author} {\bibfnamefont {R.}~\bibnamefont {Babbush}},
  \bibinfo {author} {\bibfnamefont {J.~R.}\ \bibnamefont {McClean}}, \bibinfo
  {author} {\bibfnamefont {C.}~\bibnamefont {Hempel}}, \bibinfo {author}
  {\bibfnamefont {P.~J.}\ \bibnamefont {Love}}, \ and\ \bibinfo {author}
  {\bibfnamefont {A.}~\bibnamefont {Aspuru-Guzik}},\ }\href {\doibase
  10.1088/2058-9565/aad3e4} {\bibfield  {journal} {\bibinfo  {journal} {Quantum
  Sci. Technol.}\ }\textbf {\bibinfo {volume} {4}},\ \bibinfo {pages} {014008}
  (\bibinfo {year} {2018})}\BibitemShut {NoStop}%
\bibitem [{\citenamefont {O'Gorman}\ \emph {et~al.}(2019)\citenamefont
  {O'Gorman}, \citenamefont {Huggins}, \citenamefont {Rieffel},\ and\
  \citenamefont {Whaley}}]{ogorman_generalized_2019}%
  \BibitemOpen
  \bibfield  {author} {\bibinfo {author} {\bibfnamefont {B.}~\bibnamefont
  {O'Gorman}}, \bibinfo {author} {\bibfnamefont {W.~J.}\ \bibnamefont
  {Huggins}}, \bibinfo {author} {\bibfnamefont {E.~G.}\ \bibnamefont
  {Rieffel}}, \ and\ \bibinfo {author} {\bibfnamefont {K.~B.}\ \bibnamefont
  {Whaley}},\ }\href {http://arxiv.org/abs/1905.05118} {\bibfield  {journal}
  {\bibinfo  {journal} {arXiv:1905.05118}\ } (\bibinfo {year}
  {2019})}\BibitemShut {NoStop}%
\bibitem [{\citenamefont {Takeshita}\ \emph {et~al.}(2020)\citenamefont
  {Takeshita}, \citenamefont {Rubin}, \citenamefont {Jiang}, \citenamefont
  {Lee}, \citenamefont {Babbush},\ and\ \citenamefont
  {McClean}}]{takeshita_increasing_2020}%
  \BibitemOpen
  \bibfield  {author} {\bibinfo {author} {\bibfnamefont {T.}~\bibnamefont
  {Takeshita}}, \bibinfo {author} {\bibfnamefont {N.~C.}\ \bibnamefont
  {Rubin}}, \bibinfo {author} {\bibfnamefont {Z.}~\bibnamefont {Jiang}},
  \bibinfo {author} {\bibfnamefont {E.}~\bibnamefont {Lee}}, \bibinfo {author}
  {\bibfnamefont {R.}~\bibnamefont {Babbush}}, \ and\ \bibinfo {author}
  {\bibfnamefont {J.~R.}\ \bibnamefont {McClean}},\ }\href {\doibase
  10.1103/PhysRevX.10.011004} {\bibfield  {journal} {\bibinfo  {journal} {Phys.
  Rev. X}\ }\textbf {\bibinfo {volume} {10}},\ \bibinfo {pages} {011004}
  (\bibinfo {year} {2020})}\BibitemShut {NoStop}%
\bibitem [{\citenamefont {Motta}\ \emph {et~al.}(2021)\citenamefont {Motta},
  \citenamefont {Ye}, \citenamefont {McClean}, \citenamefont {Li},
  \citenamefont {Minnich}, \citenamefont {Babbush},\ and\ \citenamefont
  {Chan}}]{motta_low_2021}%
  \BibitemOpen
  \bibfield  {author} {\bibinfo {author} {\bibfnamefont {M.}~\bibnamefont
  {Motta}}, \bibinfo {author} {\bibfnamefont {E.}~\bibnamefont {Ye}}, \bibinfo
  {author} {\bibfnamefont {J.~R.}\ \bibnamefont {McClean}}, \bibinfo {author}
  {\bibfnamefont {Z.}~\bibnamefont {Li}}, \bibinfo {author} {\bibfnamefont
  {A.~J.}\ \bibnamefont {Minnich}}, \bibinfo {author} {\bibfnamefont
  {R.}~\bibnamefont {Babbush}}, \ and\ \bibinfo {author} {\bibfnamefont
  {G.~K.-L.}\ \bibnamefont {Chan}},\ }\href {\doibase
  10.1038/s41534-021-00416-z} {\bibfield  {journal} {\bibinfo  {journal} {npj
  Quantum Inf}\ }\textbf {\bibinfo {volume} {7}},\ \bibinfo {pages} {83}
  (\bibinfo {year} {2021})}\BibitemShut {NoStop}%
\bibitem [{\citenamefont {Rubin}\ \emph {et~al.}(2022)\citenamefont {Rubin},
  \citenamefont {Lee},\ and\ \citenamefont {Babbush}}]{rubin_compressing_2022}%
  \BibitemOpen
  \bibfield  {author} {\bibinfo {author} {\bibfnamefont {N.~C.}\ \bibnamefont
  {Rubin}}, \bibinfo {author} {\bibfnamefont {J.}~\bibnamefont {Lee}}, \ and\
  \bibinfo {author} {\bibfnamefont {R.}~\bibnamefont {Babbush}},\ }\href
  {\doibase 10.1021/acs.jctc.1c00912} {\bibfield  {journal} {\bibinfo
  {journal} {J. Chem. Theory Comput.}\ }\textbf {\bibinfo {volume} {18}},\
  \bibinfo {pages} {1480} (\bibinfo {year} {2022})}\BibitemShut {NoStop}%
\bibitem [{\citenamefont {Kottmann}\ and\ \citenamefont
  {Aspuru-Guzik}(2022)}]{kottmann_optimized_2022}%
  \BibitemOpen
  \bibfield  {author} {\bibinfo {author} {\bibfnamefont {J.~S.}\ \bibnamefont
  {Kottmann}}\ and\ \bibinfo {author} {\bibfnamefont {A.}~\bibnamefont
  {Aspuru-Guzik}},\ }\href {\doibase 10.1103/PhysRevA.105.032449} {\bibfield
  {journal} {\bibinfo  {journal} {Phys. Rev. A}\ }\textbf {\bibinfo {volume}
  {105}},\ \bibinfo {pages} {032449} (\bibinfo {year} {2022})}\BibitemShut
  {NoStop}%
\bibitem [{\citenamefont {Babbush}\ \emph {et~al.}(2019)\citenamefont
  {Babbush}, \citenamefont {Berry}, \citenamefont {McClean},\ and\
  \citenamefont {Neven}}]{babbush_quantum_2019}%
  \BibitemOpen
  \bibfield  {author} {\bibinfo {author} {\bibfnamefont {R.}~\bibnamefont
  {Babbush}}, \bibinfo {author} {\bibfnamefont {D.~W.}\ \bibnamefont {Berry}},
  \bibinfo {author} {\bibfnamefont {J.~R.}\ \bibnamefont {McClean}}, \ and\
  \bibinfo {author} {\bibfnamefont {H.}~\bibnamefont {Neven}},\ }\href
  {\doibase 10.1038/s41534-019-0199-y} {\bibfield  {journal} {\bibinfo
  {journal} {npj Quantum Inf}\ }\textbf {\bibinfo {volume} {5}},\ \bibinfo
  {pages} {92} (\bibinfo {year} {2019})}\BibitemShut {NoStop}%
\bibitem [{\citenamefont {Yen}\ \emph {et~al.}(2020)\citenamefont {Yen},
  \citenamefont {Verteletskyi},\ and\ \citenamefont
  {Izmaylov}}]{yen_measuring_2020}%
  \BibitemOpen
  \bibfield  {author} {\bibinfo {author} {\bibfnamefont {T.-C.}\ \bibnamefont
  {Yen}}, \bibinfo {author} {\bibfnamefont {V.}~\bibnamefont {Verteletskyi}}, \
  and\ \bibinfo {author} {\bibfnamefont {A.~F.}\ \bibnamefont {Izmaylov}},\
  }\href {\doibase 10.1021/acs.jctc.0c00008} {\bibfield  {journal} {\bibinfo
  {journal} {J. Chem. Theory Comput.}\ }\textbf {\bibinfo {volume} {16}},\
  \bibinfo {pages} {2400} (\bibinfo {year} {2020})}\BibitemShut {NoStop}%
\bibitem [{\citenamefont {Izmaylov}\ \emph
  {et~al.}(2020{\natexlab{b}})\citenamefont {Izmaylov}, \citenamefont {Yen},
  \citenamefont {Lang},\ and\ \citenamefont
  {Verteletskyi}}]{izmaylov_unitary_2020}%
  \BibitemOpen
  \bibfield  {author} {\bibinfo {author} {\bibfnamefont {A.~F.}\ \bibnamefont
  {Izmaylov}}, \bibinfo {author} {\bibfnamefont {T.-C.}\ \bibnamefont {Yen}},
  \bibinfo {author} {\bibfnamefont {R.~A.}\ \bibnamefont {Lang}}, \ and\
  \bibinfo {author} {\bibfnamefont {V.}~\bibnamefont {Verteletskyi}},\ }\href
  {\doibase 10.1021/acs.jctc.9b00791} {\bibfield  {journal} {\bibinfo
  {journal} {J. Chem. Theory Comput.}\ }\textbf {\bibinfo {volume} {16}},\
  \bibinfo {pages} {190} (\bibinfo {year} {2020}{\natexlab{b}})}\BibitemShut
  {NoStop}%
\bibitem [{\citenamefont {Nielsen}\ and\ \citenamefont
  {Chuang}(2010)}]{nielsen_quantum_2010}%
  \BibitemOpen
  \bibfield  {author} {\bibinfo {author} {\bibfnamefont {M.~A.}\ \bibnamefont
  {Nielsen}}\ and\ \bibinfo {author} {\bibfnamefont {I.~L.}\ \bibnamefont
  {Chuang}},\ }\href@noop {} {\emph {\bibinfo {title} {Quantum computation and
  quantum information}}},\ \bibinfo {edition} {10th}\ ed.\ (\bibinfo
  {publisher} {Cambridge University Press},\ \bibinfo {address} {Cambridge ;
  New York},\ \bibinfo {year} {2010})\BibitemShut {NoStop}%
\bibitem [{\citenamefont {{Google AI Quantum and Collaborators}}\ \emph
  {et~al.}(2020)\citenamefont {{Google AI Quantum and Collaborators}},
  \citenamefont {Arute}, \citenamefont {Arya}, \citenamefont {Babbush},
  \citenamefont {Bacon}, \citenamefont {Bardin}, \citenamefont {Barends},
  \citenamefont {Boixo}, \citenamefont {Broughton}, \citenamefont {Buckley},
  \citenamefont {Buell}, \citenamefont {Burkett}, \citenamefont {Bushnell},
  \citenamefont {Chen}, \citenamefont {Chen}, \citenamefont {Chiaro},
  \citenamefont {Collins}, \citenamefont {Courtney}, \citenamefont {Demura},
  \citenamefont {Dunsworth}, \citenamefont {Farhi}, \citenamefont {Fowler},
  \citenamefont {Foxen}, \citenamefont {Gidney}, \citenamefont {Giustina},
  \citenamefont {Graff}, \citenamefont {Habegger}, \citenamefont {Harrigan},
  \citenamefont {Ho}, \citenamefont {Hong}, \citenamefont {Huang},
  \citenamefont {Huggins}, \citenamefont {Ioffe}, \citenamefont {Isakov},
  \citenamefont {Jeffrey}, \citenamefont {Jiang}, \citenamefont {Jones},
  \citenamefont {Kafri}, \citenamefont {Kechedzhi}, \citenamefont {Kelly},
  \citenamefont {Kim}, \citenamefont {Klimov}, \citenamefont {Korotkov},
  \citenamefont {Kostritsa}, \citenamefont {Landhuis}, \citenamefont {Laptev},
  \citenamefont {Lindmark}, \citenamefont {Lucero}, \citenamefont {Martin},
  \citenamefont {Martinis}, \citenamefont {McClean}, \citenamefont {McEwen},
  \citenamefont {Megrant}, \citenamefont {Mi}, \citenamefont {Mohseni},
  \citenamefont {Mruczkiewicz}, \citenamefont {Mutus}, \citenamefont {Naaman},
  \citenamefont {Neeley}, \citenamefont {Neill}, \citenamefont {Neven},
  \citenamefont {Niu}, \citenamefont {O{\textquoteright}Brien}, \citenamefont
  {Ostby}, \citenamefont {Petukhov}, \citenamefont {Putterman}, \citenamefont
  {Quintana}, \citenamefont {Roushan}, \citenamefont {Rubin}, \citenamefont
  {Sank}, \citenamefont {Satzinger}, \citenamefont {Smelyanskiy}, \citenamefont
  {Strain}, \citenamefont {Sung}, \citenamefont {Szalay}, \citenamefont
  {Takeshita}, \citenamefont {Vainsencher}, \citenamefont {White},
  \citenamefont {Wiebe}, \citenamefont {Yao}, \citenamefont {Yeh},\ and\
  \citenamefont
  {Zalcman}}]{google_ai_quantum_and_collaborators_hartree-fock_2020}%
  \BibitemOpen
  \bibfield  {author} {\bibinfo {author} {\bibnamefont {{Google AI Quantum and
  Collaborators}}}, \bibinfo {author} {\bibfnamefont {F.}~\bibnamefont
  {Arute}}, \bibinfo {author} {\bibfnamefont {K.}~\bibnamefont {Arya}},
  \bibinfo {author} {\bibfnamefont {R.}~\bibnamefont {Babbush}}, \bibinfo
  {author} {\bibfnamefont {D.}~\bibnamefont {Bacon}}, \bibinfo {author}
  {\bibfnamefont {J.~C.}\ \bibnamefont {Bardin}}, \bibinfo {author}
  {\bibfnamefont {R.}~\bibnamefont {Barends}}, \bibinfo {author} {\bibfnamefont
  {S.}~\bibnamefont {Boixo}}, \bibinfo {author} {\bibfnamefont
  {M.}~\bibnamefont {Broughton}}, \bibinfo {author} {\bibfnamefont {B.~B.}\
  \bibnamefont {Buckley}}, \bibinfo {author} {\bibfnamefont {D.~A.}\
  \bibnamefont {Buell}}, \bibinfo {author} {\bibfnamefont {B.}~\bibnamefont
  {Burkett}}, \bibinfo {author} {\bibfnamefont {N.}~\bibnamefont {Bushnell}},
  \bibinfo {author} {\bibfnamefont {Y.}~\bibnamefont {Chen}}, \bibinfo {author}
  {\bibfnamefont {Z.}~\bibnamefont {Chen}}, \bibinfo {author} {\bibfnamefont
  {B.}~\bibnamefont {Chiaro}}, \bibinfo {author} {\bibfnamefont
  {R.}~\bibnamefont {Collins}}, \bibinfo {author} {\bibfnamefont
  {W.}~\bibnamefont {Courtney}}, \bibinfo {author} {\bibfnamefont
  {S.}~\bibnamefont {Demura}}, \bibinfo {author} {\bibfnamefont
  {A.}~\bibnamefont {Dunsworth}}, \bibinfo {author} {\bibfnamefont
  {E.}~\bibnamefont {Farhi}}, \bibinfo {author} {\bibfnamefont
  {A.}~\bibnamefont {Fowler}}, \bibinfo {author} {\bibfnamefont
  {B.}~\bibnamefont {Foxen}}, \bibinfo {author} {\bibfnamefont
  {C.}~\bibnamefont {Gidney}}, \bibinfo {author} {\bibfnamefont
  {M.}~\bibnamefont {Giustina}}, \bibinfo {author} {\bibfnamefont
  {R.}~\bibnamefont {Graff}}, \bibinfo {author} {\bibfnamefont
  {S.}~\bibnamefont {Habegger}}, \bibinfo {author} {\bibfnamefont {M.~P.}\
  \bibnamefont {Harrigan}}, \bibinfo {author} {\bibfnamefont {A.}~\bibnamefont
  {Ho}}, \bibinfo {author} {\bibfnamefont {S.}~\bibnamefont {Hong}}, \bibinfo
  {author} {\bibfnamefont {T.}~\bibnamefont {Huang}}, \bibinfo {author}
  {\bibfnamefont {W.~J.}\ \bibnamefont {Huggins}}, \bibinfo {author}
  {\bibfnamefont {L.}~\bibnamefont {Ioffe}}, \bibinfo {author} {\bibfnamefont
  {S.~V.}\ \bibnamefont {Isakov}}, \bibinfo {author} {\bibfnamefont
  {E.}~\bibnamefont {Jeffrey}}, \bibinfo {author} {\bibfnamefont
  {Z.}~\bibnamefont {Jiang}}, \bibinfo {author} {\bibfnamefont
  {C.}~\bibnamefont {Jones}}, \bibinfo {author} {\bibfnamefont
  {D.}~\bibnamefont {Kafri}}, \bibinfo {author} {\bibfnamefont
  {K.}~\bibnamefont {Kechedzhi}}, \bibinfo {author} {\bibfnamefont
  {J.}~\bibnamefont {Kelly}}, \bibinfo {author} {\bibfnamefont
  {S.}~\bibnamefont {Kim}}, \bibinfo {author} {\bibfnamefont {P.~V.}\
  \bibnamefont {Klimov}}, \bibinfo {author} {\bibfnamefont {A.}~\bibnamefont
  {Korotkov}}, \bibinfo {author} {\bibfnamefont {F.}~\bibnamefont {Kostritsa}},
  \bibinfo {author} {\bibfnamefont {D.}~\bibnamefont {Landhuis}}, \bibinfo
  {author} {\bibfnamefont {P.}~\bibnamefont {Laptev}}, \bibinfo {author}
  {\bibfnamefont {M.}~\bibnamefont {Lindmark}}, \bibinfo {author}
  {\bibfnamefont {E.}~\bibnamefont {Lucero}}, \bibinfo {author} {\bibfnamefont
  {O.}~\bibnamefont {Martin}}, \bibinfo {author} {\bibfnamefont {J.~M.}\
  \bibnamefont {Martinis}}, \bibinfo {author} {\bibfnamefont {J.~R.}\
  \bibnamefont {McClean}}, \bibinfo {author} {\bibfnamefont {M.}~\bibnamefont
  {McEwen}}, \bibinfo {author} {\bibfnamefont {A.}~\bibnamefont {Megrant}},
  \bibinfo {author} {\bibfnamefont {X.}~\bibnamefont {Mi}}, \bibinfo {author}
  {\bibfnamefont {M.}~\bibnamefont {Mohseni}}, \bibinfo {author} {\bibfnamefont
  {W.}~\bibnamefont {Mruczkiewicz}}, \bibinfo {author} {\bibfnamefont
  {J.}~\bibnamefont {Mutus}}, \bibinfo {author} {\bibfnamefont
  {O.}~\bibnamefont {Naaman}}, \bibinfo {author} {\bibfnamefont
  {M.}~\bibnamefont {Neeley}}, \bibinfo {author} {\bibfnamefont
  {C.}~\bibnamefont {Neill}}, \bibinfo {author} {\bibfnamefont
  {H.}~\bibnamefont {Neven}}, \bibinfo {author} {\bibfnamefont {M.~Y.}\
  \bibnamefont {Niu}}, \bibinfo {author} {\bibfnamefont {T.~E.}\ \bibnamefont
  {O{\textquoteright}Brien}}, \bibinfo {author} {\bibfnamefont
  {E.}~\bibnamefont {Ostby}}, \bibinfo {author} {\bibfnamefont
  {A.}~\bibnamefont {Petukhov}}, \bibinfo {author} {\bibfnamefont
  {H.}~\bibnamefont {Putterman}}, \bibinfo {author} {\bibfnamefont
  {C.}~\bibnamefont {Quintana}}, \bibinfo {author} {\bibfnamefont
  {P.}~\bibnamefont {Roushan}}, \bibinfo {author} {\bibfnamefont {N.~C.}\
  \bibnamefont {Rubin}}, \bibinfo {author} {\bibfnamefont {D.}~\bibnamefont
  {Sank}}, \bibinfo {author} {\bibfnamefont {K.~J.}\ \bibnamefont {Satzinger}},
  \bibinfo {author} {\bibfnamefont {V.}~\bibnamefont {Smelyanskiy}}, \bibinfo
  {author} {\bibfnamefont {D.}~\bibnamefont {Strain}}, \bibinfo {author}
  {\bibfnamefont {K.~J.}\ \bibnamefont {Sung}}, \bibinfo {author}
  {\bibfnamefont {M.}~\bibnamefont {Szalay}}, \bibinfo {author} {\bibfnamefont
  {T.~Y.}\ \bibnamefont {Takeshita}}, \bibinfo {author} {\bibfnamefont
  {A.}~\bibnamefont {Vainsencher}}, \bibinfo {author} {\bibfnamefont
  {T.}~\bibnamefont {White}}, \bibinfo {author} {\bibfnamefont
  {N.}~\bibnamefont {Wiebe}}, \bibinfo {author} {\bibfnamefont {Z.~J.}\
  \bibnamefont {Yao}}, \bibinfo {author} {\bibfnamefont {P.}~\bibnamefont
  {Yeh}}, \ and\ \bibinfo {author} {\bibfnamefont {A.}~\bibnamefont
  {Zalcman}},\ }\href {\doibase 10.1126/science.abb9811} {\bibfield  {journal}
  {\bibinfo  {journal} {Science}\ }\textbf {\bibinfo {volume} {369}},\ \bibinfo
  {pages} {1084} (\bibinfo {year} {2020})}\BibitemShut {NoStop}%
\bibitem [{\citenamefont {Yen}\ and\ \citenamefont
  {Izmaylov}(2021)}]{yen_cartan_2021}%
  \BibitemOpen
  \bibfield  {author} {\bibinfo {author} {\bibfnamefont {T.-C.}\ \bibnamefont
  {Yen}}\ and\ \bibinfo {author} {\bibfnamefont {A.~F.}\ \bibnamefont
  {Izmaylov}},\ }\href {\doibase 10.1103/PRXQuantum.2.040320} {\bibfield
  {journal} {\bibinfo  {journal} {PRX Quantum}\ }\textbf {\bibinfo {volume}
  {2}},\ \bibinfo {pages} {040320} (\bibinfo {year} {2021})}\BibitemShut
  {NoStop}%
\bibitem [{\citenamefont {McArdle}\ \emph {et~al.}(2019)\citenamefont
  {McArdle}, \citenamefont {Yuan},\ and\ \citenamefont
  {Benjamin}}]{mcardle_error-mitigated_2019}%
  \BibitemOpen
  \bibfield  {author} {\bibinfo {author} {\bibfnamefont {S.}~\bibnamefont
  {McArdle}}, \bibinfo {author} {\bibfnamefont {X.}~\bibnamefont {Yuan}}, \
  and\ \bibinfo {author} {\bibfnamefont {S.}~\bibnamefont {Benjamin}},\ }\href
  {\doibase 10.1103/PhysRevLett.122.180501} {\bibfield  {journal} {\bibinfo
  {journal} {Phys. Rev. Lett.}\ }\textbf {\bibinfo {volume} {122}},\ \bibinfo
  {pages} {180501} (\bibinfo {year} {2019})}\BibitemShut {NoStop}%
\bibitem [{\citenamefont {Arute}\ \emph {et~al.}(2020)\citenamefont {Arute},
  \citenamefont {Arya}, \citenamefont {Babbush}, \citenamefont {Bacon},
  \citenamefont {Bardin}, \citenamefont {Barends}, \citenamefont {Bengtsson},
  \citenamefont {Boixo}, \citenamefont {Broughton}, \citenamefont {Buckley},
  \citenamefont {Buell}, \citenamefont {Burkett}, \citenamefont {Bushnell},
  \citenamefont {Chen}, \citenamefont {Chen}, \citenamefont {Chen},
  \citenamefont {Chiaro}, \citenamefont {Collins}, \citenamefont {Cotton},
  \citenamefont {Courtney}, \citenamefont {Demura}, \citenamefont {Derk},
  \citenamefont {Dunsworth}, \citenamefont {Eppens}, \citenamefont {Eckl},
  \citenamefont {Erickson}, \citenamefont {Farhi}, \citenamefont {Fowler},
  \citenamefont {Foxen}, \citenamefont {Gidney}, \citenamefont {Giustina},
  \citenamefont {Graff}, \citenamefont {Gross}, \citenamefont {Habegger},
  \citenamefont {Harrigan}, \citenamefont {Ho}, \citenamefont {Hong},
  \citenamefont {Huang}, \citenamefont {Huggins}, \citenamefont {Ioffe},
  \citenamefont {Isakov}, \citenamefont {Jeffrey}, \citenamefont {Jiang},
  \citenamefont {Jones}, \citenamefont {Kafri}, \citenamefont {Kechedzhi},
  \citenamefont {Kelly}, \citenamefont {Kim}, \citenamefont {Klimov},
  \citenamefont {Korotkov}, \citenamefont {Kostritsa}, \citenamefont
  {Landhuis}, \citenamefont {Laptev}, \citenamefont {Lindmark}, \citenamefont
  {Lucero}, \citenamefont {Marthaler}, \citenamefont {Martin}, \citenamefont
  {Martinis}, \citenamefont {Marusczyk}, \citenamefont {McArdle}, \citenamefont
  {McClean}, \citenamefont {McCourt}, \citenamefont {McEwen}, \citenamefont
  {Megrant}, \citenamefont {Mejuto-Zaera}, \citenamefont {Mi}, \citenamefont
  {Mohseni}, \citenamefont {Mruczkiewicz}, \citenamefont {Mutus}, \citenamefont
  {Naaman}, \citenamefont {Neeley}, \citenamefont {Neill}, \citenamefont
  {Neven}, \citenamefont {Newman}, \citenamefont {Niu}, \citenamefont
  {O'Brien}, \citenamefont {Ostby}, \citenamefont {Pat{\'o}}, \citenamefont
  {Petukhov}, \citenamefont {Putterman}, \citenamefont {Quintana},
  \citenamefont {Reiner}, \citenamefont {Roushan}, \citenamefont {Rubin},
  \citenamefont {Sank}, \citenamefont {Satzinger}, \citenamefont {Smelyanskiy},
  \citenamefont {Strain}, \citenamefont {Sung}, \citenamefont {Schmitteckert},
  \citenamefont {Szalay}, \citenamefont {Tubman}, \citenamefont {Vainsencher},
  \citenamefont {White}, \citenamefont {Vogt}, \citenamefont {Yao},
  \citenamefont {Yeh}, \citenamefont {Zalcman},\ and\ \citenamefont
  {Zanker}}]{arute_observation_2020}%
  \BibitemOpen
  \bibfield  {author} {\bibinfo {author} {\bibfnamefont {F.}~\bibnamefont
  {Arute}}, \bibinfo {author} {\bibfnamefont {K.}~\bibnamefont {Arya}},
  \bibinfo {author} {\bibfnamefont {R.}~\bibnamefont {Babbush}}, \bibinfo
  {author} {\bibfnamefont {D.}~\bibnamefont {Bacon}}, \bibinfo {author}
  {\bibfnamefont {J.~C.}\ \bibnamefont {Bardin}}, \bibinfo {author}
  {\bibfnamefont {R.}~\bibnamefont {Barends}}, \bibinfo {author} {\bibfnamefont
  {A.}~\bibnamefont {Bengtsson}}, \bibinfo {author} {\bibfnamefont
  {S.}~\bibnamefont {Boixo}}, \bibinfo {author} {\bibfnamefont
  {M.}~\bibnamefont {Broughton}}, \bibinfo {author} {\bibfnamefont {B.~B.}\
  \bibnamefont {Buckley}}, \bibinfo {author} {\bibfnamefont {D.~A.}\
  \bibnamefont {Buell}}, \bibinfo {author} {\bibfnamefont {B.}~\bibnamefont
  {Burkett}}, \bibinfo {author} {\bibfnamefont {N.}~\bibnamefont {Bushnell}},
  \bibinfo {author} {\bibfnamefont {Y.}~\bibnamefont {Chen}}, \bibinfo {author}
  {\bibfnamefont {Z.}~\bibnamefont {Chen}}, \bibinfo {author} {\bibfnamefont
  {Y.-A.}\ \bibnamefont {Chen}}, \bibinfo {author} {\bibfnamefont
  {B.}~\bibnamefont {Chiaro}}, \bibinfo {author} {\bibfnamefont
  {R.}~\bibnamefont {Collins}}, \bibinfo {author} {\bibfnamefont {S.~J.}\
  \bibnamefont {Cotton}}, \bibinfo {author} {\bibfnamefont {W.}~\bibnamefont
  {Courtney}}, \bibinfo {author} {\bibfnamefont {S.}~\bibnamefont {Demura}},
  \bibinfo {author} {\bibfnamefont {A.}~\bibnamefont {Derk}}, \bibinfo {author}
  {\bibfnamefont {A.}~\bibnamefont {Dunsworth}}, \bibinfo {author}
  {\bibfnamefont {D.}~\bibnamefont {Eppens}}, \bibinfo {author} {\bibfnamefont
  {T.}~\bibnamefont {Eckl}}, \bibinfo {author} {\bibfnamefont {C.}~\bibnamefont
  {Erickson}}, \bibinfo {author} {\bibfnamefont {E.}~\bibnamefont {Farhi}},
  \bibinfo {author} {\bibfnamefont {A.}~\bibnamefont {Fowler}}, \bibinfo
  {author} {\bibfnamefont {B.}~\bibnamefont {Foxen}}, \bibinfo {author}
  {\bibfnamefont {C.}~\bibnamefont {Gidney}}, \bibinfo {author} {\bibfnamefont
  {M.}~\bibnamefont {Giustina}}, \bibinfo {author} {\bibfnamefont
  {R.}~\bibnamefont {Graff}}, \bibinfo {author} {\bibfnamefont {J.~A.}\
  \bibnamefont {Gross}}, \bibinfo {author} {\bibfnamefont {S.}~\bibnamefont
  {Habegger}}, \bibinfo {author} {\bibfnamefont {M.~P.}\ \bibnamefont
  {Harrigan}}, \bibinfo {author} {\bibfnamefont {A.}~\bibnamefont {Ho}},
  \bibinfo {author} {\bibfnamefont {S.}~\bibnamefont {Hong}}, \bibinfo {author}
  {\bibfnamefont {T.}~\bibnamefont {Huang}}, \bibinfo {author} {\bibfnamefont
  {W.}~\bibnamefont {Huggins}}, \bibinfo {author} {\bibfnamefont {L.~B.}\
  \bibnamefont {Ioffe}}, \bibinfo {author} {\bibfnamefont {S.~V.}\ \bibnamefont
  {Isakov}}, \bibinfo {author} {\bibfnamefont {E.}~\bibnamefont {Jeffrey}},
  \bibinfo {author} {\bibfnamefont {Z.}~\bibnamefont {Jiang}}, \bibinfo
  {author} {\bibfnamefont {C.}~\bibnamefont {Jones}}, \bibinfo {author}
  {\bibfnamefont {D.}~\bibnamefont {Kafri}}, \bibinfo {author} {\bibfnamefont
  {K.}~\bibnamefont {Kechedzhi}}, \bibinfo {author} {\bibfnamefont
  {J.}~\bibnamefont {Kelly}}, \bibinfo {author} {\bibfnamefont
  {S.}~\bibnamefont {Kim}}, \bibinfo {author} {\bibfnamefont {P.~V.}\
  \bibnamefont {Klimov}}, \bibinfo {author} {\bibfnamefont {A.~N.}\
  \bibnamefont {Korotkov}}, \bibinfo {author} {\bibfnamefont {F.}~\bibnamefont
  {Kostritsa}}, \bibinfo {author} {\bibfnamefont {D.}~\bibnamefont {Landhuis}},
  \bibinfo {author} {\bibfnamefont {P.}~\bibnamefont {Laptev}}, \bibinfo
  {author} {\bibfnamefont {M.}~\bibnamefont {Lindmark}}, \bibinfo {author}
  {\bibfnamefont {E.}~\bibnamefont {Lucero}}, \bibinfo {author} {\bibfnamefont
  {M.}~\bibnamefont {Marthaler}}, \bibinfo {author} {\bibfnamefont
  {O.}~\bibnamefont {Martin}}, \bibinfo {author} {\bibfnamefont {J.~M.}\
  \bibnamefont {Martinis}}, \bibinfo {author} {\bibfnamefont {A.}~\bibnamefont
  {Marusczyk}}, \bibinfo {author} {\bibfnamefont {S.}~\bibnamefont {McArdle}},
  \bibinfo {author} {\bibfnamefont {J.~R.}\ \bibnamefont {McClean}}, \bibinfo
  {author} {\bibfnamefont {T.}~\bibnamefont {McCourt}}, \bibinfo {author}
  {\bibfnamefont {M.}~\bibnamefont {McEwen}}, \bibinfo {author} {\bibfnamefont
  {A.}~\bibnamefont {Megrant}}, \bibinfo {author} {\bibfnamefont
  {C.}~\bibnamefont {Mejuto-Zaera}}, \bibinfo {author} {\bibfnamefont
  {X.}~\bibnamefont {Mi}}, \bibinfo {author} {\bibfnamefont {M.}~\bibnamefont
  {Mohseni}}, \bibinfo {author} {\bibfnamefont {W.}~\bibnamefont
  {Mruczkiewicz}}, \bibinfo {author} {\bibfnamefont {J.}~\bibnamefont {Mutus}},
  \bibinfo {author} {\bibfnamefont {O.}~\bibnamefont {Naaman}}, \bibinfo
  {author} {\bibfnamefont {M.}~\bibnamefont {Neeley}}, \bibinfo {author}
  {\bibfnamefont {C.}~\bibnamefont {Neill}}, \bibinfo {author} {\bibfnamefont
  {H.}~\bibnamefont {Neven}}, \bibinfo {author} {\bibfnamefont
  {M.}~\bibnamefont {Newman}}, \bibinfo {author} {\bibfnamefont {M.~Y.}\
  \bibnamefont {Niu}}, \bibinfo {author} {\bibfnamefont {T.~E.}\ \bibnamefont
  {O'Brien}}, \bibinfo {author} {\bibfnamefont {E.}~\bibnamefont {Ostby}},
  \bibinfo {author} {\bibfnamefont {B.}~\bibnamefont {Pat{\'o}}}, \bibinfo
  {author} {\bibfnamefont {A.}~\bibnamefont {Petukhov}}, \bibinfo {author}
  {\bibfnamefont {H.}~\bibnamefont {Putterman}}, \bibinfo {author}
  {\bibfnamefont {C.}~\bibnamefont {Quintana}}, \bibinfo {author}
  {\bibfnamefont {J.-M.}\ \bibnamefont {Reiner}}, \bibinfo {author}
  {\bibfnamefont {P.}~\bibnamefont {Roushan}}, \bibinfo {author} {\bibfnamefont
  {N.~C.}\ \bibnamefont {Rubin}}, \bibinfo {author} {\bibfnamefont
  {D.}~\bibnamefont {Sank}}, \bibinfo {author} {\bibfnamefont {K.~J.}\
  \bibnamefont {Satzinger}}, \bibinfo {author} {\bibfnamefont {V.}~\bibnamefont
  {Smelyanskiy}}, \bibinfo {author} {\bibfnamefont {D.}~\bibnamefont {Strain}},
  \bibinfo {author} {\bibfnamefont {K.~J.}\ \bibnamefont {Sung}}, \bibinfo
  {author} {\bibfnamefont {P.}~\bibnamefont {Schmitteckert}}, \bibinfo {author}
  {\bibfnamefont {M.}~\bibnamefont {Szalay}}, \bibinfo {author} {\bibfnamefont
  {N.~M.}\ \bibnamefont {Tubman}}, \bibinfo {author} {\bibfnamefont
  {A.}~\bibnamefont {Vainsencher}}, \bibinfo {author} {\bibfnamefont
  {T.}~\bibnamefont {White}}, \bibinfo {author} {\bibfnamefont
  {N.}~\bibnamefont {Vogt}}, \bibinfo {author} {\bibfnamefont {Z.~J.}\
  \bibnamefont {Yao}}, \bibinfo {author} {\bibfnamefont {P.}~\bibnamefont
  {Yeh}}, \bibinfo {author} {\bibfnamefont {A.}~\bibnamefont {Zalcman}}, \ and\
  \bibinfo {author} {\bibfnamefont {S.}~\bibnamefont {Zanker}},\ }\href
  {http://arxiv.org/abs/2010.07965} {\bibfield  {journal} {\bibinfo  {journal}
  {arXiv:2010.07965 [quant-ph]}\ } (\bibinfo {year} {2020})}\BibitemShut
  {NoStop}%
\bibitem [{\citenamefont {O{\textquoteright}Brien}\ \emph
  {et~al.}(2021)\citenamefont {O{\textquoteright}Brien}, \citenamefont {Polla},
  \citenamefont {Rubin}, \citenamefont {Huggins}, \citenamefont {McArdle},
  \citenamefont {Boixo}, \citenamefont {McClean},\ and\ \citenamefont
  {Babbush}}]{obrien_error_2021}%
  \BibitemOpen
  \bibfield  {author} {\bibinfo {author} {\bibfnamefont {T.~E.}\ \bibnamefont
  {O{\textquoteright}Brien}}, \bibinfo {author} {\bibfnamefont
  {S.}~\bibnamefont {Polla}}, \bibinfo {author} {\bibfnamefont {N.~C.}\
  \bibnamefont {Rubin}}, \bibinfo {author} {\bibfnamefont {W.~J.}\ \bibnamefont
  {Huggins}}, \bibinfo {author} {\bibfnamefont {S.}~\bibnamefont {McArdle}},
  \bibinfo {author} {\bibfnamefont {S.}~\bibnamefont {Boixo}}, \bibinfo
  {author} {\bibfnamefont {J.~R.}\ \bibnamefont {McClean}}, \ and\ \bibinfo
  {author} {\bibfnamefont {R.}~\bibnamefont {Babbush}},\ }\href {\doibase
  10.1103/PRXQuantum.2.020317} {\bibfield  {journal} {\bibinfo  {journal} {PRX
  Quantum}\ }\textbf {\bibinfo {volume} {2}},\ \bibinfo {pages} {020317}
  (\bibinfo {year} {2021})}\BibitemShut {NoStop}%
\bibitem [{\citenamefont {Dukelsky}\ \emph {et~al.}(2004)\citenamefont
  {Dukelsky}, \citenamefont {Pittel},\ and\ \citenamefont
  {Sierra}}]{dukelsky_colloquium:_2004}%
  \BibitemOpen
  \bibfield  {author} {\bibinfo {author} {\bibfnamefont {J.}~\bibnamefont
  {Dukelsky}}, \bibinfo {author} {\bibfnamefont {S.}~\bibnamefont {Pittel}}, \
  and\ \bibinfo {author} {\bibfnamefont {G.}~\bibnamefont {Sierra}},\ }\href
  {\doibase 10.1103/RevModPhys.76.643} {\bibfield  {journal} {\bibinfo
  {journal} {Rev. Mod. Phys.}\ }\textbf {\bibinfo {volume} {76}},\ \bibinfo
  {pages} {643} (\bibinfo {year} {2004})}\BibitemShut {NoStop}%
\bibitem [{noa(2019)}]{noauthor_qiskit_2019}%
  \BibitemOpen
  \href {\doibase 10.5281/zenodo.2562111} {\enquote {\bibinfo {title} {Qiskit:
  {An} {Open}-source {Framework} for {Quantum} {Computing}},}\ } (\bibinfo
  {year} {2019})\BibitemShut {NoStop}%
\bibitem [{\citenamefont {Tsuchimochi}\ \emph {et~al.}(2022)\citenamefont
  {Tsuchimochi}, \citenamefont {Taii}, \citenamefont {Nishimaki},\ and\
  \citenamefont {Ten-no}}]{tsuchimochi_adaptive_2022}%
  \BibitemOpen
  \bibfield  {author} {\bibinfo {author} {\bibfnamefont {T.}~\bibnamefont
  {Tsuchimochi}}, \bibinfo {author} {\bibfnamefont {M.}~\bibnamefont {Taii}},
  \bibinfo {author} {\bibfnamefont {T.}~\bibnamefont {Nishimaki}}, \ and\
  \bibinfo {author} {\bibfnamefont {S.~L.}\ \bibnamefont {Ten-no}},\ }\href
  {https://arxiv.org/abs/2205.07097} {\bibfield  {journal} {\bibinfo  {journal}
  {arXiv.2205.07097}\ } (\bibinfo {year} {2022})}\BibitemShut {NoStop}%
\bibitem [{\citenamefont {Bertels}\ \emph {et~al.}(2022)\citenamefont
  {Bertels}, \citenamefont {Grimsley}, \citenamefont {Economou}, \citenamefont
  {Barnes},\ and\ \citenamefont {Mayhall}}]{bertels_symmetry_2022}%
  \BibitemOpen
  \bibfield  {author} {\bibinfo {author} {\bibfnamefont {L.~W.}\ \bibnamefont
  {Bertels}}, \bibinfo {author} {\bibfnamefont {H.~R.}\ \bibnamefont
  {Grimsley}}, \bibinfo {author} {\bibfnamefont {S.~E.}\ \bibnamefont
  {Economou}}, \bibinfo {author} {\bibfnamefont {E.}~\bibnamefont {Barnes}}, \
  and\ \bibinfo {author} {\bibfnamefont {N.~J.}\ \bibnamefont {Mayhall}},\
  }\href {https://arxiv.org/abs/2207.03063} {\bibfield  {journal} {\bibinfo
  {journal} {arxiv.2207.03063}\ } (\bibinfo {year} {2022})}\BibitemShut
  {NoStop}%
\bibitem [{\citenamefont {Baran}\ and\ \citenamefont
  {Dukelsky}(2021)}]{baran_variational_2021}%
  \BibitemOpen
  \bibfield  {author} {\bibinfo {author} {\bibfnamefont {V.~V.}\ \bibnamefont
  {Baran}}\ and\ \bibinfo {author} {\bibfnamefont {J.}~\bibnamefont
  {Dukelsky}},\ }\href {\doibase 10.1103/PhysRevC.103.054317} {\bibfield
  {journal} {\bibinfo  {journal} {Phys. Rev. C}\ }\textbf {\bibinfo {volume}
  {103}},\ \bibinfo {pages} {054317} (\bibinfo {year} {2021})}\BibitemShut
  {NoStop}%
\bibitem [{\citenamefont {McClean}\ \emph {et~al.}(2018)\citenamefont
  {McClean}, \citenamefont {Boixo}, \citenamefont {Smelyanskiy}, \citenamefont
  {Babbush},\ and\ \citenamefont {Neven}}]{mcclean_barren_2018}%
  \BibitemOpen
  \bibfield  {author} {\bibinfo {author} {\bibfnamefont {J.~R.}\ \bibnamefont
  {McClean}}, \bibinfo {author} {\bibfnamefont {S.}~\bibnamefont {Boixo}},
  \bibinfo {author} {\bibfnamefont {V.~N.}\ \bibnamefont {Smelyanskiy}},
  \bibinfo {author} {\bibfnamefont {R.}~\bibnamefont {Babbush}}, \ and\
  \bibinfo {author} {\bibfnamefont {H.}~\bibnamefont {Neven}},\ }\href
  {\doibase 10.1038/s41467-018-07090-4} {\bibfield  {journal} {\bibinfo
  {journal} {Nat. Commun.}\ }\textbf {\bibinfo {volume} {9}},\ \bibinfo {pages}
  {1} (\bibinfo {year} {2018})}\BibitemShut {NoStop}%
\bibitem [{\citenamefont {Wimmer}(2012)}]{wimmer_algorithm_2012}%
  \BibitemOpen
  \bibfield  {author} {\bibinfo {author} {\bibfnamefont {M.}~\bibnamefont
  {Wimmer}},\ }\href {\doibase 10.1145/2331130.2331138} {\bibfield  {journal}
  {\bibinfo  {journal} {ACM Trans. Math. Softw.}\ }\textbf {\bibinfo {volume}
  {38}},\ \bibinfo {pages} {1} (\bibinfo {year} {2012})}\BibitemShut {NoStop}%
\bibitem [{\citenamefont {Sun}\ \emph {et~al.}(2018)\citenamefont {Sun},
  \citenamefont {Berkelbach}, \citenamefont {Blunt}, \citenamefont {Booth},
  \citenamefont {Guo}, \citenamefont {Li}, \citenamefont {Liu}, \citenamefont
  {McClain}, \citenamefont {Sayfutyarova}, \citenamefont {Sharma},
  \citenamefont {Wouters},\ and\ \citenamefont {Chan}}]{sun_p_2018}%
  \BibitemOpen
  \bibfield  {author} {\bibinfo {author} {\bibfnamefont {Q.}~\bibnamefont
  {Sun}}, \bibinfo {author} {\bibfnamefont {T.~C.}\ \bibnamefont {Berkelbach}},
  \bibinfo {author} {\bibfnamefont {N.~S.}\ \bibnamefont {Blunt}}, \bibinfo
  {author} {\bibfnamefont {G.~H.}\ \bibnamefont {Booth}}, \bibinfo {author}
  {\bibfnamefont {S.}~\bibnamefont {Guo}}, \bibinfo {author} {\bibfnamefont
  {Z.}~\bibnamefont {Li}}, \bibinfo {author} {\bibfnamefont {J.}~\bibnamefont
  {Liu}}, \bibinfo {author} {\bibfnamefont {J.~D.}\ \bibnamefont {McClain}},
  \bibinfo {author} {\bibfnamefont {E.~R.}\ \bibnamefont {Sayfutyarova}},
  \bibinfo {author} {\bibfnamefont {S.}~\bibnamefont {Sharma}}, \bibinfo
  {author} {\bibfnamefont {S.}~\bibnamefont {Wouters}}, \ and\ \bibinfo
  {author} {\bibfnamefont {G.~K.-L.}\ \bibnamefont {Chan}},\ }\href {\doibase
  10.1002/wcms.1340} {\bibfield  {journal} {\bibinfo  {journal} {WIREs Comput.
  Mol. Sci.}\ }\textbf {\bibinfo {volume} {8}} (\bibinfo {year} {2018}),\
  10.1002/wcms.1340}\BibitemShut {NoStop}%
\bibitem [{\citenamefont {Stair}\ and\ \citenamefont
  {Evangelista}(2022)}]{stair_qforte_2022}%
  \BibitemOpen
  \bibfield  {author} {\bibinfo {author} {\bibfnamefont {N.~H.}\ \bibnamefont
  {Stair}}\ and\ \bibinfo {author} {\bibfnamefont {F.~A.}\ \bibnamefont
  {Evangelista}},\ }\href {\doibase 10.1021/acs.jctc.1c01155} {\bibfield
  {journal} {\bibinfo  {journal} {J. Chem. Theory Comput.}\ }\textbf {\bibinfo
  {volume} {18}},\ \bibinfo {pages} {1555} (\bibinfo {year}
  {2022})}\BibitemShut {NoStop}%
\bibitem [{\citenamefont {Virtanen}\ \emph {et~al.}(2020)\citenamefont
  {Virtanen}, \citenamefont {Gommers}, \citenamefont {Oliphant}, \citenamefont
  {Haberland}, \citenamefont {Reddy}, \citenamefont {Cournapeau}, \citenamefont
  {Burovski}, \citenamefont {Peterson}, \citenamefont {Weckesser},
  \citenamefont {Bright}, \citenamefont {van~der Walt}, \citenamefont {Brett},
  \citenamefont {Wilson}, \citenamefont {Millman}, \citenamefont {Mayorov},
  \citenamefont {Nelson}, \citenamefont {Jones}, \citenamefont {Kern},
  \citenamefont {Larson}, \citenamefont {Carey}, \citenamefont {Polat},
  \citenamefont {Feng}, \citenamefont {Moore}, \citenamefont {VanderPlas},
  \citenamefont {Laxalde}, \citenamefont {Perktold}, \citenamefont {Cimrman},
  \citenamefont {Henriksen}, \citenamefont {Quintero}, \citenamefont {Harris},
  \citenamefont {Archibald}, \citenamefont {Ribeiro}, \citenamefont
  {Pedregosa}, \citenamefont {van Mulbregt}, \citenamefont {{SciPy 1.0
  Contributors}}, \citenamefont {Vijaykumar}, \citenamefont {Bardelli},
  \citenamefont {Rothberg}, \citenamefont {Hilboll}, \citenamefont {Kloeckner},
  \citenamefont {Scopatz}, \citenamefont {Lee}, \citenamefont {Rokem},
  \citenamefont {Woods}, \citenamefont {Fulton}, \citenamefont {Masson},
  \citenamefont {H{\"a}ggstr{\"o}m}, \citenamefont {Fitzgerald}, \citenamefont
  {Nicholson}, \citenamefont {Hagen}, \citenamefont {Pasechnik}, \citenamefont
  {Olivetti}, \citenamefont {Martin}, \citenamefont {Wieser}, \citenamefont
  {Silva}, \citenamefont {Lenders}, \citenamefont {Wilhelm}, \citenamefont
  {Young}, \citenamefont {Price}, \citenamefont {Ingold}, \citenamefont
  {Allen}, \citenamefont {Lee}, \citenamefont {Audren}, \citenamefont {Probst},
  \citenamefont {Dietrich}, \citenamefont {Silterra}, \citenamefont {Webber},
  \citenamefont {Slavi{\v c}}, \citenamefont {Nothman}, \citenamefont
  {Buchner}, \citenamefont {Kulick}, \citenamefont {Sch{\"o}nberger},
  \citenamefont {de~Miranda~Cardoso}, \citenamefont {Reimer}, \citenamefont
  {Harrington}, \citenamefont {Rodr{\'i}guez}, \citenamefont {Nunez-Iglesias},
  \citenamefont {Kuczynski}, \citenamefont {Tritz}, \citenamefont {Thoma},
  \citenamefont {Newville}, \citenamefont {K{\"u}mmerer}, \citenamefont
  {Bolingbroke}, \citenamefont {Tartre}, \citenamefont {Pak}, \citenamefont
  {Smith}, \citenamefont {Nowaczyk}, \citenamefont {Shebanov}, \citenamefont
  {Pavlyk}, \citenamefont {Brodtkorb}, \citenamefont {Lee}, \citenamefont
  {McGibbon}, \citenamefont {Feldbauer}, \citenamefont {Lewis}, \citenamefont
  {Tygier}, \citenamefont {Sievert}, \citenamefont {Vigna}, \citenamefont
  {Peterson}, \citenamefont {More}, \citenamefont {Pudlik}, \citenamefont
  {Oshima}, \citenamefont {Pingel}, \citenamefont {Robitaille}, \citenamefont
  {Spura}, \citenamefont {Jones}, \citenamefont {Cera}, \citenamefont {Leslie},
  \citenamefont {Zito}, \citenamefont {Krauss}, \citenamefont {Upadhyay},
  \citenamefont {Halchenko},\ and\ \citenamefont
  {V{\'a}zquez-Baeza}}]{virtanen_scipy_2020}%
  \BibitemOpen
  \bibfield  {author} {\bibinfo {author} {\bibfnamefont {P.}~\bibnamefont
  {Virtanen}}, \bibinfo {author} {\bibfnamefont {R.}~\bibnamefont {Gommers}},
  \bibinfo {author} {\bibfnamefont {T.~E.}\ \bibnamefont {Oliphant}}, \bibinfo
  {author} {\bibfnamefont {M.}~\bibnamefont {Haberland}}, \bibinfo {author}
  {\bibfnamefont {T.}~\bibnamefont {Reddy}}, \bibinfo {author} {\bibfnamefont
  {D.}~\bibnamefont {Cournapeau}}, \bibinfo {author} {\bibfnamefont
  {E.}~\bibnamefont {Burovski}}, \bibinfo {author} {\bibfnamefont
  {P.}~\bibnamefont {Peterson}}, \bibinfo {author} {\bibfnamefont
  {W.}~\bibnamefont {Weckesser}}, \bibinfo {author} {\bibfnamefont
  {J.}~\bibnamefont {Bright}}, \bibinfo {author} {\bibfnamefont {S.~J.}\
  \bibnamefont {van~der Walt}}, \bibinfo {author} {\bibfnamefont
  {M.}~\bibnamefont {Brett}}, \bibinfo {author} {\bibfnamefont
  {J.}~\bibnamefont {Wilson}}, \bibinfo {author} {\bibfnamefont {K.~J.}\
  \bibnamefont {Millman}}, \bibinfo {author} {\bibfnamefont {N.}~\bibnamefont
  {Mayorov}}, \bibinfo {author} {\bibfnamefont {A.~R.~J.}\ \bibnamefont
  {Nelson}}, \bibinfo {author} {\bibfnamefont {E.}~\bibnamefont {Jones}},
  \bibinfo {author} {\bibfnamefont {R.}~\bibnamefont {Kern}}, \bibinfo {author}
  {\bibfnamefont {E.}~\bibnamefont {Larson}}, \bibinfo {author} {\bibfnamefont
  {C.~J.}\ \bibnamefont {Carey}}, \bibinfo {author} {\bibfnamefont
  {{\.I}.}~\bibnamefont {Polat}}, \bibinfo {author} {\bibfnamefont
  {Y.}~\bibnamefont {Feng}}, \bibinfo {author} {\bibfnamefont {E.~W.}\
  \bibnamefont {Moore}}, \bibinfo {author} {\bibfnamefont {J.}~\bibnamefont
  {VanderPlas}}, \bibinfo {author} {\bibfnamefont {D.}~\bibnamefont {Laxalde}},
  \bibinfo {author} {\bibfnamefont {J.}~\bibnamefont {Perktold}}, \bibinfo
  {author} {\bibfnamefont {R.}~\bibnamefont {Cimrman}}, \bibinfo {author}
  {\bibfnamefont {I.}~\bibnamefont {Henriksen}}, \bibinfo {author}
  {\bibfnamefont {E.~A.}\ \bibnamefont {Quintero}}, \bibinfo {author}
  {\bibfnamefont {C.~R.}\ \bibnamefont {Harris}}, \bibinfo {author}
  {\bibfnamefont {A.~M.}\ \bibnamefont {Archibald}}, \bibinfo {author}
  {\bibfnamefont {A.~H.}\ \bibnamefont {Ribeiro}}, \bibinfo {author}
  {\bibfnamefont {F.}~\bibnamefont {Pedregosa}}, \bibinfo {author}
  {\bibfnamefont {P.}~\bibnamefont {van Mulbregt}}, \bibinfo {author}
  {\bibnamefont {{SciPy 1.0 Contributors}}}, \bibinfo {author} {\bibfnamefont
  {A.}~\bibnamefont {Vijaykumar}}, \bibinfo {author} {\bibfnamefont {A.~P.}\
  \bibnamefont {Bardelli}}, \bibinfo {author} {\bibfnamefont {A.}~\bibnamefont
  {Rothberg}}, \bibinfo {author} {\bibfnamefont {A.}~\bibnamefont {Hilboll}},
  \bibinfo {author} {\bibfnamefont {A.}~\bibnamefont {Kloeckner}}, \bibinfo
  {author} {\bibfnamefont {A.}~\bibnamefont {Scopatz}}, \bibinfo {author}
  {\bibfnamefont {A.}~\bibnamefont {Lee}}, \bibinfo {author} {\bibfnamefont
  {A.}~\bibnamefont {Rokem}}, \bibinfo {author} {\bibfnamefont {C.~N.}\
  \bibnamefont {Woods}}, \bibinfo {author} {\bibfnamefont {C.}~\bibnamefont
  {Fulton}}, \bibinfo {author} {\bibfnamefont {C.}~\bibnamefont {Masson}},
  \bibinfo {author} {\bibfnamefont {C.}~\bibnamefont {H{\"a}ggstr{\"o}m}},
  \bibinfo {author} {\bibfnamefont {C.}~\bibnamefont {Fitzgerald}}, \bibinfo
  {author} {\bibfnamefont {D.~A.}\ \bibnamefont {Nicholson}}, \bibinfo {author}
  {\bibfnamefont {D.~R.}\ \bibnamefont {Hagen}}, \bibinfo {author}
  {\bibfnamefont {D.~V.}\ \bibnamefont {Pasechnik}}, \bibinfo {author}
  {\bibfnamefont {E.}~\bibnamefont {Olivetti}}, \bibinfo {author}
  {\bibfnamefont {E.}~\bibnamefont {Martin}}, \bibinfo {author} {\bibfnamefont
  {E.}~\bibnamefont {Wieser}}, \bibinfo {author} {\bibfnamefont
  {F.}~\bibnamefont {Silva}}, \bibinfo {author} {\bibfnamefont
  {F.}~\bibnamefont {Lenders}}, \bibinfo {author} {\bibfnamefont
  {F.}~\bibnamefont {Wilhelm}}, \bibinfo {author} {\bibfnamefont
  {G.}~\bibnamefont {Young}}, \bibinfo {author} {\bibfnamefont {G.~A.}\
  \bibnamefont {Price}}, \bibinfo {author} {\bibfnamefont {G.-L.}\ \bibnamefont
  {Ingold}}, \bibinfo {author} {\bibfnamefont {G.~E.}\ \bibnamefont {Allen}},
  \bibinfo {author} {\bibfnamefont {G.~R.}\ \bibnamefont {Lee}}, \bibinfo
  {author} {\bibfnamefont {H.}~\bibnamefont {Audren}}, \bibinfo {author}
  {\bibfnamefont {I.}~\bibnamefont {Probst}}, \bibinfo {author} {\bibfnamefont
  {J.~P.}\ \bibnamefont {Dietrich}}, \bibinfo {author} {\bibfnamefont
  {J.}~\bibnamefont {Silterra}}, \bibinfo {author} {\bibfnamefont {J.~T.}\
  \bibnamefont {Webber}}, \bibinfo {author} {\bibfnamefont {J.}~\bibnamefont
  {Slavi{\v c}}}, \bibinfo {author} {\bibfnamefont {J.}~\bibnamefont
  {Nothman}}, \bibinfo {author} {\bibfnamefont {J.}~\bibnamefont {Buchner}},
  \bibinfo {author} {\bibfnamefont {J.}~\bibnamefont {Kulick}}, \bibinfo
  {author} {\bibfnamefont {J.~L.}\ \bibnamefont {Sch{\"o}nberger}}, \bibinfo
  {author} {\bibfnamefont {J.~V.}\ \bibnamefont {de~Miranda~Cardoso}}, \bibinfo
  {author} {\bibfnamefont {J.}~\bibnamefont {Reimer}}, \bibinfo {author}
  {\bibfnamefont {J.}~\bibnamefont {Harrington}}, \bibinfo {author}
  {\bibfnamefont {J.~L.~C.}\ \bibnamefont {Rodr{\'i}guez}}, \bibinfo {author}
  {\bibfnamefont {J.}~\bibnamefont {Nunez-Iglesias}}, \bibinfo {author}
  {\bibfnamefont {J.}~\bibnamefont {Kuczynski}}, \bibinfo {author}
  {\bibfnamefont {K.}~\bibnamefont {Tritz}}, \bibinfo {author} {\bibfnamefont
  {M.}~\bibnamefont {Thoma}}, \bibinfo {author} {\bibfnamefont
  {M.}~\bibnamefont {Newville}}, \bibinfo {author} {\bibfnamefont
  {M.}~\bibnamefont {K{\"u}mmerer}}, \bibinfo {author} {\bibfnamefont
  {M.}~\bibnamefont {Bolingbroke}}, \bibinfo {author} {\bibfnamefont
  {M.}~\bibnamefont {Tartre}}, \bibinfo {author} {\bibfnamefont
  {M.}~\bibnamefont {Pak}}, \bibinfo {author} {\bibfnamefont {N.~J.}\
  \bibnamefont {Smith}}, \bibinfo {author} {\bibfnamefont {N.}~\bibnamefont
  {Nowaczyk}}, \bibinfo {author} {\bibfnamefont {N.}~\bibnamefont {Shebanov}},
  \bibinfo {author} {\bibfnamefont {O.}~\bibnamefont {Pavlyk}}, \bibinfo
  {author} {\bibfnamefont {P.~A.}\ \bibnamefont {Brodtkorb}}, \bibinfo {author}
  {\bibfnamefont {P.}~\bibnamefont {Lee}}, \bibinfo {author} {\bibfnamefont
  {R.~T.}\ \bibnamefont {McGibbon}}, \bibinfo {author} {\bibfnamefont
  {R.}~\bibnamefont {Feldbauer}}, \bibinfo {author} {\bibfnamefont
  {S.}~\bibnamefont {Lewis}}, \bibinfo {author} {\bibfnamefont
  {S.}~\bibnamefont {Tygier}}, \bibinfo {author} {\bibfnamefont
  {S.}~\bibnamefont {Sievert}}, \bibinfo {author} {\bibfnamefont
  {S.}~\bibnamefont {Vigna}}, \bibinfo {author} {\bibfnamefont
  {S.}~\bibnamefont {Peterson}}, \bibinfo {author} {\bibfnamefont
  {S.}~\bibnamefont {More}}, \bibinfo {author} {\bibfnamefont {T.}~\bibnamefont
  {Pudlik}}, \bibinfo {author} {\bibfnamefont {T.}~\bibnamefont {Oshima}},
  \bibinfo {author} {\bibfnamefont {T.~J.}\ \bibnamefont {Pingel}}, \bibinfo
  {author} {\bibfnamefont {T.~P.}\ \bibnamefont {Robitaille}}, \bibinfo
  {author} {\bibfnamefont {T.}~\bibnamefont {Spura}}, \bibinfo {author}
  {\bibfnamefont {T.~R.}\ \bibnamefont {Jones}}, \bibinfo {author}
  {\bibfnamefont {T.}~\bibnamefont {Cera}}, \bibinfo {author} {\bibfnamefont
  {T.}~\bibnamefont {Leslie}}, \bibinfo {author} {\bibfnamefont
  {T.}~\bibnamefont {Zito}}, \bibinfo {author} {\bibfnamefont {T.}~\bibnamefont
  {Krauss}}, \bibinfo {author} {\bibfnamefont {U.}~\bibnamefont {Upadhyay}},
  \bibinfo {author} {\bibfnamefont {Y.~O.}\ \bibnamefont {Halchenko}}, \ and\
  \bibinfo {author} {\bibfnamefont {Y.}~\bibnamefont {V{\'a}zquez-Baeza}},\
  }\href {\doibase 10.1038/s41592-019-0686-2} {\bibfield  {journal} {\bibinfo
  {journal} {Nat. Methods.}\ }\textbf {\bibinfo {volume} {17}},\ \bibinfo
  {pages} {261} (\bibinfo {year} {2020})}\BibitemShut {NoStop}%
\bibitem [{\citenamefont {Byrd}\ \emph {et~al.}(1995)\citenamefont {Byrd},
  \citenamefont {Lu}, \citenamefont {Nocedal},\ and\ \citenamefont
  {Zhu}}]{byrd_limited_1995}%
  \BibitemOpen
  \bibfield  {author} {\bibinfo {author} {\bibfnamefont {R.~H.}\ \bibnamefont
  {Byrd}}, \bibinfo {author} {\bibfnamefont {P.}~\bibnamefont {Lu}}, \bibinfo
  {author} {\bibfnamefont {J.}~\bibnamefont {Nocedal}}, \ and\ \bibinfo
  {author} {\bibfnamefont {C.}~\bibnamefont {Zhu}},\ }\href {\doibase
  10.1137/0916069} {\bibfield  {journal} {\bibinfo  {journal} {SIAM J. Sci.
  Comput.}\ }\textbf {\bibinfo {volume} {16}},\ \bibinfo {pages} {1190}
  (\bibinfo {year} {1995})}\BibitemShut {NoStop}%
\end{thebibliography}%

\end{document}